\documentclass[acmsmall]{acmart}

\def\BibTeX{{\rm B\kern-.05em{\sc i\kern-.025em b}\kern-.08emT\kern-.1667em\lower.7ex\hbox{E}\kern-.125emX}}
    
\setcopyright{none}

\usepackage{amssymb}
\usepackage{xcolor}
\usepackage{wrapfig}

\newcommand{\eg}{\emph{e.g.}} 
\newcommand{\ie}{\emph{i.e.}} 
\newcommand{\etc}{\emph{etc.}} 
\newcommand{\etal}{\emph{et al.}} 
\newcommand{\ala}{\emph{\`{a} la}} 
\newcommand{\apriori}{\emph{a priori}} 

\renewcommand{\eqref}[1]{Eqn.~\ref{eq:#1}}

\newcommand{\secref}[1]{Sec.~\ref{sec:#1}}

\newcommand{\figref}[1]{Fig.~\ref{fig:#1}}

\newcommand{\figloc}[1]{\emph{#1}}

\begin{document}

%
\title{A Survey of Algorithms for Geodesic Paths and Distances}

%
\author{Keenan Crane}
\affiliation{%
  \institution{Carnegie Mellon University}
  \city{Pittsburgh}
  \country{USA}}
\email{kmcrane@cs.cmu.edu}

\author{Marco Livesu}
\affiliation{%
  \institution{CNR IMATI}
  \city{Genoa}
  \country{Italy}}
\email{marco.livesu@gmail.com}

\author{Enrico Puppo}
\affiliation{%
  \institution{University of Genoa}
  \city{Genoa}
  \country{Italy}
}

\author{Yipeng Qin}
\affiliation{%
 \institution{Cardiff University}
 \city{Cardiff}
 \country{UK}}
 
%
\renewcommand{\shortauthors}{Crane, et al.}

%
\begin{abstract}
Numerical computation of shortest paths or \emph{geodesics} on curved domains, as well as the associated \emph{geodesic distance}, arises in a broad range of applications across digital geometry processing, scientific computing, computer graphics, and computer vision.  Relative to Euclidean distance computation, these tasks are complicated by the influence of curvature on the behavior of shortest paths, as well as the fact that the representation of the domain may itself be approximate.  In spite of the difficulty of this problem, recent literature has developed a wide variety of sophisticated methods that enable rapid queries of geodesic information, even on relatively large models.  This survey reviews the major categories of approaches to the computation of geodesic paths and distances, highlighting common themes and opportunities for future improvement.
\end{abstract}

%
%


%

%

%
\maketitle



\newcommand{\cino}     			[1]{{\color{magenta}	Cino: #1}}
\newcommand{\enrico}			[1]{{\color{orange}		Enrico: #1}}
\newcommand{\keenan}			[1]{{\color{teal}			Keenan: #1}}
\newcommand{\yipeng}		    [1]{{\color{cyan}			Yipeng: #1}}

\newcommand{\com} 				[1]{{}} 
\newcommand{\edit}				[1]{{\color{red}			#1}} 

\newcommand{\T}{\mathsf{T}} 
\newcommand{\Asf}{\mathsf{A}}
\newcommand{\Bsf}{\mathsf{B}}
\newcommand{\Csf}{\mathsf{C}}
\newcommand{\Dsf}{\mathsf{D}}
\newcommand{\Esf}{\mathsf{E}}
\newcommand{\Fsf}{\mathsf{F}}
\newcommand{\Gsf}{\mathsf{G}}
\newcommand{\Hsf}{\mathsf{H}}
\newcommand{\Isf}{\mathsf{I}}
\newcommand{\Jsf}{\mathsf{J}}
\newcommand{\Ksf}{\mathsf{K}}
\newcommand{\Lsf}{\mathsf{L}}
\newcommand{\Msf}{\mathsf{M}}
\newcommand{\Nsf}{\mathsf{N}}
\newcommand{\Osf}{\mathsf{O}}
\newcommand{\Psf}{\mathsf{P}}
\newcommand{\Qsf}{\mathsf{Q}}
\newcommand{\Rsf}{\mathsf{R}}
\newcommand{\Ssf}{\mathsf{S}}
\newcommand{\Tsf}{\mathsf{T}}
\newcommand{\Usf}{\mathsf{U}}
\newcommand{\Vsf}{\mathsf{V}}
\newcommand{\Wsf}{\mathsf{W}}
\newcommand{\Xsf}{\mathsf{X}}
\newcommand{\Ysf}{\mathsf{Y}}
\newcommand{\Zsf}{\mathsf{Z}}
\newcommand{\asf}{\mathsf{a}}
\newcommand{\bsf}{\mathsf{b}}
\newcommand{\csf}{\mathsf{c}}
\newcommand{\dsf}{\mathsf{d}}
\newcommand{\esf}{\mathsf{e}}
\newcommand{\fsf}{\mathsf{f}}
\newcommand{\gsf}{\mathsf{g}}
\newcommand{\hsf}{\mathsf{h}}
\newcommand{\isf}{\mathsf{i}}
\newcommand{\jsf}{\mathsf{j}}
\newcommand{\ksf}{\mathsf{k}}
\newcommand{\lsf}{\mathsf{l}}
\newcommand{\msf}{\mathsf{m}}
\newcommand{\nsf}{\mathsf{n}}
\newcommand{\osf}{\mathsf{o}}
\newcommand{\psf}{\mathsf{p}}
\newcommand{\qsf}{\mathsf{q}}
\newcommand{\rsf}{\mathsf{r}}
\newcommand{\ssf}{\mathsf{s}}
\newcommand{\tsf}{\mathsf{t}}
\newcommand{\usf}{\mathsf{u}}
\newcommand{\vsf}{\mathsf{v}}
\newcommand{\wsf}{\mathsf{w}}
\newcommand{\xsf}{\mathsf{x}}
\newcommand{\ysf}{\mathsf{y}}
\newcommand{\zsf}{\mathsf{z}}


\section{Introduction}

The ability to rapidly compute shortest paths and/or distances between pairs or sets of points is a fundamental operation in computational science.  This survey considers algorithms for computing \emph{geodesic} paths and distances, \ie, paths through curved 
domains that arise in a broad range of tasks across digital geometry processing, scientific computing, computer graphics, and computer vision.  Relative to computing distance on Euclidean domains, this problem is complicated by the influence of curvature, 
 as well as the fact that the domain itself may not be known exactly.  

Several problems arise in this context, which have different applications, and are best tackled with different techniques: 
from tracing a geodesic line given a starting point and a direction; to computing the shortest path between a pair of points; to computing the distance function over a whole surface, with respect to a source point or set; to provide a framework to rapidly extract the shortest path between any pair of points.  Broadly speaking, there are two major classes of methods: those rooted in computational geometry, which view a polyhedral surface as an exact description of the geometry, and those rooted in scientific computing, which view a polyhedron as an approximation of a smooth surface.  Neither class of approaches is universally ``better'': each may be best-suited to a particular task (\eg{}, finding geodesics on a CAD model, versus approximating geodesic distance on a scanned surface), and in a particular setting, according to a wide variety of trade offs in terms of accuracy, storage cost, run time performance, and scalability.

In general one might wish to compute geodesics and geodesic distances on many different data structures (point clouds, voxelizations, \etc{}), though in this survey we focus primarily on \emph{polyhedral surfaces}, represented as triangle meshes.  The survey is organized according to the different geodesic distance problems and the attendant classes of approaches to their solution (see Table \ref{tab:summary} for a summary).  After introducing the basic problems in \secref{problems} and providing the necessary background in \secref{background}, we review major classes of algorithms in Secs. \ref{sec:fem}, \ref{sec:cg} and \ref{sec:tracing}.  In \secref{meshing}, we examine the relationship between mesh quality and geodesic computation.  In \secref{eval}, we provide a partial evaluation of the reviewed methods, on the basis of available implementations.  Finally, in \secref{conc}, we draw some conclusions and we highlight common themes and opportunities for future improvement in geodesic algorithms.

\textbf{Previous surveys.}  Our focus in this survey is on practical algorithms and their behavior on real-world datasets.
A relatively recent survey by Bose et al.\ \cite{Bose:2011:SGP} deals primarily with theoretical aspects of one particular problem (the polyhedral shortest path problem), with a focus on asymptotic time complexity and approximation bounds, as well as special cases such as convex polyhedra.  Less attention is devoted to broader problems (such as geodesic distance transforms) or issues such as real-world performance, mesh quality, \etc{}; in this respect, such a survey is complementary to ours.
Patan\'e \cite{Patane:2016fr} provides a detailed survey on the specific topic of Laplacian spectral kernels and distances, which is likewise complementary to our account of PDE-based methods.  Peyr{\'e} and colleagues~\cite{peyre2009geodesic,peyre2010geodesic} provide a nice overview of several aspects of PDE-based methods, though the last few years have seen major advancements in PDE-based methods, which we discuss in \secref{fem}. 

\subsection{Geodesic Queries}
\label{sec:problems}
Tasks in geometry processing may require a variety of different queries about geodesics and geodesic distances; though seemingly similar, these queries must be answered via very different algorithms.  We review several important cases here.

\noindent\textbf{Initial Value Problems.} Perhaps the simplest type of problem is the \emph{initial value problem}, which traces out a geodesic starting at a given point in a given direction.  In the Euclidean case, the solution is simply a straight ray through space,  though in general one must of course account for the influence of curvature; in the discrete setting one also encounters some difficulty in defining which ray is most ``straight'' (see Section \ref{sec:polyhedral}).  Computationally this problem, which we will refer to as \emph{Geodesic Tracing (GT)}, tends to be among the easiest of geodesic queries; we examine it in detail in \secref{tracing}.

\noindent\textbf{Boundary Value Problems.} In \emph{boundary value problems}, a set of points on the domain is given as input, and one seeks (for instance) geodesic paths between these points, or geodesic distances to all these points.  This type of problem is generally not as easy as an initial value problem since, for instance, one cannot simply shoot a geodesic ray from one point and hope to hit a particular target point.  We consider in particular the following problems:
\begin{itemize}
   \item Point-to-Point Geodesic Path (PPGP): given two distinct points \(p,q\), find any shortest geodesic \(\gamma\) between them. (Note that in general the solution may not be unique---see \secref{background}.)
   \item Single Source Geodesic Distance / Shortest Path (SSGD/SSSP): given a source point \(p\), compute a scalar function \(\varphi(q)\) that provides the length of the shortest path from \(p\) to each point \(q\).  In SSGD, only the distance function is provided; in SSSP, the shortest paths are also explicitly computed.
   \item Multiple Source Geodesic Distance / Shortest Path (MSGD/MSSP): in this case the source is no longer just a single point \(p\), but rather a collection \(S\) of points, curves, \etc.  The result is a function \(\phi(p)\) which for each point \(p\) gives the distance to the \emph{closest} point in the set; this query is also sometimes called a \emph{distance transform}.  In MSSP, paths are also provided.
   \item All-Pairs Geodesic Distances / Shortest Paths (APGD/APSP): given a source set \(S\), compute the shortest distance/paths between all pairs of points in \(S\).
\end{itemize}
In Sections \ref{sec:fem} and \ref{sec:cg}, we review the methods for solving these problems, categorized according to their basic algorithmic approach they take.  Most methods that solve the SSGD/SSSP problems also provide a solution to PPGP as a byproduct; several easily generalize to MSGD/MSSP; only few methods address the APGD/APSP problems.  In Section \ref{sec:local}, we review some algorithms that are specifically developed to solve PPGP. 

Note that most methods in the literature assume that initial points or boundary conditions are specified at vertices of an input triangulation, though some methods allow direct evaluation of geodesic queries at arbitrary points of the domain.  For algorithms that only support queries at vertices, a pragmatic approach is to locally refine the mesh by splitting faces or edges at the query point.



\subsection{Polyhedral vs. Smooth Geodesic Problem}
\label{sec:PolyhedralVsSmooth}

When thinking about algorithms for computing geodesics, it is important to consider what our domain represents: does it exactly represent the geometry of interest, or is it merely an approximation of the true domain?  The answer to this question depends on context.  For instance, in problems like animation or CAD/CAM, where surfaces are designed by artists or engineers, we may have an exact boundary representation and want to compute exact paths along this surface.  On the other hand, when a mesh is obtained via, say, 3D scanning or numerical simulation, it can only provide an approximation of the true domain, \ie, the real physical surface of interest.  In this case, the accuracy of geodesic computation is fundamentally limited by the accuracy of the domain representation itself.  The choice of application therefore influences how we talk about error, and also which algorithms are best suited to a given task.

\setlength{\columnsep}{1em}
\setlength{\intextsep}{.5em}
\begin{wrapfigure}{r}{71px}
    \includegraphics[width=\linewidth]{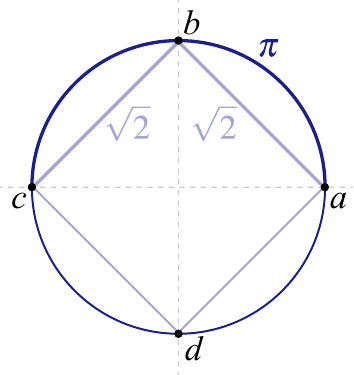}
\end{wrapfigure}
In the context of polyhedral surfaces, a common misconception is that the ``correct'' answer is obtained by considering piecewise linear paths through the domain -- though of course these paths may only be approximations of true (smooth) geodesic curves.  A simple 1D example is illustrated in the inset.  What is the ``correct'' distance between the point \(a\) and the point \(c\)?  If we view the polygon as an exact representation of the geometry (\ie, if we wish to compute distance along the square itself), then the geodesic distance is obtained by summing the two edge lengths \(|ab| + |bc| = 2\sqrt{2} \approx 2.82\).  On the other hand, if we view the polygon as an approximation of the smooth circle, the distance should just be the arc length between \(a\) and \(c\), \ie, \(\pi \approx 3.14\).  Not surprisingly, the polygonal geodesic distance is an underestimate of the smooth geodesic distance, since each straight segment takes a ``short cut'' from point to point.  This same phenomenon occurs on triangulated surfaces: the polyhedral geodesic distance generally underestimates the smooth geodesic distance (for instance, the distance along a cube provides a poor approximation of the distance along the circumscribed sphere).  
To make the distinction clear in this survey, we therefore distinguish between two versions of the geodesic problem:

\begin{itemize}
   \item \textbf{Polyhedral Geodesic Problem} --- view a discrete surface as an exact description of the geometry, and aim to compute exact geodesics or exact distances along this polyhedral surface.
   \item \textbf{Smooth Geodesic Problem} --- view a discrete surface as an approximation of the true geometry, and aim to compute the most accurate possible approximation of geodesics or geodesic distances.
\end{itemize}

\begin{figure}
   \centering
   \includegraphics{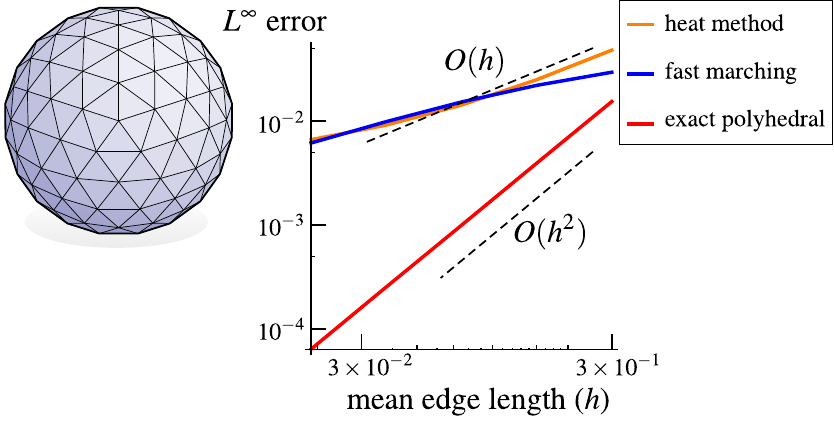}
   \caption{When a polyhedral surface is an approximation of a smooth surface, the ``exact'' polyhedral distance still does not recover the true distance of the smooth surface.  For instance, even for very nice triangulations of the smooth sphere, algorithms from computational geometry improve accuracy by only one order relative to PDE-based methods (from linear to quadratic).\label{fig:ConvergenceRate}}
\end{figure}

The polyhedral problem captures the standard viewpoint of computational geometry; the smooth problem encapsulates the standard viewpoint of scientific computing.  These two problems interact of course: for instance, one can use the exact polyhedral distance to approximate the true smooth distance.  Surprisingly enough, however, the best strategy for approximating the smooth geodesic distance may \emph{not} be to simply compute the exact polyhedral distance.  For instance, even on a nicely triangulated sphere, the polyhedral distance gives only an \(O(h^2)\) approximation of the smooth distance (\figref{ConvergenceRate}); greater accuracy can be achieved by considering the distance along a spline or subdivision surface that approximates the same sampling of the domain~\cite{DeGoes:2016im,Nguyen:2016:CFE}.  The fact that one can do better than the exact polyhedral distance should not come as a surprise: higher-order geometric elements (\eg, spline or subdivision patches) use a much larger stencil of information than individual triangles, which correspond to linear elements (consider, for instance, approximating the circle by four B\'{e}zier curves rather than four straight segments).  On the other hand, higher-order interpolation provides no benefit for the polyhedral geodesic problem, where one is interested in the exact distance along a particular piecewise linear surface.

\section{Background}
\label{sec:background}

In this section we provide some basic concepts and definitions that will facilitate our discussion of algorithms---for a more thorough introduction, see \cite{do1976differential} in the smooth setting (especially Chapter 4, Sections 4--4, 4--6, and 4--7), and \cite{Crane:2013:DGP} for the discrete setting.  At a high level, geodesics can be characterized as curves that are in some sense \emph{straightest} and \emph{(locally) shortest}, though one must be careful about the relationship between these two characterizations (\secref{global}), especially on polyhedra where they may lead to different discrete definitions~\cite{polthier1998straightest}.  We first give some standard background in the smooth setting (\secref{smooth_theory}), followed by a discussion of geodesics on polyhedral surfaces (\secref{polyhedral}).

\subsection{Smooth Setting}
\label{sec:smooth_theory}

\begin{figure}
   \centering
   \includegraphics{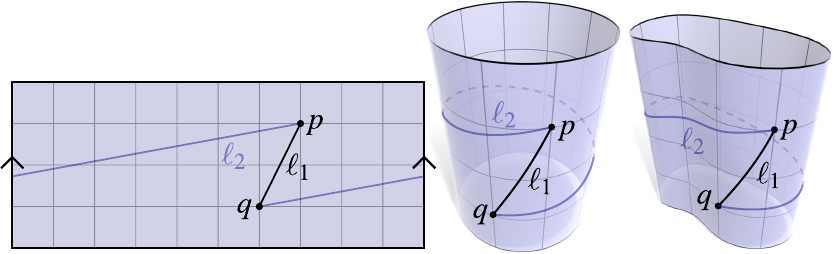}
   \caption{A geodesic is any ``straight'' curve between two points \(p\) and \(q\), and are a feature of the intrinsic geometry of the surface: they do not change if we bend the surface, so long as there is no stretching, shearing, or ripping.  Though the shortest path will always be a geodesic (here, \(\ell_1\)), a geodesic is not necessarily the shortest path (consider \(\ell_2\)).\label{fig:TubeGeodesics}}
\end{figure}

\textbf{Intrinsic vs. Extrinsic Geometry.} When studying geodesics, how should we describe the shape of the domain?  An important feature of geodesics is that they depend only on distances \emph{along} the surface, and not at all how the surface sits in space.  As a concrete example, consider a piece of paper rolled up into a tube (\figref{TubeGeodesics}): a straight line \(\ell_1\) drawn between two nearby points \(p\) and \(q\) on the flat piece of paper is still a shortest curve on the tube (see \figref{TubeGeodesics}).  One can find an even shorter path by drawing a straight line through space, but this line is no longer a path \emph{along} the surface.  Likewise, there are many other ways we could bend the tube without changing the shape or length of shortest paths---for this reason, when studying geodesics we really want to ignore the \emph{extrinsic} geometry of the surface (how it sits in space), and focus purely on its \emph{intrinsic} geometry (only those things that can be measured by walking along the surface).  The desire to compute shortest intrinsic (rather than extrinsic) paths is of course quite natural: for instance, the ``shortest'' route between two cities is the one that best avoids mountains and valleys---not the one that tunnels straight through the Earth!

\textbf{Shortest vs. Straightest Curves.} The tube example also helps to illustrate another feature of geodesics: shortest implies straight, but straight does not necessarily imply shortest.  Consider, for instance, a long straight line \(\ell_2\) from \(p\) that ``goes off the edge of the map'' before returning to \(q\).  This curve is still a geodesic, but it is not a \emph{shortest geodesic}---in particular, it is not as short as \(\ell_1\).  Finally, the word ``shortest'' does not imply that there is a \emph{unique} geodesic of minimal length.  For instance, there may be two equally short ways to go around a hill; in fact, there may be \emph{infinitely many} shortest paths between two given points, such as the north and south pole of a sphere.

Unfortunately, finding geodesics is not as simple as just ``unrolling'' a smooth surface into the plane and finding straight lines (as we did with the tube), since most surfaces cannot be flattened without distorting distances.  Consider for instance the many different map projections used by cartographers (Mercator, Robinson, \etc{}), \emph{none} of which preserve geodesic distances.  However, we can nonetheless reason about shortness and straightness of curves: for instance, a curve on an embedded surface is ``straight'' if there is no acceleration as we travel along the curve, except in the normal direction---in other words, if we only turn by the bare minimum amount needed to remain on the surface.  Alternatively, we can adopt the perspective of \emph{Riemannian geometry}, which allows one to reason about intrinsic geometry without thinking about how it is embedded in space.

\subsubsection{Smooth Surfaces}
\label{sec:SmoothSurfaces}

In the intrinsic picture, our main object of study is a \emph{smooth surface} \(M\) with \emph{Riemannian metric} \(g\).  The fact that \(M\) is \emph{smooth} implies that at each point \(p \in M\) we have a \emph{tangent space} \(T_p M\), where each \emph{tangent vector} \(X \in T_p M\) specifies a vector pointing ``along'' the surface, \ie, all possible directions we can travel away from \(p\).  Since we have no information about how the surface sits in space, the one and only way to measure lengths and angles of tangent vectors is via the Riemannian metric, which for tangent space \(T_p M\) provides a positive-definite inner product \(g_p: T_p M \times T_p M \to \mathbb{R}_{\geq 0}\).  For instance, the length of any tangent vector \(X \in T_p M\) is given by \(|X| := \sqrt{g_p(X,X)}\); the angle between any two tangent vectors \(X,Y \in T_p M\) is given by \(\arccos(g_p(X,Y)/|X||Y|)\), just as in ordinary Euclidean \(\mathbb{R}^n\).  In fact, whenever the surface \(M\) \emph{is} sitting in space, the Riemannian metric and Euclidean inner product coincide.  When the meaning is understood from context, or when working with a \emph{vector field} \(X\) (\ie, a choice of vector in each tangent space), we can drop the subscript \(p\).

\subsubsection{Shortest Curves}
\label{sec:ShortestCurves}

The Riemannian metric enables us to easily define geodesic curves in terms of conditions on length.  In particular, consider any differentiable map \(\gamma\) from an interval \([a,b]\) of the real line into the surface \(M\), and let \(\dot{\gamma}(t) := \tfrac{d}{dt}\gamma(t)\); we say that \(\gamma\) is \emph{arc-length parameterized} if \(|\dot{\gamma}(t)| = 1\) for all \(t \in [a,b]\).  We can express the length of any curve \(\gamma\) as
\[
   L(\gamma) := \int_a^b |\dot{\gamma}(t)|\ dt = \int_a^b \sqrt{g_{\gamma(t)}\left( \dot{\gamma}(t), \dot{\gamma}(t) \right)}\ dt.
\]
The \emph{geodesic distance} \(\phi(p,q)\) between any two points \(p, q \in M\) is the infimum of length over all curves \(\gamma\) such that \(\gamma(a) = p\) and \(\gamma(b) = q\).  An arc-length parameterized curve \(\gamma\) is a \emph{shortest geodesic} if it achieves this infimal length.  An arc-length parameterized curve \(\gamma\) is a \emph{geodesic} if it is ``locally shortest,'' \ie, if there is some \(\delta > 0\) such that \(\gamma\) is a shortest geodesic when restricted to any interval \([t_1,t_2] \subset [a,b]\) such that \(t_2 - t_1 \leq \delta\).  Intuitively, a shortest geodesic is a curve that cannot be made any shorter (without moving its endpoints); \emph{a geodesic is a curve that cannot be made shorter by adjusting any small piece of it.}

\subsubsection{Straightest Curves}
\label{sec:StraightestCurves}

We can also characterize geodesics in terms of \emph{straightness} rather than \emph{shortness}.  This perspective is best understood from a dynamic point of view: imagine we are driving along a road at constant speed, and never turn our wheel left or right.  Although we may encounter hills and valleys along the way, our trajectory is in some sense as straight as it possibly could be.  In the extrinsic setting the only acceleration we experience is in the direction \emph{normal} to the surface, \ie, the minimal possible acceleration needed to keep our vehicle on the ground.  In the intrinsic setting, we do not need to worry about this normal acceleration since our surface is not sitting in space---instead, we just say that \emph{a geodesic is a curve of zero acceleration.}

How can we measure the acceleration of a curve \(\gamma(t)\) in a surface \(M\)?  In the Euclidean plane, we can just take the second derivative of \(\gamma\) with respect to \(t\).  In the Riemannian setting, however, the velocity vectors \(\gamma^\prime(t_1)\) and \(\gamma^\prime(t_2)\) at two different times \(t_1, t_2\) will in general belong to two different tangent spaces \(T_{\gamma(t_1)} M, T_{\gamma(t_2)}M\), and hence cannot be compared directly.  Instead, we can use the \emph{Levi-Civita connection} or \emph{covariant derivative} \(\nabla_X Y\), which (intuitively) uses the metric \(g\) to measure how much \(Y\) is turning as we move along the direction \(X\).  In particular, a curve \(\gamma\) is a \emph{geodesic} if its tangent \(\dot{\gamma}\) exhibits zero turning as we move along the tangent direction, \ie, if
\[
   \nabla_{\dot{\gamma}(t)} \dot{\gamma}(t) = 0
\]
at all times \(t \in [a,b]\).  (Since the covariant derivative is local, it does not help us define \emph{shortest} geodesics.)  More generally, for any arc-length parameterized curve \(\gamma(s)\) we have
\[
   \nabla_{\dot{\gamma}(s)} \dot{\gamma}(s) = \kappa_g(s) n(s),
\]
where for each time \(t\), the vector \(n(t) \in T_{\gamma(t)} M\) is the \emph{unit normal} to the curve (\ie, a unit vector orthogonal to the tangent), and the scalar \(\kappa_g(t) \in \mathbb{R}\) is the \emph{geodesic curvature}, which measures the rate at which the curve is bending.  A geodesic is then \emph{a curve with zero geodesic curvature}.

\begin{figure}[tb]
	\centering
        \includegraphics{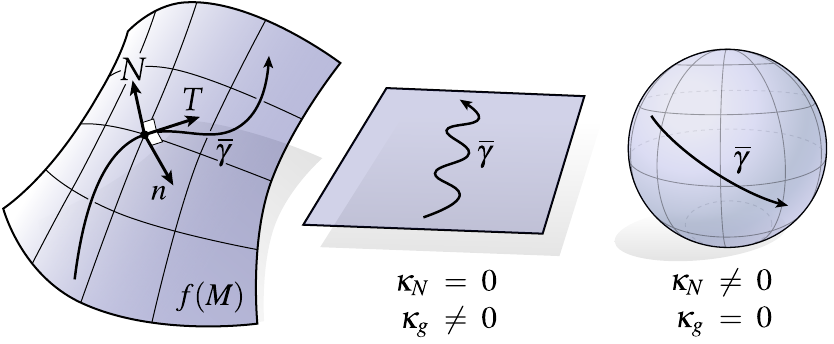}
        \caption{Extrinsically, the curvature of a curve \(\bar{\gamma}\) can be decomposed into the normal and geodesic curvatures \(\kappa_N, \kappa_g\), which measure the change in the tangent in the direction of the surface normal \(N\) and the curve normal \(n\), respectively (left).  A ``wiggly'' but planar curve will have zero normal curvature and nonzero geodesic curvature (center), whereas a great arc on a sphere has zero geodesic curvature but nonzero normal curvature (right).\label{fig:GeodesicVsNormalCurvature}}
\end{figure}

We can also understand geodesic curvature from an extrinsic point of view.  Suppose we have a map \(f: M \to \mathbb{R}^3\) assigning each point of the surface \(M\) a location in space, and let \(N: M \to S^2 \subset \mathbb{R}^3\) be the corresponding \emph{Gauss map} giving the unit normal at each point.  A curve \(\gamma: [a,b] \to M\) on the surface can now be realized as a curve \(\bar{\gamma} := f \circ \gamma\) in Euclidean space.  Assuming \(\bar{\gamma}\) is parameterized by its arc-length \(s\), its unit tangent is given by \(\bar{T}(s) := \tfrac{d}{ds}\bar{\gamma}(s)\); we will use \(n := T \times N\) to denote the normal to the curve (as opposed to the normal \(N\) of the surface).  The derivative of \(T\) in turn describes the curvature of the curve, which can be decomposed into two scalar components:
\[
   \begin{array}{rcl}
      \kappa_g &:=& \langle n, \tfrac{d}{ds} T \rangle, \\
      \kappa_N &:=& \langle N, \tfrac{d}{ds} T \rangle. \\
   \end{array}
\]
Note that there is no derivative in the \(T\) direction, since \(\bar{\gamma}\) is arc-length parameterized and hence its tangent doesn't change length. In other words, the geodesic curvature \(\kappa_g\) describes how much the curve \(\gamma\) is bending ``in plane''; the \emph{normal curvature} \(\kappa_N\) describes how much \(\gamma\) is bending ``out of plane.'' Geodesics are again curves of zero geodesic curvature, \ie, those that go straight \emph{along} the surface (and bend purely to remain \emph{on} the surface).  An illustration is provided in \figref{GeodesicVsNormalCurvature}.

\subsubsection{Exponential Map}
\label{sec:ExponentialMap}


At any point \(p \in M\) and any unit vector \(X \in T_p M\), there is a unique geodesic traveling away from \(p\) in the direction \(X\), \ie, such that \(\dot{\gamma} = X\) (this perspective leads to the tracing algorithms discussed in \secref{tracing}).  More generally, the \emph{exponential map} \(\exp_p: T_p M \to M\) takes any tangent vector \(X\) at \(p\) to the point \(q\) of \(M\) reached by walking in the direction \(X/|X|\) a distance \(|X|\).  In general, however, this map is not invertible: there can be distinct points \(q_1 \ne q_2\) that are reached by starting at \(p\) and walking along geodesics in different directions (or for different distances).  Any neighborhood \(U\) of \(p\) over which \(\exp_p\) is invertible is called a \emph{normal neighborhood} of \(p\); on any such neighborhood, the inverse of the exponential map defines the \emph{logarithmic map} \(\log_p: U \to T_p M\).  The logarithmic map provides local coordinates around \(p\) called \emph{normal coordinates}, by effectively ``flattening out'' a small patch of the surface onto its tangent space.  Polar coordinates on the tangent space are sometimes called \emph{geodesic polar coordinates}; here, geodesics can be characterized (locally) as lines of constant angle.  Circles in this tangent plane are mapped by the exponential map to geodesic circles, which orthogonally intersect geodesic rays from \(p\) (Gauss' lemma).

Away from a normal neighborhood, geodesics may start behaving in less intuitive and undesirable ways. 
We review some global aspects of geodesics in \secref{global}.

\subsubsection{Geodesic Distance}
\label{sec:GeodesicDistance}

Many of the algorithms discussed in this survey aim not to compute individual geodesic curves, but rather the \emph{geodesic distance function} \(\phi: M \times M \to \mathbb{R}\) over the entire surface.  In particular, for any two points \(x,y \in M\), \(\phi(x,y)\) is the smallest length of any geodesic between \(x\) and \(y\).  This function satisfies all the properties of a distance metric, \ie, nonnegativity (\(\phi(x,y) \geq 0\)), nondegeneracy (\(\phi(x,y) = 0 \iff x=y\)), symmetry (\(\phi(x,y) = \phi(y,x)\)) and triangle inequality (\(\phi(x,y) + \phi(y,z) \geq \phi(x,z)\)); these properties will be important to keep in mind when studying numerical approximations of geodesic distance.  A \emph{geodesic ball} \(B_p(r)\) is the set of all points \(q \in M\) such that \(\phi(p,q) < r\).

\paragraph*{Cut locus.} For a given point \(p\), the \emph{injectivity radius} is the radius \(r > 0\) of the largest geodesic ball \(B_p(r)\) such that there is a unique shortest path from any point \(q \in B_p(r)\) back to \(p\).  Outside this radius, there are points \(q\) with two or more shortest paths to \(p\).  The collection of all such points is called the \emph{cut locus}.  More generally, for any source set \(\Omega\) (\eg, a collection of isolated points, or a network of curves, surfaces, \etc{}) the cut locus is the set of points \(q\) for which there is not a unique shortest path to some point in \(\Omega\).  The geodesic distance function \(\phi(x,y)\) is smooth away from the cut locus.  In most situations, it is not particularly important that geodesic distance algorithms accurately reconstruct the cut locus: only that the distance values or paths computed at each point are accurate.  However, some applications (\eg, computing the \emph{medial axis} of a shape) may require an accurate cut locus. 

\subsection{Polyhedral Setting}
\label{sec:polyhedral}

Geodesics on polyhedral surfaces behave somewhat differently from geodesics on smooth surfaces---especially in the vicinity of vertices, where the surface fails to be smooth.  Methods coming from computational geometry (\secref{cg}) must therefore carefully consider the specific behavior of polyhedral geodesic paths; methods based on the finite element viewport (\secref{fem}) largely side-step these questions by approaching geodesic computation from the perspective of \emph{function approximation} rather than explicit path tracing.

\setlength{\columnsep}{.5em}
\setlength{\intextsep}{.5em}
\begin{wrapfigure}{r}{82px}
    \includegraphics{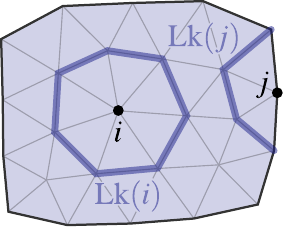}
\end{wrapfigure}
A polyhedral surface is, roughly speaking, a collection of Euclidean (planar) polygons glued together to form a surface.  Since any planar polygon can be triangulated (and the choice of triangulation has no effect on geodesics), we will assume that the combinatorics of any polyhedron are encoded by a simplicial complex \(\Msf\) with vertices \(\Vsf\), edges \(\Esf\), and faces \(\Fsf\).  For simplicity we also assume that this complex is \emph{manifold}, \ie, the link \(\mathsf{Lk}(i)\) of every vertex \(i \in \Vsf\) interior vertex is a single closed loop; the link of every boundary vertex is a single path (see inset figure).

\begin{figure}[b!]
   \centering
   \includegraphics{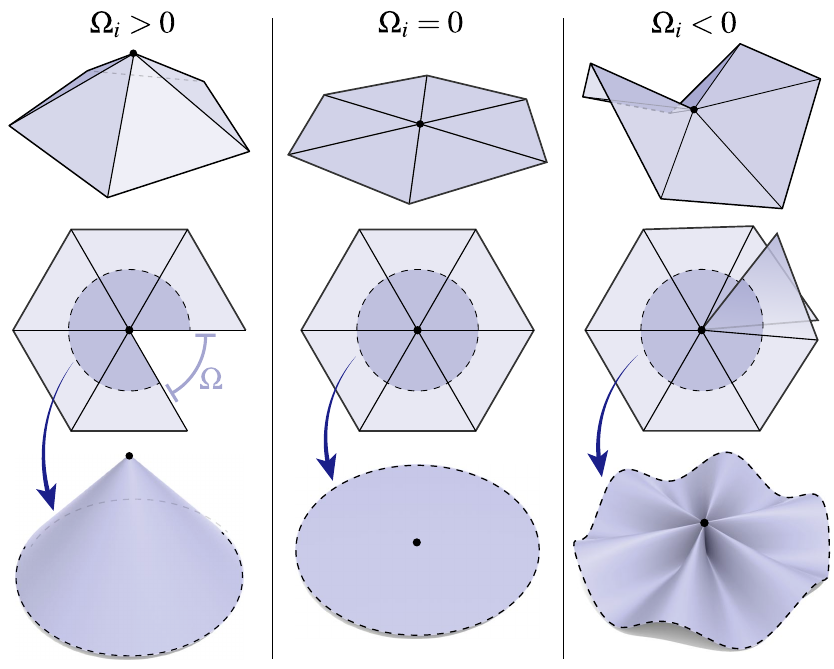}
   \caption{Intrinsically, a vertex \(i\) of a polyhedral surface looks like either a circular cone, a flat disk, or a saddle, depending on the sign of the angle defect \(\Omega_i\).\label{fig:IntrinsicVertex}}
\end{figure}

The geometry of a polyhedral surface can be specified in one of two ways: either \emph{extrinsically}, using vertex coordinates \(f: \Vsf \to \mathbb{R}^n\) which are linearly interpolated over each triangle, or \emph{intrinsically}, by specifying positive edge lengths \(\ell: \Esf \to \mathbb{R}_{>0}\) that satisfy the triangle inequality in each face: \(\ell_{ij} + \ell_{jk} \geq \ell_{ki}\) for all \(ijk \in \Fsf\).  Of course, one can always obtain edge lengths from vertex coordinates (\(\ell_{ij} := |f_j - f_i|\)), though the vast majority of geodesic algorithms can in principle be implemented without reference to vertex coordinates---reflecting the fact that geodesics are a purely intrinsic object.  Instead, quantities like triangle areas \(A_{ijk}\) and interior angles \(\theta_i^{jk}\) at triangle corners can be computed directly from edge lengths (using expressions like \emph{Heron's formula} and the law of cosines).  This fact is especially useful given that in some geometric problems (\eg, manifold learning) one may have distances between points, but not an explicit embedding in \(\mathbb{R}^n\).

\noindent~For each vertex \(i \in \Vsf\), the \emph{angle defect} is the quantity
\[
   \Omega_i := 2\pi - \sum_{ijk \in F} \theta_i^{jk},
\]
where \(\theta_i^{jk}\) denotes the interior angle at vertex \(i\) of triangle \(ijk\), and the sum is taken over all triangles \(ijk\) containing vertex \(i\).  Intuitively, this quantity captures the ``flatness'' of the vertex, and is often viewed as a discrete analogue of the Gaussian curvature.  An important mental image is provided in \figref{IntrinsicVertex}: imagine that the triangles around a vertex \(i \in \Vsf\) are flat pieces of paper glued together along edges.  Depending on the value of \(\Omega_i\), these triangles can be bent into a smooth circular cone, a flat circular disk, or a saddle-like figure, all without changing geodesic distance.  We will therefore refer to vertices with positive, zero, and negative angle defect as \emph{spherical} or \emph{cone-like}, \emph{Euclidean}, and \emph{hyperbolic} or \emph{saddle-like}, respectively. This local picture nicely encapsulates the intrinsic geometry of any polyhedron: it is smooth and intrinsically flat away from the vertices---even the location of \emph{edges} is irrelevant when it comes to thinking about geodesics and geodesic distance, since the edges are effectively invisible to an intrinsic observer.  The sign of \(\Omega\) plays an especially important role in the context of polyhedral geodesics, since shortest geodesics will often pass through vertices where \(\Omega_i < 0\), but can \emph{never} pass through vertices where \(\Omega_i > 0\) (as will be discussed in \secref{ShortestVsStraightest}).

Working with polyhedral rather than smooth surfaces has some interesting geometric consequences.  On the one hand, since each individual triangle is flat, we can study geodesics by ``unfolding'' local neighborhoods into the plane, \ie, by finding an arrangement of vertices in \(\mathbb{R}^2\) that agrees with the intrinsic edge lengths---several examples are shown in \figref{IntrinsicVertex}.  This picture makes it clear that, \emph{locally}, geodesics on polyhedral surfaces can be constructed by simply drawing line segments in the Euclidean plane.  The main computational challenge, therefore, is answering more \emph{global} questions: for example, which sequence of triangles must we unfold to find the \emph{shortest} such line?  An analogous perspective is not generally not available for smooth surfaces, since any local flattening will invariably distort lengths (\ie, geodesics are rarely straight lines in local coordinates).  On the other hand, the fact that our surface is no longer smooth makes the definition of polyhedral geodesics somewhat more nuanced---especially in the vicinity of vertices.

\subsubsection{Shortest vs. Straightest}
\label{sec:ShortestVsStraightest}

In the smooth setting we had two equivalent characterizations of geodesic curves: they are both \emph{straightest} (\secref{StraightestCurves}) and \emph{locally shortest} (\secref{ShortestCurves}).  As studied by Polthier \& Schmies~\cite{polthier1998straightest}, these two characterizations are no longer equivalent in the polyhedral setting.  This situation reflects a common ``no free lunch'' scenario in the discretization of objects from differential geometry~\cite{Crane:2017:GID}, namely that one typically cannot find a single definition (in this case, for discrete geodesics) that exactly captures all the key properties of the original smooth object (in this case, both straightest and locally shortest).

Locally, polyhedral geodesics essentially behave the same as in the smooth setting.  Consider for instance a pair of points \(p, q\) contained in the same triangle---here geodesics are just ordinary line segments, which are both shortest and straightest.  Likewise, for two points \(p,q\) close to a common edge (and away from any vertex) we can simply unfold the two incident triangles into the plane and connect them by the unique shortest, straightest segment (see \figref{PathUnfold}, left).  Globally, however, the situation is more complicated due to non-smooth points at vertices.

\noindent\textbf{Straightest.} To find the \emph{straightest} path leaving a point \(p \in \Msf\) in a direction \(u \in T_p \Msf\), we can simply apply the local observations made above: inside a given triangle the shortest path is found by extending a straight ray along \(u\); to continue this path into the next triangle we can imagine unfolding neighboring triangles into the plane and extending this ray into the next triangle.  The resulting path corresponds to a straight line along a strip of planar triangles, as depicted in \figref{PathUnfold}, right.  (Note that for very long paths we may encounter the same triangle more than once, in which case we would have multiple copies of this triangle in the unfolding.)  This tracing operation effectively defines a discrete version of the exponential map discussed in \secref{ExponentialMap}.

\setlength{\columnsep}{.5em}
\setlength{\intextsep}{.5em}
\begin{wrapfigure}{r}{65pt}
    \includegraphics{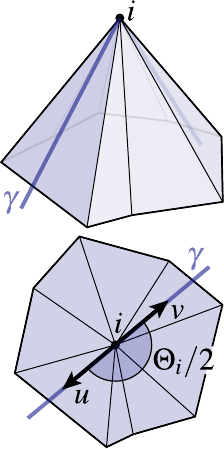}
\end{wrapfigure}
What should we do if our path enters a vertex \(i \in \Vsf\)?  In particular, which outgoing direction describes the ``straightest'' curve?  Unless the angle defect \(\Omega_i\) is equal to zero, we cannot simply unfold triangles into the plane.  An idea considered by Polthier \& Schmies~\cite{polthier1998straightest} is to instead pick the outgoing direction such that there is ``equal angle'' on either side.  More precisely, let \(u\) be the incoming direction; we can define the outgoing direction \(v\) as the one such that the total angle between \(u\) and \(v\) is exactly \emph{half} the sum \(\smash{\Theta_i := \sum_{ijk} \theta_i^{jk}}\) of interior angles \(\smash{\theta_i^{jk}}\) at vertex \(i\) (see inset figure).  Equivalently, one can work in a local polar coordinate system where angles \(\smash{\theta_i^{jk}}\) are normalized to sum to \(2\pi\);
this viewpoint has been carefully studied by Troyanov~\cite{Troyanov:1986:LSE}, and was later adopted in geometry processing for problems involving polyhedral geodesics~\cite{polthier1998straightest} and tangent vector fields at vertices~\cite{Zhang:2006:VFD}; Sharp \etal~\cite[Section 5.2]{Sharp:2018:VHM} provides a concise description.  As in the smooth case, this definition of straightness yields a unique solution to the initial value problem, even for paths through vertices with nonzero angle defect.

\noindent\textbf{Shortest.} In contrast, the behavior of \emph{shortest} curves on a polyhedral surface depends critically on the sign of the angle defect \(\Omega_i\).  Consider for instance two points \(a,b\) directly opposite a cone-like vertex (\(\Omega_i > 0\)), as depicted in \figref{ConeGeodesics}.  By symmetry, one might expect that the shortest route between these points is to walk along the straightest possible path \(\gamma\) from \(a\) to \(i\), then from \(i\) to \(b\).  However, one can find an even shorter path by walking ``around'' the cone---just as one might find a shorter path by walking \emph{around} a hill, rather than walking over it.  In particular, if we cut the cone from the boundary through the point \(a\) and up to the vertex \(i\), then the straight line segment from \(b\) to either copy of \(a\) gives us a shortest path.  Hence, \emph{straightest} curves on a polyhedral surface are not always shortest.  In this case, \(\gamma\) is not even \emph{locally} shortest, since even in a \emph{tiny} region around \(i\) would can make it slightly shorter by going around vertex \(i\), instead of through it.  In general, it is \emph{never} be advantageous (in terms of path length) to pass through a spherical vertex.  Therefore, the only shortest geodesics passing through a positive vertex \(i\) are those terminating at \(i\).

\setlength{\columnsep}{.5em}
\setlength{\intextsep}{.5em}
\begin{wrapfigure}{l}{74pt}
    \includegraphics{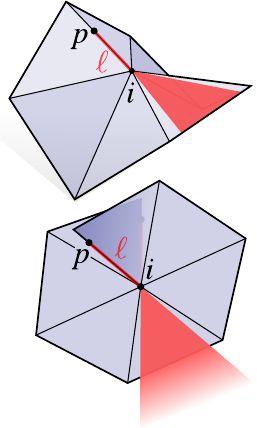}
\end{wrapfigure}
On the other hand, if \(i\) is a saddle vertex (\(\Omega_i < 0\)), one can often find much shorter paths by passing through vertices.  Some basic intuition is provided by \figref{IntrinsicVertex}, bottom right: to go from one side of a saddle region to another, it is much quicker to go straight through the vertex than to travel across numerous ``ripples.''  In fact, starting with a straight line \(\ell\) from a point \(p \in \Msf\) to a saddle vertex \(i\), there will be infinitely many outgoing directions \(v\) that yield a shortest path; these directions form a wedge-like region of angle \(|\Omega_i|\) around the incoming direction \(u\) (see inset figure).  The union of the incoming path \(\ell\) with the wedge is sometimes called a \emph{funnel}, and is the starting point for a large family of algorithms reviewed in \secref{global}.  Again the ``shortest'' and ``straightest'' pictures disagree: a straightest geodesic passing through a saddle vertex must bisect the funnel, whereas a shortest geodesic can include \emph{any} from \(i\) contained in the funnel.

\begin{figure}[t!]
\centering
\includegraphics{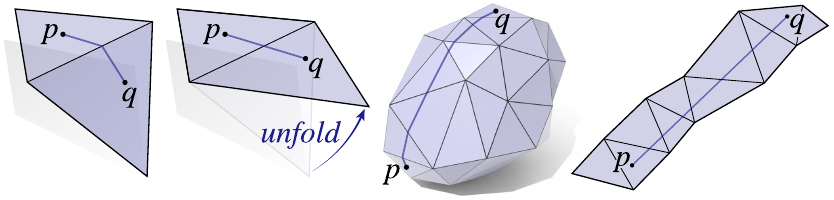}
   \caption{Any pair of adjacent triangles can be unfolded into the plane without distorting distance (left).  A geodesic on a polyhedral surface is therefore equivalent to a straight line along some planar triangle strip---so long as it does not pass through any vertices.\label{fig:PathUnfold}}
\end{figure}

\begin{figure}
   \centering
   \includegraphics{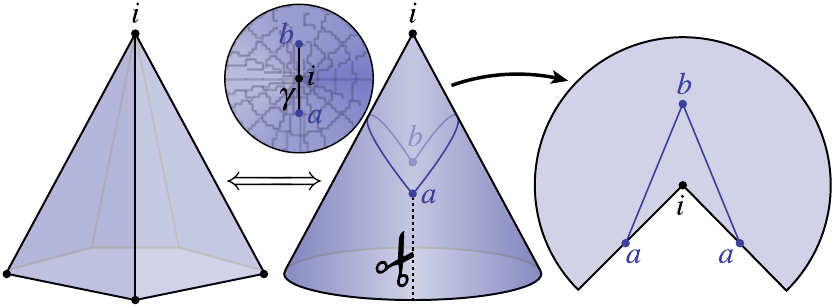}
   \caption{The triangles around any vertex \(i\) with positive angle defect \(\Omega_i > 0\) (left) are intrinsically no different from a smooth cone (center).  Therefore, the shortest path between two points \(a,b\) on a polyhedral surface will never pass through a positive vertex: we can always cut open the cone in a way that lets us draw a straight line from \(a\) to \(b\) without passing through \(i\).  When \(i\) sits directly between \(a\) and \(b\), this path will not even be unique (right).\label{fig:ConeGeodesics}}
\end{figure}

\subsubsection{Discrete Exponential Map}
\label{sec:DiscreteExponentialMap}

The analysis above has important implications for the exponential map on a polyhedral surface.  Consider for instance tracing straightest geodesics in every possible direction from a vertex \(i \in \Vsf\).  When paths hit a spherical vertex \(j\), they split into two groups which meet discontinuously on the opposite side of \(j\).  When paths hit a saddle

\setlength{\columnsep}{.5em}
\setlength{\intextsep}{.5em}
\begin{wrapfigure}{r}{80pt}
    \includegraphics{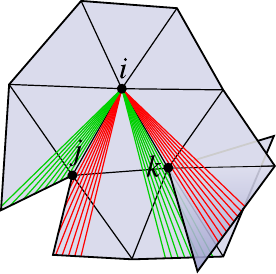}
\end{wrapfigure}
\noindent~vertex \(k\), they again split into two groups, which do not completely cover the funnel opposite \(k\).  In either case, the exponential map fails to be injective as soon as we hit \emph{any} non-flat vertex; in other words, the injectivity radius is simply the distance to the closest vertex.  Moreover, every spherical vertex of a polyhedral surface is contained in the cut locus.  As explored by Itoh \& Sinclair~\cite{Itoh:2004:TTA}, this means that algorithms which exactly compute distance on a polyhedral surface will (surprisingly enough) do a very poor job of approximating the cut locus of the smooth surface it approximates.  For instance, the cut locus on a polyhedral sphere will contain \emph{every single vertex}, whereas the cut locus of a smooth sphere consists of just a single point.



\section{PDE-based Methods}
\label{sec:fem}

%

A large number of methods for computing geodesic distance are based on formulating the problem in terms of partial differential equations (PDEs) on a smooth manifold, then discretizing and solving these PDEs via, \eg, finite element methods (FEM) or other numerical techniques.  These methods are generally suitable for computing the single or multiple source geodesic distance (SSGD/MSGD); explicit geodesic can subsequently be obtained by, \eg, tracing curves through the gradient of the distance function (though such tracing requires careful consideration in order to achieve accurate results; see \secref{tracing}).  Some of these methods are also quite attractive for solving all-pairs geodesic distance problems (APGD), since their solutions are well-approximated by precomputing a relatively small collection of eigenfunctions (\secref{SpectralDistances}).  PDE-based methods are attractive because they are built on top of well-established techniques from scientific computing (such as FEM), as opposed to computational geometry methods (\secref{cg}) which must be built ``from the ground up.''  As a result, such techniques often come with a clearer picture of, \eg, convergence rates under refinement, making them particularly well-suited for the smooth geodesic problem (Section \secref{PolyhedralVsSmooth}).  Moreover, they easily generalize to problems involving multiple or curved sources, and can often be implemented on data structures beyond triangulations (as shown in \figref{EllipticDiscretizations}).  On the other hand, since these methods compute only the geodesic \emph{distance} \(\phi\), additional work must be done if one wishes to extract geodesic \emph{paths}, as discussed in \secref{tracing}.

PDE-based methods can be categorized according to the type of equation used to characterize geodesic distance.  Different starting points will lead to different numerical treatments, which subsequently have different computational trade offs (\eg, different mesh or solver requirements).  At a high level there are two basic classes of methods:
\begin{itemize}
   \item \textbf{Wavefront-based.} The basic principle behind wavefront-based methods resembles (in spirit) classical methods like Dijkstra's algorithm, or the window-based methods discussed in Section \ref{sec:cg}: information about distance propagates outward from a given source point.  In the continuous setting, this perspective corresponds to \emph{hyperbolic PDEs}, \ie, those that describe the evolution of a wavefront emanating from the source.
   \item \textbf{Diffusion-based.} Diffusion-based methods more closely resemble problems arising in, say, spectral graph theory: information about distance is obtained by way of functions associated with a discrete \emph{Laplace operator}, computed via a process that looks more like repeated local averaging rather than wavefront propagation.  In the continuous setting, this perspective corresponds to \emph{elliptic and parabolic PDEs} such as Poisson equations and heat diffusion.
\end{itemize}

\paragraph*{Trade offs.} Historically, wavefront-based methods were developed prior to most diffusion-based algorithms; as a result, a wide variety of higher-order accurate strategies have been developed for regular grids on Euclidean domains (\eg{}, \cite{Ahmed:2011:TOA}), motivated in large part by accurate numerical solvers for \emph{level set equations}~\cite{Osher:2003:LSM}.  However, most of these techniques do not immediately generalize to the setting of curved surfaces, where one typically does not have a regular, uniform tessellation of the domain needed to support (for instance) larger finite difference stencils.  Diffusion-based algorithms tend to have some nice performance advantages, since they are largely based on solving easy linear elliptic equations rather than more difficult nonlinear hyperbolic ones, though solutions can sometimes be over-regularized (\eg, near the cut locus of the geodesic distance function).  A particularly nice feature of linear methods is that one can often pre-factor the associated matrices, which in practice yields about an order of magnitude improvement in amortized performance using modern direct solvers~\cite{Chen:2008:ACS,Schenk:2001:PHS}.

Meshing requirements for elliptic methods also tend to be less stringent than for hyperbolic ones; see \secref{meshing} for further discussion.  As discussed in \secref{PolyhedralVsSmooth}, the accuracy of all PDE-based methods (as well as all methods from computational geometry) is fundamentally limited by the accuracy of the domain representation itself (\eg, the use of linear elements to approximate curved surfaces).  The only real resolution to this issue is to compute geodesic distance on higher-order surface representations such as splines and subdivision surfaces; several authors have recently considered this approach~\cite{DeGoes:2016im,Nguyen:2016:CFE}.

Overall, there is a rather nice viewpoint that unifies all current PDE-based methods: they are all effectively trying to minimize the residual of the eikonal equation \(\int_M (|\nabla|^2 - 1)^2\ dA\), albeit by very different means (see especially the discussion at the end of \secref{HeatMethod}.  Since this energy is nonlinear and nonconvex, it is not surprising to find that many different computational strategies have been proposed: hyperbolic methods like fast marching deal with nonlinearity by taking advantage of special update orderings, whereas elliptic methods like the heat method decompose the problem into two linear pieces connected by a nonlinear change of variables.  Very little work has been done on synthesizing these perspectives (\eg, combining fast linear approximations with carefully-ordered updates), though is likely to be fruitful, especially given the diverse range of application problems and machine architectures on which geodesic distance queries are needed.

\begin{figure}
   \label{fig:EllipticDiscretizations}
   \includegraphics[width=\columnwidth]{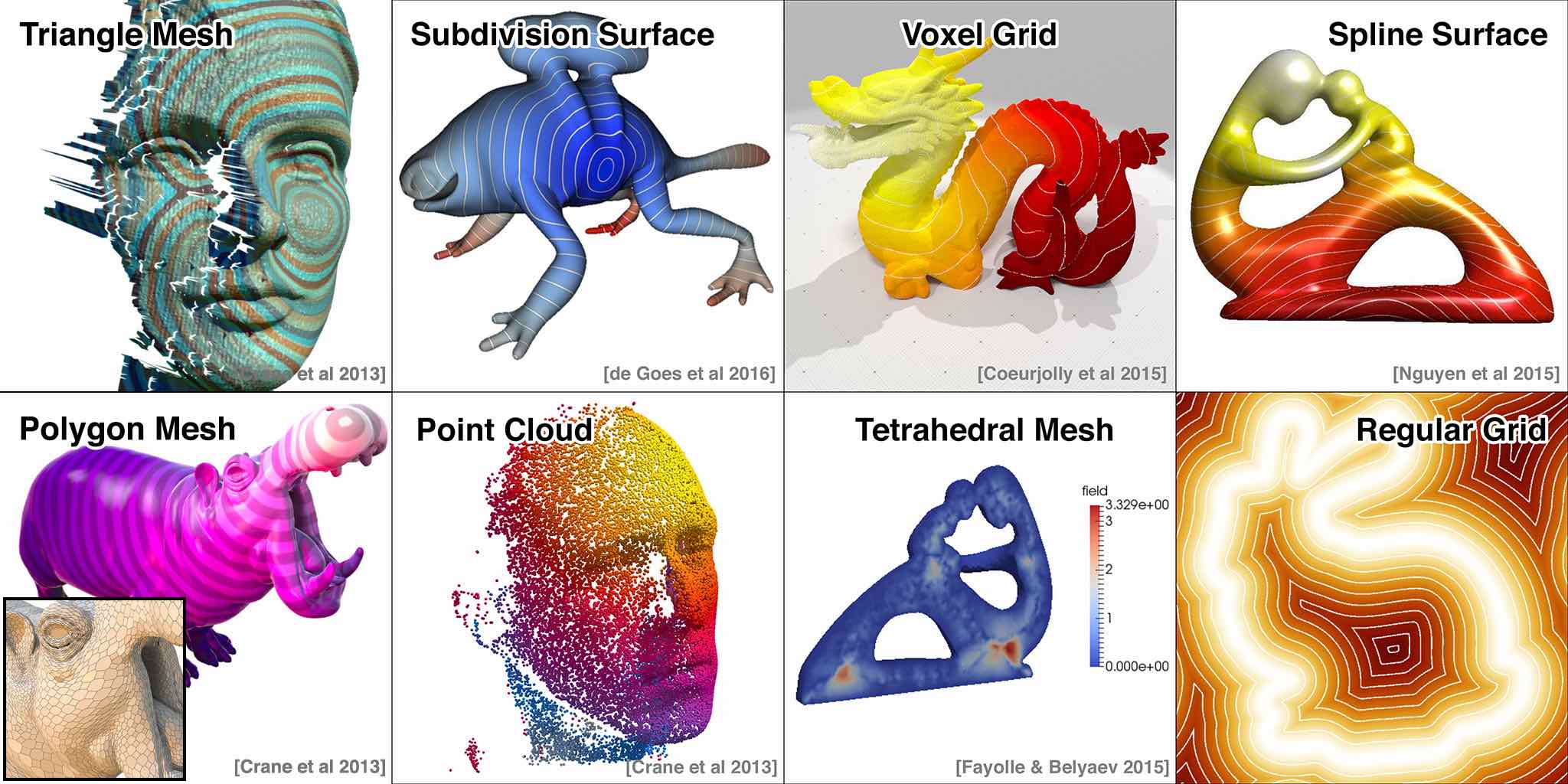}
   \caption{Unlike methods from computational geometry, which are specially tailored to triangulations, PDE-based methods can easily be implemented on a variety of data structures by applying standard discrete differential operators.  Here, we show a variety of implementations of the heat method using different discrete Laplace operators.\label{fig:EllipticDiscretizations}}
\end{figure}

\subsection{Laplace-Beltrami and Cotan Laplace}
\label{sec:LaplaceBeltramiAndCotanLaplace}

PDE-based methods for geodesic distance computation all have a close relationship to the \emph{Laplace-Beltrami operator} \(\Delta\), which is discretized by a \emph{(weighted) graph Laplacian}.  In particular, for a graph \(G = (V,E)\) with edge weights \(w: E \to \mathbb{R}\), the graph Laplacian is encoded by a real symmetric matrix \(L \in \mathbb{R}^{|V| \times |V|}\) with off-diagonal entries
\[
   L_{ij} = \begin{cases}
               -w_{ij}, & ij \in E, \\
               0, & otherwise,
            \end{cases}
\]
and diagonal entries
\[
   L_{ii} = \sum_{ij \in E} w_{ij}.
\]
A simple choice of edge weights is just \(w_{ij} = 1\), though of course these weights do not capture much geometric information.  When the vertices \(i \in V\) have associated vertex positions \(f_i \in \mathbb{R}^3\), the edge weights can be determined by some function of the edge lengths; one common choice is to use Gaussian edge weights \(w_{ij} = e^{-|f_j-f_i|^2/t}\) (for some small parameter \(t > 0\)), which is motivated by the close connection between the Laplace-Beltrami operator and the Euclidean \emph{heat kernel} (see \secref{Diffusion}).  Finally when the edges of the graph come from the edges of a triangulated surface, one typically adopts the \emph{cotangent weights}
\[
   w_{ij} = \tfrac{1}{2}(\cot\alpha_{ij} + \cot\beta_{ij}),
\]
where \(\alpha_{ij},\beta_{ij}\) are the interior angles opposite edge \(ij\) (or zero, for edges on the domain boundary).  More broadly, these discrete Laplacians are only the tip of the iceberg: discrete Laplace operators have been developed for a large variety of geometric data structures, providing a wide variety of options for implementing PDE-based geodesic schemes (see especially the discussion in \secref{HeatMethod}).  Some methods, such as those based on diffusion (\secref{Diffusion}) easily generalize to these settings by just ``plugging in'' a different Laplace matrix; other methods, such as fast marching (\secref{FastMarching}) do not immediately generalize since they may depend on particular features of a discretization (\eg, the fact that values at the vertices of a simplex uniquely determine an interpolating affine function).  Note that since the Laplace-Beltrami operator depends only on the intrinsic geometry of the domain (\ie, its metric) and not the way it sits in space, any distance algorithm defined in terms of the Laplacian (including all the methods we consider here) will automatically be isometry invariant.

\subsection{Wavefront Propagation}
\label{sec:WavefrontPropagation}

The basic prototype for wavefront-based methods is the linear hyperbolic \emph{wave equation}
\begin{equation}
   \label{eq:WaveEquation}
   \tfrac{d^2}{dt^2}u_t = \Delta u_t,
\end{equation}
where for each time \(t\), \(u_t\) is a real-valued function on a manifold \(M\), and \(\Delta\) is the \emph{Laplace-Beltrami} operator on \(M\) (for \(M = \mathbb{R}^n\), \(\Delta\) is just the standard Laplacian).  The intuitive connection to geodesic distance is that a perturbation at some initial point (\eg, a small stone dropped in a pond) will send out a wavefront that maintains a constant distance from the source point.  Hence, the arrival time of the wavefront is correlated with the geodesic distance.  Though a few methods extract distance information from the classical wave equation~\cite{Sinha:2016:GWC} or the quantum mechanical (Schr\"{o}dinger) wave equation~\cite{Gurumoorthy:2009:SEF}, by far the most common approach is to consider the nonlinear hyperbolic \emph{eikonal equation}
\begin{equation}
   \label{eq:eikonal}
   \begin{array}{rl}
      |\nabla\phi|^2 = 1, & \text{on}\ M \\
      \phi = 0            & \text{on}\ \partial M \\
   \end{array}
\end{equation}
which directly characterizes the distance function \(\phi: M \to \mathbb{R}\) in terms of the norm of its gradient \(|\nabla\phi|\).  Intuitively, this equation says something very straightforward: on the boundary of the domain \(\partial M\) (\ie, at any point in the source set), the distance \(\phi\) should be zero.  At every other point of the domain, the \emph{change in distance per unit distance} along the direction of greatest increase should simply be 1.  However, actually \emph{solving} this equation is not as straightforward since it is nonlinear and hence cannot be approximated using standard linear finite element methods.  Instead, the general strategy is to iteratively update the solution using techniques like (nonlinear) Gauss-Seidel; for carefully-crafted update orders (and suitable conditions on the mesh geometry) such strategies can actually converge in a single iteration, corresponding to well-established algorithms such as \emph{fast marching} (this perspective is nicely explained by Bornemann and Rasch~\cite{Bornemann:2004:FED}).

\subsubsection{Fast Marching}
\label{sec:FastMarching}

The \emph{fast marching method} was originally developed for distance transforms on Euclidean domains discretized as regular grids; the basic principles of this method and its variants (\eg, \emph{fast sweeping}~\cite{Luo:2013ff}) are shared by a broad class of schemes used in \emph{level set methods}~\cite{OsherFedkiw} and to solve \emph{Hamilton-Jacobi equations} (of which the eikonal equation is one example).  Kimmel and Sethian developed a level set method for curved domains represented as triangulated surfaces~\cite{Kimmel98computinggeodesic}.  The basic strategy of this method is very similar to Dijkstra's shortest path algorithm: set the distance at the source point (or set) to zero, and set the tentative distance to infinity; then use a region-growing strategy to update the remaining distances in an ``upwind'' order, \ie, consider the node with smallest distance first.  Like Dijkstra, the running time is therefore \(O(|V| \log |V|)\), since for a triangulation, \(|E| \in O(|V|)\).

\setlength{\columnsep}{.5em}
\setlength{\intextsep}{.5em}
\begin{wrapfigure}{r}{94px}
    \includegraphics{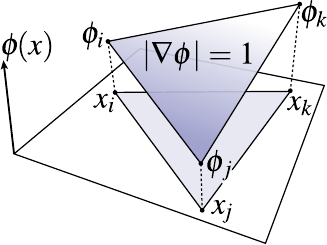}
\end{wrapfigure}
Relative to Dijkstra, the key modification is that distances are not updated according to paths along edges; instead, one updates the distance value at a vertex by solving for the linear function that satisfies the eikonal equation \cite[Section 4.1]{Kimmel98computinggeodesic}.  In particular, if the values \(\phi_i,\phi_j\) at two corners are known, one simply needs to pick a third value \(\phi_k\) so that the slope of a triangle \(|\nabla\phi|\) passing through all three values equals \(+1\)  (as illustrated in the inset figure).

We can connect the fast marching method to other PDE-based methods by observing that the term \(|\nabla\phi|^2\) is the integrand of the \emph{Dirichlet energy}, which in turn can be expressed in terms of the \emph{cotan Laplace operator}. 
In particular, consider a triangle \(ijk\) where the distance values at vertices \(i\) and \(j\) are known, and the distance at vertex \(k\) remains to be determined.  Suppose we encode these distance values as a column vector \(\phi = [ \phi_i\ \phi_j\ \phi_k ]^T\), and let \(L \in \mathbb{R}^{3 \times 3}\) be the local cotan matrix given by
\[
   \begin{array}{rcl}
      2L_{ii} &=& \cot\alpha_j + \cot\alpha_k, \\
      2L_{ij} &=& -\cot\alpha_k
   \end{array}
\]
(see for instance \cite[Section 6.2]{Crane:2013:DGP}).  A linear function \(\phi\) interpolating the distance values at the three vertices will then satisfy the eikonal equation if \(\phi^T L  \phi = 1\), which we can view as an ordinary quadratic equation for the unknown distance \(\phi_k\).  Solving this equation yields (in general) two solutions corresponding to positive and negative slope; the update will always use the larger value, since distance increases monotonically as we move away from the source.

\begin{figure}
   \includegraphics{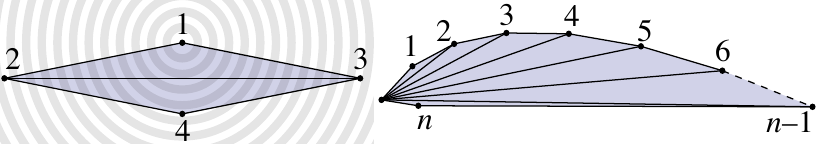}
   \caption{Methods based on solving hyperbolic equations may fail to respect causality in the presence of obtuse angles.  For instance, in the left figure a source at 1 will get propagated to 2 and 3 before 4, even though 4 is closer.  In the worst case this phenomenon can be highly nonlocal, as seen on the right where the node \(n\) closest to the source node 1 is the last one to be updated.\label{fig:FMM_mesh}}
\end{figure}

A basic difficulty with wavefront propagation methods is that the order in which vertices are updated may violate \emph{causality}, \ie, even if a vertex \(i\) is closer to the source than a vertex \(j\), the distance at \(j\) may be finalized prior to the distance at \(i\); hence, the solution \(\phi\) will fail to be monotonic (two examples of how this can happen are shown in Figure~\ref{fig:FMM_mesh}).  Sethian and Kimmel suggest to resolve this issue by an ``unfolding'' procedure, but this procedure is nonlocal and may not terminate before reaching the boundary.  In general, one therefore has to either apply an iterative strategy (as discussed by Bornemann and Rasch~\cite{Bornemann:2004:FED}), or perform remeshing, as discussed in \secref{meshing}.  From a computational point of view, the basic challenge with any method based on wavefront propagation (including Dijkstra's algorithm) is that it is difficult to parallelize: the upwind strategy effectively induces a serial ordering on computation.  Since the order of computation is data-dependent (\ie, it depends on earlier distance values), it can also lead to significant dynamic branching and incoherent memory access compared to linear methods (\secref{Diffusion}) which more typically depend on a fixed traversal order (\eg, using a fixed matrix factorization).

\subsection{Diffusion}
\label{sec:Diffusion}

The basic prototype for diffusion-based methods is the linear parabolic \emph{heat equation}
\begin{equation}
   \label{eq:HeatEquation}
   \tfrac{d}{dt}u_t = \Delta u_t,
\end{equation}
where for each time \(t\) the function \(u_t\) is a real-valued function on the domain.  Despite the very similar form of \eqref{HeatEquation} and \eqref{WaveEquation}, they have very different behavior and hence lead to very different computational methods.  Whereas small perturbations to \(u\) yield high-frequency oscillations in solutions to the wave equation, such perturbations are diffused or smoothed out by the heat equation, hinting at greater stability (\eg, small computational errors are not propagated forward in time).  In particular, if \(\delta_x\) is a Dirac delta centered at a point \(x\), then the \emph{heat kernel at \(x\)} is the solution to the heat equation
\begin{equation}
   \label{eq:HeatKernel}
   \begin{array}{rcl}
      \frac{d}{dt} k_t &=& \Delta k_t, \\
      k_0 &=& \delta_x.
   \end{array}
\end{equation}
More generally we will use \(k_t(x,y)\) to denote the heat kernel at time \(t\), centered at the point \(x\), and evaluated at the point \(y\).  In the Euclidean case the heat kernel is just a Gaussian of increasing width and decaying magnitude, though in general it has no closed form (see for instance \cite[Section 3.2]{Sharp:2018:VHM}).  Any other solution to the heat equation can be expressed as a convolution with the heat kernel, \ie, a gradual ``blurring out'' of the initial data.  The heat equation also has an important statistical interpretation: if one views the initial function \(u_0\) as describing the spatial distribution of a large number of discrete particles, then the solution \(u_t\) describes the distribution after these particles have taken ``random walks'' (\ie, Brownian motion) for time \(t\)~\cite{Lawler:2010:Random}.

Diffusion-based methods were initially used to compute smooth distance-like functions (\secref{PoissonBasedMethods}) or functions that satisfy the axioms of a distance metrics, but do not actually provide the geometric distance between points (\secref{SpectralDistances}); more recently, a diffusion-based strategy was introduced that provides the true geodesic distance (\secref{HeatMethod}).  Computationally, the appeal of all such methods is they amount to sparse linear systems that can be efficiently solved using standard techniques from numerical linear algebra.  As a result, one does not have to develop solvers specially tailored to the task of computing geodesic distance, as with the hyperbolic methods discussed in \secref{WavefrontPropagation} or the polyhedral strategies in \secref{cg}.  Instead, any development in numerical linear algebra (such as more accurate linear solvers, or fast parallel implementations) can immediately be used to improve computation of geodesic distances.  Diffusion-based methods have also become popular due to their ease of implementation, which generally involves only three basic ingredients:
\begin{itemize}
   \item Building a weighted graph Laplacian \(L\) associated with the mesh;
   \item Solving one or more sparse linear algebra problems involving \(L\);
   \item Evaluating simple per-triangle or per-vertex expressions.
\end{itemize}
Per-element operations might be performed either before or after solving linear problems (or both), but generally amount to evaluations of, \eg, simple closed-form sums (this pattern is in fact shared by all methods we will discuss in this section).  The bulk of the complexity is therefore taken care of by the linear solver, which can be treated as a ``black box'': one does not have to implement complex topological data structures, or even maintain a priority queue.  On the flip side, this formulation assumes that a suitable mesh is already provided as input; see \secref{meshing} for further discussion.

\subsubsection{Spectral Distances}
\label{sec:SpectralDistances}

The original motivation for diffusion-based distances comes from a desire to embed an abstract surface or manifold into a Euclidean space -- such an embedding allows one to approximate distances on the original manifold by measuring the ordinary Euclidean distance between points in \(\mathbb{R}^n\), though this approximation may only roughly resemble the true geodesic distance.  This point of view was originally studied by B\'{e}rard \etal{} for a different purpose: to study the precompactness of smooth manifolds \(M\) with bounded diameter and curvature~\cite[Chapter VI E.53]{Berard:1986:SGD},\cite{Berard:1994:ERM}.  Later, this same point of view became a common theme in the context of data analysis, machine learning, and dimensionality reduction, where the manifold \(M\) and Laplace-Beltrami operator \(\Delta\) are replaced with a discrete graph \(G = (V,E)\) and graph Laplacian \(L \in \mathbb{R}^{|V| \times |V|}\).  In particular, let \(\psi_i \in \mathbb{R}^{|V|}\) and \(\lambda_i \in \mathbb{R}\) be the eigenvectors and eigenvalues of \(L\), so that
\begin{equation}
   \label{eq:LaplaceEigenproblem}
   L\psi_i = \lambda_i\psi_i,\ i \in V.
\end{equation}
Belkin \& Niyogi propose an embedding by \emph{Laplacian eigenmaps} that maps each vertex \(v \in V\) to the coordinates \(f_i := (\psi_1(v),\ldots,\psi_k(v))\) (for some choice of \(n \leq |V|\)), arguing that this embedding minimizes the \(\ell_2\) distortion of edge lengths~\cite{Belkin:2003:LED}.  A notion of distance is then given by \(\phi_E(i,j) := |f_j-f_i|\), where \(|\cdot|\) is just the Euclidean norm.  Gobel \& Jagers study a notion of distance that is closely connected to the random walk interpretation of the heat equation -- in particular, they interpret a graph as a Markov chain with probabilities determined by edge weights, and show that the expected time to walk from a vertex \(i\) to a vertex \(j\) and then back to \(i\) determines a metric on this graph~\cite{Gobel:1974:RWG}.  This \emph{commute time distance} \(\phi_C\) is equivalent to the effective resistance between \(i\) and \(j\) in an electrical network~\cite{Klein:1993:RD}, which is also connected to the Laplacian \(L\)~\cite{Saerens:2004:PCA,Fouss:2007:RWC}.  Lipman~\etal{} define a closely related \emph{biharmonic distance} \(\phi_B\) based on the bi-Laplacian \(\Delta^2\) rather than the Laplacian \(\Delta\)~\cite{Lipman:2010:BD}.  Coifman \& Lafon instead define a \emph{diffusion distance} \(\phi_D\) which captures the amount of information that diffuses from a vertex \(i\) to a vertex \(j\) after time \(t\)~\cite{Coifman:2006:DM}.  In fact, all of these distances can be linked back to a discrete diffusion process -- consider in particular the \emph{(discrete) harmonic Green's function} \(G_x: V \to \mathbb{R}_{>0}\), which is the solution to the equation \(LG_x = \delta_x\) (where \(\delta_x\) is the Kronecker delta centered at \(x \in V\)). This function can also be viewed as the stationary solution to the discrete heat equation with fixed (Dirichlet) boundary conditions at vertex \(x\).  Letting \(H_x\) be the corresponding function for the bi-Laplacian, we can write the three distances mentioned above as
\[
   \begin{array}{rcl}
      \phi^2_C(x,y) &=& G_{x,x} - 2G_{x,y} + G_{y,y} \\
      \phi^2_B(x,y) &=& H_{x,x} - 2H_{x,y} + H_{y,y} \\
      \phi^2_D(x,y) &=& || k_{t,x} - k_{t,y} ||^2,
   \end{array}
\]
where \(k_{t,x}: V \to \mathbb{R}_{>0}\) denotes the solution to the discrete heat equation at time \(t\) and with initial conditions \(u_0 = \delta_x\).  Evaluating each of these distances for a given pair of vertices \(x,y \in V\) therefore amounts to solving a small number of linear equations.  If, however, one wishes to compute the distance between a large number of vertex pairs (\eg, to solve the APGD problem), it is often more efficient to use a \emph{spectral expansion} of these distance functions, \ie, to write them in terms of the eigenvalues and eigenvectors of the discrete Laplace operator (Equation~\ref{eq:LaplaceEigenproblem}), leading to a unified view of all four distances defined above:
\[
   \begin{array}{rrcr}
      \text{\emph{(eigenmap)}}     & \phi^2_E(x,y) &=& \sum_{i=1}^n (\psi_i(x) - \psi_i(y))^2, \\
      \text{\emph{(commute time)}} & \phi^2_C(x,y) &=& \sum_{i=1}^n \tfrac{1}{\lambda_i}(\psi_i(x) - \psi_i(y))^2, \\
      \text{\emph{(biharmonic)}}   & \phi^2_B(x,y) &=& \sum_{i=1}^n \tfrac{1}{\lambda_i^2}(\psi_i(x) - \psi_i(y))^2, \\
      \text{\emph{(diffusion)}}    & \phi^2_D(x,y) &=& \sum_{i=1}^n e^{-2t\lambda_k}(\psi_i(x) - \psi_i(y))^2. \\
   \end{array}
\]
Note that the \emph{global point signature} of Rustamov~\cite{Rustamov:2007:LED} yields an identical distance to the commute time distance \(\phi_C\). 
At this point it becomes apparent that all of these distances are variations on a common theme: embed the surface in \(\mathbb{R}^n\) according to the eigenfunctions; compute a weighted Euclidean distance in the embedding space.  The different choice of weights will impact the regularity and other features of the resulting distance function---for instance, the diffusion distance exhibits a degree of anisotropy around the source that depends on the diffusion time \(t\), as shown in \cite[Figure 2]{Lipman:2010:BD}.

Notably, \emph{none} of these distances give a good approximation of geodesic distance: on the whole they look more like highly regularized versions of the true geodesic distance, and cannot resolve the cut locus.  However, they do provide an extremely efficient way to obtain a rough proxy for distance: since all of these functions are quite smooth, one can obtain a good approximation by using a truncated series where \(n\) is much smaller than the number of vertices \(|V|\) (in practice, around 150--200).  In this case one can precompute the eigenfunctions, at which point it becomes extremely efficient to compute the point-to-point geodesic distance (by just evaluating a small sum), and by extension, solving the APGD problem for a small subset of vertices (\eg, a collection of landmark points) becomes relatively inexpensive.  However, computing the distance to a large set of vertices (such as the boundary of the domain) may be expensive, since one needs to evaluate a large sum for each point; here, other algorithms discussed in this section may be more appropriate.  Finally, since these distances are ultimately derived from distances in Euclidean \(\mathbb{R}^n\), they exactly satisfy key properties of a metric: nonnegativity, symmetry, and triangle inequality.  Note that \emph{any} embedding into \(\mathbb{R}^n\) will immediately satisfy these properties; likewise, as with any algorithm built on top of the Laplace-Beltrami operator, these distances are also isometry invariant.

Further connections between distance and spectral approximation are detailed in the recent survey by Patan\'e~\cite{Patane:2016fr}.

\subsubsection{Poisson-based Methods}
\label{sec:PoissonBasedMethods}

There is a large class of methods for generating smoothed distance-like functions to the boundary \(\partial M\) of a domain \(M\) by solving linear elliptic equations; these methods are largely motivated by problems from image processing and computer vision, such as extracting medial axes or other ``skeletons'' from image regions.  For instance, Tari~\etal \cite{Tari:1997:ESS} solve the \emph{screened Poisson equation}
\begin{equation}
   \label{eq:ScreenedPoisson}
   \begin{array}{rcll}
      (\Delta - \tfrac{1}{\rho^2}\mathrm{id}) u &=& 0 & \text{on}\ M, \\
      u &=& 1 & \text{on}\ \partial M,
   \end{array}
\end{equation}
and Gorelick~\etal~\cite{Gorelick:2006:SRC} solve a Poisson equation
\begin{equation}
   \label{eq:PoissonSkeleton}
   \begin{array}{rcll}
      \Delta u &=& -1 & \text{on}\ M, \\
      u &=& 0 & \text{on}\ \partial M;
   \end{array}
\end{equation}
derivatives of such functions can then be used to extract an approximate cut locus/medial axis to serve as a ``skelton'' of the region \(M\).  Many variants of this strategy have been considered that yield more distance-like functions, such as applying various pointwise transformations or ``normalizations'' to the solution \(u\) of PDEs like Equation~\ref{eq:ScreenedPoisson} or \ref{eq:PoissonSkeleton}; such as the \emph{Spalding-Tucker} transformation \(u \mapsto \sqrt{|\nabla u|^2 + 2u} - |\nabla u|\)~\cite[Section 6]{belyaev2015variational}.  Much like the functions studied in \secref{SpectralDistances}, however, none of these smoothed functions correspond to the actual geodesic distance \(\phi\).  Moreover, unlike the functions from \secref{SpectralDistances}, they do not provide a well-defined distance metric \(d: M \times M \to \mathbb{R}\), but rather just compute the distance transform of the domain boundary \(\partial M\).

\subsubsection{Heat Method}
\label{sec:HeatMethod}

The diffusion-based perspective can also be used to compute the true geodesic distance, rather than a distance-like function, via the \emph{heat method}~\cite{Crane:2017:HMD}.  This method is ultimately connected to both the eikonal equation discussed in \secref{WavefrontPropagation} as well as the Poisson-based approaches from \secref{PoissonBasedMethods}, though the starting point is an important observation about the close relationship between the geodesic distance function \(\phi\) (\secref{GeodesicDistance}) and the short-time heat kernel \(k_t\) (Equation~\ref{eq:HeatKernel}) known as \emph{Varadhan's formula}~\cite{Varadhan:1967:OTB}:
\begin{equation}
   \label{eq:VaradhanFormula}
   \phi(x,y) = \lim_{t \to 0} \sqrt{-4t \log k_t(x,y)}.
\end{equation}
This formula effectively says that if a ``pin prick'' of heat centered at the point \(x\) diffuses for a \emph{very} short time \(t\), then the resulting heat distribution looks nearly identical to the geodesic distance function, up to a simple transformation of the value at each point.  However, it is quite challenging to compute a numerical approximation of the heat kernel that decays at exactly the right rate; Crane~\etal{} sidestep this issue by recalling the eikonal equation (Equation~\ref{eq:eikonal}), which says that the gradient must have unit magnitude (\(|\nabla\phi|=1\)).  Therefore, any function \(u\) that is a monotonic function of distance can be used to obtain the distance itself, by normalizing the gradient and recovering the corresponding scalar potential.  In particular, to compute the distance to a point \(x\) one can apply the following procedure:
\begin{enumerate}
   \item Solve the heat equation \(\tfrac{d}{dt} u_t = \Delta u_t\), \(u_0 = \delta_x\);
   \item Evaluate the vector field \(X = -\nabla u/|\nabla u|\);
   \item Solve the Poisson equation \(\Delta \phi = \nabla \cdot X\).
\end{enumerate}
Step 1 is approximated by a single step of backward Euler, \ie, by solving the linear elliptic equation
\begin{equation}
   \label{eq:HeatBackwardEuler}
   (\mathrm{id} - t\Delta)u_t = \delta_x
\end{equation}
for a small time step \(t > 0\).  The Poisson equation in the third step corresponds to minimization of the energy \(\int_M |\nabla\phi - X|^2\ dA\), \ie, it finds the function \(\phi\) whose gradient is as close as possible to \(X\) (in the \(L^2\) sense).  Note that all of these steps are described as operations in the smooth setting; as with many PDE-based methods, a final algorithm is obtained by picking a particular discretization of the domain and corresponding discrete differential operators.  For instance, the heat method has been implemented on triangle meshes, polygonal meshes, point clouds~\cite{crane2013geodesics}, images~\cite{Yang:2015:GDC}, subdivision surfaces~\cite{deGoes:2014:DTF}, tetrahedral meshes~\cite{belyaev2015variational}, spline surfaces~\cite{Nguyen:2016:CFE}, and voxelizations~\cite{Caissard:2017:HKL}.  A variant of the heat method that replaces the Laplace operator \(\Delta\) with the \emph{connection Laplacian} \(\Delta^\nabla\) also computes parallel transport of vectors along shortest geodesics~\cite{Sharp:2018:VHM}, as well as the logarithmic map discussed in \secref{smooth_theory}.  An interesting consequence of using higher-order elements of splines and subdivision surfaces is that one can in principal obtain estimates of geodesic distance that are even more accurate than the exact polyhedral distance---see \cite{Nguyen:2016:CFE} and in particular \cite{DeGoes:2016im}, who explore higher-order schemes.

One can use the heat method to connect ideas from the various PDE-based methods considered so far.  From the perspective of \secref{PoissonBasedMethods}, Equation~\ref{eq:HeatBackwardEuler} is a screened Poisson equation (where the boundary is just a single point \(\partial M = \{x\}\)), and Varadhan's formula can be viewed as one possible ``normalization.''  However, as shown in Crane~\etal~\cite[Figure 6]{crane2013geodesics}, this simple transformation does not produce an accurate distance approximation, motivating the need for a more sophisticated normalization strategy.  The heat method can also be given a variational interpretation which connects it to fast marching methods~\cite{belyaev2015variational}.  In particular, consider the energy
\[
   E(u) := \int_M (|\nabla\phi| - 1)^2\ dA,
\]
\ie, the \(L^2\) residual of the eikonal equation; the corresponding Euler-Lagrange equations are given by the nonlinear equation
\[
   \Delta\phi = \nabla \cdot (\nabla\phi/|\nabla\phi|).
\]
The heat method can therefore be viewed as the first iteration of a fixed point scheme, where the solution to the heat equation provides an initial guess \(\phi_0\) for \(\phi\).  Belyaev and Fayolle show that improved accuracy can be achieved by applying successive iterations \(\Delta\phi_{k+1} \gets \nabla \cdot (\nabla\phi_i/|\nabla\phi_i|)\), and more broadly, by applying other optimization strategies to minimize \(E(u)\), albeit at higher computational cost.  This point of view provides a clear connection between the heat method and the fast marching method, which can likewise be viewed as a numerical method for minimizing \(E(u)\) (as discussed in \secref{FastMarching}).  Finally, Litman \& Bronstein note that the heat kernel \(k_t\) is well-approximated via a spectral expansion akin to those seen in \secref{SpectralDistances}, enabling them to dramatically accelerate the heat method via precomputation of Laplacian eigenvectors~\cite{Litman:2016:SAS}.

\section{Computational Geometry Methods}
\label{sec:cg}
The methods reviewed in this section are aimed at resolving the boundary problems in the polyhedral setting. 
We review both exact and approximated methods, by classifying them according to the approach taken to the solution. 
 
\subsection{Global methods}
\label{sec:global}
Global methods compute globally shortest paths.
We distinguish between: methods that work on the whole polyhedral domain and provide an exact solution (Section \ref{sec:exact}); approximations of such methods that trade-off accuracy for speed (Section \ref{sec:approx}); and methods that work on a discretization of the domain on graphs, thus providing necessarily approximated solutions (Section \ref{sec:graph}). 

\subsubsection{Polyhedral Methods -- Exact}
\label{sec:exact}

Methods for solving the polyhedral geodesic distance problem are built on the piecewise flatness of polyhedral surfaces.
This property enables the planar unfolding of triangle strips, which simplifies the computation from 3D polyhedral surfaces to 2D unfolded planes (see Section \ref{sec:polyhedral}).
Within the unfolded triangle strips, locally shortest paths are computed as straight line segments.
By enumerating all possible locally shortest paths between two points, globally shortest paths can be obtained by finding the ones with minimum length. 
However, without an efficient pruning strategy, the number of such locally shortest paths grows exponentially with the size of $\Sigma$ and quickly becomes infeasible to compute \cite{balasubramanian2009exact}.
Therefore, the key challenge is to remove the computational redundancy as much as possible.

Addressing this challenge, O'Rourke et al.\cite{O'Rourke1984} first proposed an algorithm to solve the PPGP problem in $O(n^5)$ time, where $n$ is the number of vertices on $\Sigma$. 
Their method can be viewed as a theoretical milestone from that it shows the PPGP problem can be solved in polynomial time.
However, their method is still too time-consuming for practical applications.
Thus, our discussion starts from \cite{Mitchell:1987}, which laid the foundation of practical polyhedral geodesic algorithms.

\begin{figure}[h]
	\centering
	\includegraphics[width=0.8\linewidth]{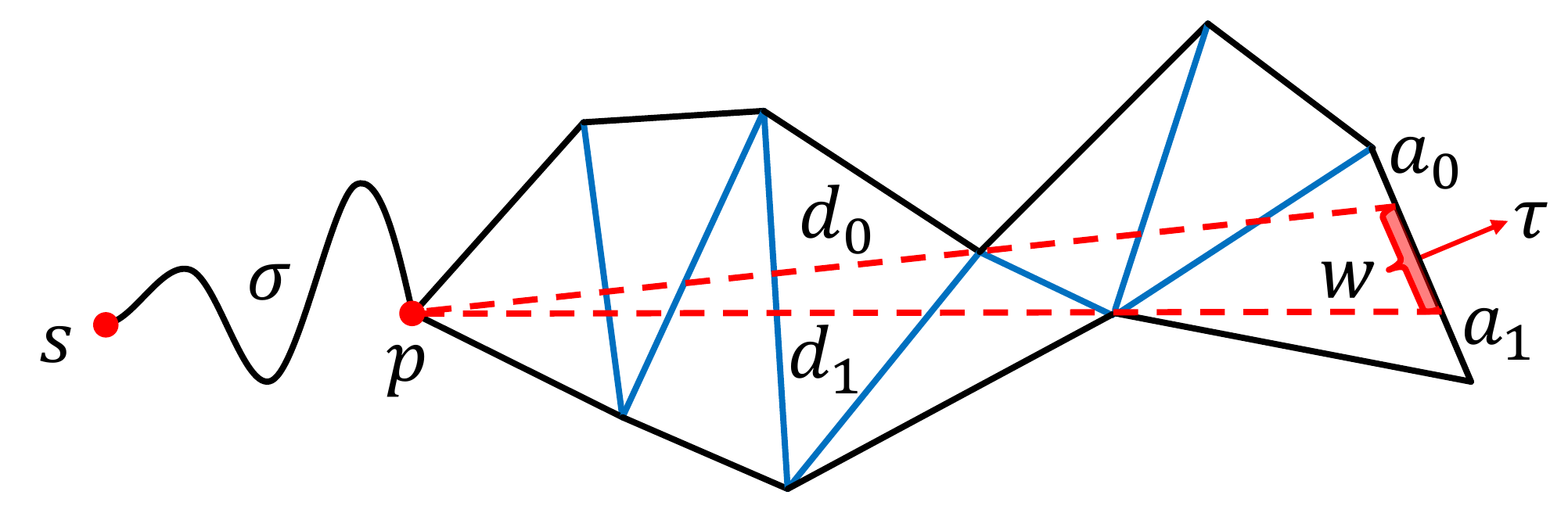}
	\caption{\label{fig:exact_window_structure}
		Illustration of window structure. The shortest paths within an unfolded triangle strip can be encoded as a window $w = (a_0, a_1, d_0, d_1, \sigma, \tau)$, where $a_0$, $a_1$ are the endpoints of $w$; $d_0$, $d_1$ are the corresponding distances from $a_0$, $a_1$ to the pseudosource $p$ (saddle vertex); $\sigma$ denotes the geodesic distance from $p$ to source $s$; and $\tau$ denotes the propagation direction. Adapted from \cite{Surazhsky:2005,qin2017fast}.
	}
\end{figure}

\begin{figure}[h]
	\centering
	\includegraphics[width=0.9\linewidth]{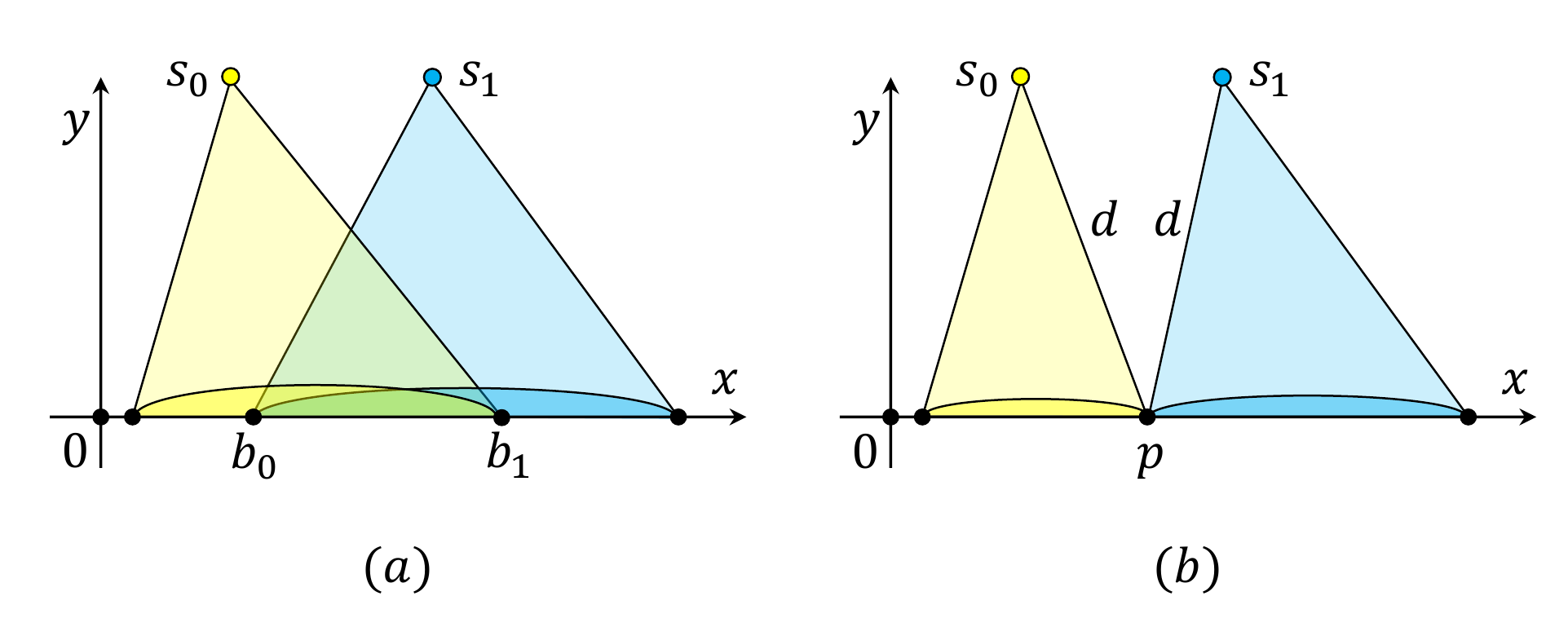}
	\caption{\label{fig:exact_MMP_trimming}
		Illustration of window trimming. 
		(a) Suppose there are two overlapping windows whose unfolded
		pseudosources are $s_0$ and $s_1$ and their intersection $\delta = [b_0, b_1]$. 
		(b) By solving a quadratic equation, the two windows are trimmed into disjoint ones. An illustration of special case $\sigma_0$ = $\sigma_1$ and then $\| s_0 - p\| = \|s_1 - p\| = d$ is shown. From \cite{Surazhsky:2005}.
	}
\end{figure}

Mitchell et al. \cite{Mitchell:1987} proposed the first practical algorithm for geodesic computation on polyhedral surfaces, which is commonly referred to as the MMP (Mitchell-Mount-Papadimitriou) algorithm.
Their main idea can be summarized as the \textit{continuous Dijkstra} technique, which extends the famous Dijkstra's algorithm \cite{Dijkstra:1959} from graphs to polyhedral surfaces.
As an analogy, they propose to view edges of polyhedral surfaces as nodes of a graph. 
However, an edge contains infinite points and its distance cannot be represented by a scalar value.
To handle this problem, a dedicated data structure named \textit{window}\footnote{In \cite{Mitchell:1987}, Mitchell et al. used the term \textit{interval}. 
While in the following works \cite{Surazhsky:2005,xin2009improving,XuWLL015,Qin:2016}, an alternative term \textit{window} gains popularity from its intuitiveness.
Thus, the term \textit{window} is employed in this survey.} is introduced, which encodes all locally shortest paths in an unfolded triangle strip with a tuple (Figure \ref{fig:exact_window_structure}).
Then, the polyhedral geodesic distances can be computed by finding the optimal windows on edges of $\Sigma$.
The optimization of windows on an edge is accomplished by trimming overlapping windows into disjoint ones according to the smaller distance in the overlapping part (Figure \ref{fig:exact_MMP_trimming}). 
More specifically, windows are trimmed in a pairwise manner after being propagated to edges and stored in an ordered list according to their positions. 

It can be observed that MMP's time cost is positively correlated to the number of windows arriving each edge.
To minimize it, the MMP algorithm borrows the wavefront propagation paradigm appearing in Dijkstra's algorithm and fast marching (\secref{FastMarching}) and propagates windows across $\Sigma$ from near to far by maintaining a priority queue.
Such paradigm ensures the redundant windows will be trimmed at the earliest possible stage.
When propagation is finished, each edge of $\Sigma$ will be subdivided into a list of end-to-end linked windows containing the geodesic distance field to any point on $\Sigma$.
Mitchell et al. proved that the algorithm creates at most $O(n^2)$ windows.
Thus, it can be easily derived that MMP can solve the SSGD problem in $O(n^2\log{n})$ time and $O(n^2)$ space, where $n$ is the number of vertices of $\Sigma$. 
The key component of their proof is the analysis of the maximal number of windows arrived at each edge, which is $O(n)$.
Nevertheless, such analysis is too pessimistic and is inconsistent with MMP's practical performance.

The MMP algorithm \cite{Mitchell:1987} is commonly viewed as a landmark in the research of polyhedral geodesic algorithms. 
Its distinct contribution is the \textit{window propagation framework}, which contains three major components: window propagation, window pruning (e.g. trimming) and window management (e.g. priority queue).
This framework is employed by most of the following works \cite{Chen:1990,Surazhsky:2005,xin2009improving,XuWLL015,Qin:2016} and they differ from their unique techniques used in the three components. 

\begin{figure}[h]
	\centering
	\includegraphics[width=0.8\linewidth]{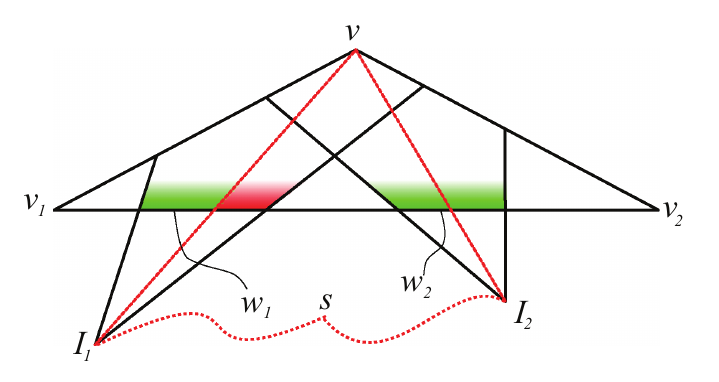}
	\caption{\label{fig:exact_CH_one_angle_one_split}
		Vertex $v$ splits any window spanning it into two halves.
		The ``one angle, one split'' rule shows that for any two windows passing through $v$, only three of the four split windows can provide shortest paths and continue propagation.
		As illustrated, the three green windows are valid and continue propagation.
		The red one contains no shortest paths beyond triangle $\Delta vv_1v_2$ and stops propagation. 
		From \cite{xin2009improving}.
	}
\end{figure}

Although it is straightforward for ``continuous Dijkstra'' technique to use a priority queue to manage windows, an extra $O(\log n)$ time cost is introduced, which slows down the computation. 
Addressing this issue, Chen and Han \cite{Chen:1990} proposed the CH (Chen-Han) algorithm. 
Their algorithm manages windows with a First-In-First-Out (FIFO) queue, whose overhead is $O(1)$.
For window pruning, they proposed a very simple rule named ``one angle, one split'' to remove redundant windows around vertices of $\Sigma$ (Figure \ref{fig:exact_CH_one_angle_one_split}).
This rule only cares about the redundancy around vertices and is tolerant of overlapping windows. Thus, it is much less powerful than the trimming rule used by the MMP algorithm \cite{Mitchell:1987}.
However, it can still be proved that the number of windows generated in CH algorithm is at most $O(n^2)$. 
This analysis also supports that the worst-case analysis of the MMP algorithm is too pessimistic and that its practical performance should be much better.
Since the CH algorithm uses a $O(1)$ FIFO queue to manage windows and has a $O(n^2)$ window complexity, its overall time complexity is $O(n^2)$.
In addition, the CH algorithm does not store propagated windows on edges to perform window pruning. Thus, its space cost is $O(n)$.


Theoretically, the CH algorithm achieves the best asymptotic complexities so far.
However, its practical performance is poor. 
Surazhsky et al. \cite{Surazhsky:2005} reported that their implementation of the MMP algorithm runs in sub-quadratic time and is many times faster than Kaneva and O'Rourke's implementation of the CH algorithm \cite{Kaneva:2000}.
Since then, several methods have been proposed to improve the practical performance of the MMP algorithm. 
Liu et al. \cite{Liu:2007} observed that floating point error may cause the degeneration of window propagations to frequently occur when applying the MMP algorithm to real world models. 
To make the MMP algorithm more robust, they conducted a systematic analysis on all the degenerated cases and proposed techniques to handle them accordingly.
Observing that the half-edge data structure used in Surazhsky et al.'s implementation \cite{Surazhsky:2005} may generate redundant windows, Liu \cite{Liu:2013} proposed to implement the MMP algorithm with an edge-based data structure. 
Experimental results show that on average the edge-based version runs 44\% faster and uses 29\% less memory.

In summary, the MMP algorithm's practical success comes from its effective wavefront propagation paradigm which enables the removal of redundant windows at the earliest stage.
However, the removal of redundant windows is a complex and expensive process which involves inserting a newly propagated window into a ordered list and trimming overlapping parts by solving quadratic equations.

\begin{figure}[h]
	\centering
	\includegraphics[width=0.9\linewidth]{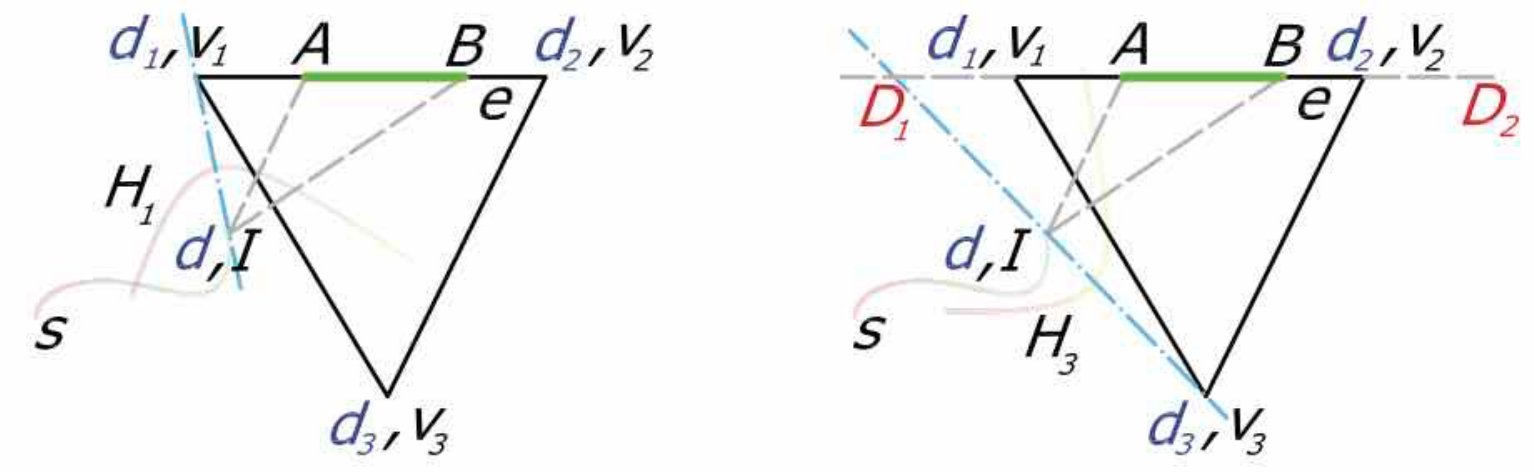}
	\caption{\label{fig:exact_ICH_window_filtering}
		Illustration of ICH's window filtering rules.
		The key idea is to filter out redundant windows with the minimum-so-far distance $d_1, d_2, d_3$ of vertices.
		Left: $d + \|\overline{IB}\| > d_1 + \|\overline{v_1B}\|$ and its symmetric counterpart $d + \|\overline{IA}\| > d_2 + \|\overline{v_2A}\|$;
		Right: $d + \|\overline{IA}\| > d_3 + \|\overline{v_3A}\|$.
		From \cite{xin2009improving}.
	}
\end{figure}

Addressing this issue, Xin and Wang \cite{xin2009improving} proposed the ICH (Improved Chen-Han) algorithm, which combines the advantages of both the MMP algorithm and the CH algorithm.
From the MMP algorithm, they borrowed the wavefront propagation paradigm and used a priority queue to manage window propagations according to their distances.
To avoid the MMP's costly window trimming operations, they borrowed the ``one angle, one split'' rule from the CH algorithm and extended it into three novel window filtering rules (Figure \ref{fig:exact_ICH_window_filtering}).
Experimental results show that these new rules help to remove more than 99\% redundant windows during propagation.
Theoretically, introducing the priority queue increases ICH's time complexity to $O(n^2\log{n})$ and makes its space complexity no longer $O(n)$. 
However, its practical performance increased dramatically.
Experimental results show that ICH greatly outperforms the original CH algorithm.
In addition, although ICH usually propagates more windows, it runs comparable to the MMP algorithm \cite{Surazhsky:2005} while using considerably less space.
This reveals that there exists a trade-off between the effectiveness of window pruning strategies and their costs. 

\begin{figure}[h]
	\centering
	\includegraphics[width=0.9\linewidth]{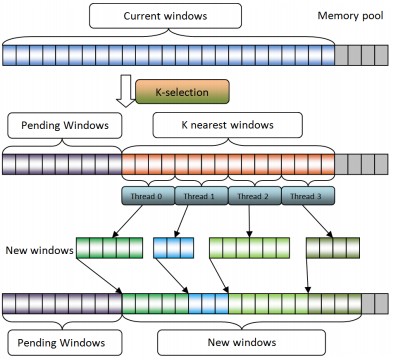}
	\caption{\label{fig:exact_PCH_k_nearest}
		The PCH algorithm selects $k$ nearest windows and assigns them to different GPU threads. 
		Each thread propagates the selected windows and filters the redundant ones independently.
		Then, the newly created windows are collected and re-organized in the memory pool. 
		From \cite{Ying:2014}.
	}
\end{figure}

One prominent difference between MMP and ICH's window pruning rules is that: MMP's trimming rule involves interaction between windows while ICH's 
``one angle, one split'' and window filtering rules do not. 
More specifically, ICH filters redundant windows by distance comparison with the minimum-so-far distances at vertices, which is a relatively independent process and is ideal for parallelization.
However, there is still one obstacle.
To remove redundant windows at the earliest stage, ICH organizes window propagations in a strict order with a priority queue, which is sequential.
Fortunately, since the correctness of window propagation algorithms is independent of the order of window propagations, this strategy can be loosened for parallelization.
Based on the above observations, Ying et al. \cite{Ying:2014} proposed the PCH (Parallel Chen-Han) algorithm, which is a parallel version of the ICH algorithm.
In their algorithm, $k$ nearest windows are selected and propagated in parallel at each iteration, where $k$ is a user-specified parameter (Figure \ref{fig:exact_PCH_k_nearest}).
Then, the propagated windows are filtered in parallel according to the minimum-so-far distances at vertices.
To avoid data conflict at vertices, the distance updates are delayed until the propagation of selected windows finished.
Experiment results show that PCH propagates slightly more windows than the ICH algorithm, but runs an order of magnitude faster.
 
\begin{figure}[t]
	\centering
	\includegraphics[width=0.6\linewidth]{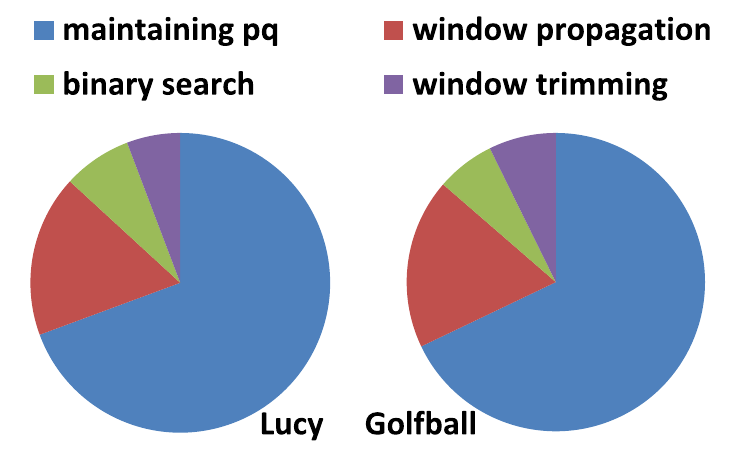}
	\caption{\label{fig:exact_FWP_priority_queue_cost}
		The running time breakdown of the MMP algorithm. 
		Note that the maintenance of priority queue is the most time-consuming part which occupies around 70\% of the running time.
		The ICH algorithm has a similar percentage.
		From \cite{XuWLL015}.
	}
\end{figure}

The performance of the PCH algorithm shows that slightly relaxing the order of propagation will not have a big impact on the window pruning effectiveness. 
Based on this spirit, \cite{XuWLL015} proposed the FWP (Fast Wavefront Propagation) technique to accelerate the MMP and ICH algorithm by replacing the strictly ordered priority queue with a loosely-ordered bucket-based FIFO queue.
More specifically, windows with similar distances are stored in the same bucket and propagated in a FIFO order.
Although straightforward, their method performs well in practice from the observation that both MMP and ICH spend roughly 70\% time on maintaining the priority queue (Figure \ref{fig:exact_FWP_priority_queue_cost}).
Experimental results show that their FWP-MMP algorithms runs 3-10 times faster than the MMP algorithm, and their FWP-CH algorithm runs 2-5 times faster than the ICH algorithm.

\begin{figure}[t]
	\centering
	\includegraphics[width=0.9\linewidth]{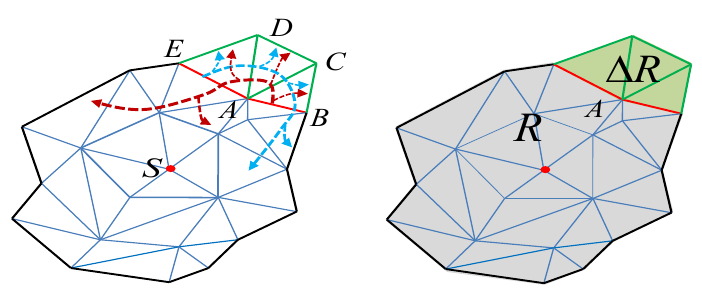}
	\caption{\label{fig:exact_VTP_triangle_growing_scheme}
		Illustration of triangle-oriented growing scheme.
		The black and red line segments denote the boundary of the traversed area.
		The green line segments denote the triangle to be added into the traversed area.
		Left: The triangles in vertex $A$'s 1-ring neighbourhood is adding into the traversed area.
		The red line segments become interior edges. Thus, the windows on them are propagated through the traversed area until they reach the updated boundary edges.
		Right: $R$ is the traversed area before growing. After growing, $\Delta R$ will be merged into $R$.
		From \cite{Qin:2016}.
	}
\end{figure}

Both PCH and FWP employ a batch processing strategy to accelerate existing algorithms.
However, their selection of window batches depends only on the distances and lacks the consideration of their geometric interrelationship.
Observing that the prunings of all propagated windows within a triangle are inter-dependent, Qin et al. \cite{Qin:2016} proposed the VTP (Vertex-oriented Triangle Propagation) algorithm. 
Their algorithm organizes window propagations with a triangle-oriented growing scheme.
In this scheme, a traversed area is defined as the union of all visited triangles.
The boundary of this traversed area forms the propagation wavefront, which contains all the windows to be propagated.
Then, this traversed area is expanded in a Dijkstra-like style by gradually enclosing adjacent triangles (Figure \ref{fig:exact_VTP_triangle_growing_scheme}).
Their algorithm terminates when the traversed area covers the entire $\Sigma$.
During expansion, local windows entering the same triangle from the same edge are organized in a batch and they are propagated simultaneously.
In such batches, redundancy checks can be intensively performed between any pair of windows.
To remove a maximal number of windows in a low cost way, they proposed three rules which are summarized from an exhaustive list of scenarios for pairwise window pruning inside a triangle.
Note that these pruning scenarios are listed under the assumption that the window trimming technique \cite{Surazhsky:2005} is not used due to its relatively high computational cost of solving quadratic equations.
Experimental results show that their algorithm runs 4-15 times faster than MMP and ICH algorithms, 2-4 times faster than FWP-MMP and FWP-CH algorithms, and also consumes the least memory, which ranks it as the best performing polyhedral geodesic algorithm to date.

\textbf{Remark:} Most polyhedral geodesic algorithms aim at solving the SSGD problem, while Balasubramanian et al. \cite{balasubramanian2009exact} proposed an algorithm to solve the APGD problem. 
The main idea of their algorithm is to build a vertex-to-vertex graph by computing the \textit{minimal-geodesic distances} between all pairs of vertices on $\Sigma$. The minimal-geodesic distance is the length of the shortest path between two vertices under the condition that the path contains no intermediate vertices.
Then, the shortest distances between any pair of vertex can be computed by searching the vertex-to-vertex graph using standard algorithms like the Dijkstra's algorithm \cite{Dijkstra:1959}.
To compute the minimal-geodesic distance, they employed a \textit{triangle chain flattening} method and proposed to reduce the redundancy through visibility.
However, the overall method is essentially the same as the standard window propagation algorithms without any pruning.
Thus, its practical performance is expected to be much worse than other state-of-the-art geodesic algorithms.
Note that in \cite{balasubramanian2009exact}, the runtime of Surazhsky et al.'s \cite{Surazhsky:2005} MMP implementation is \textit{estimated} but not tested by experiments. Thus, the comparison may be inaccurate.

\begin{figure}[h]
	\centering
	\includegraphics[width=0.7\linewidth]{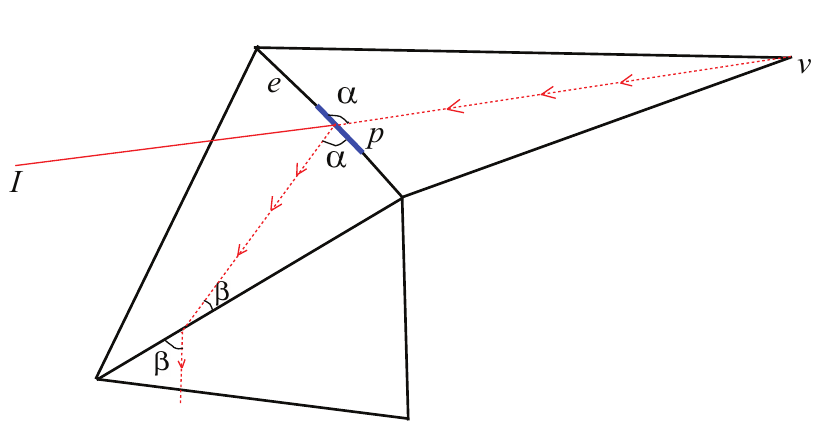}
	\caption{\label{fig:Exact_backtracing_vertex}
		Illustration of backtracing from a vertex of $\Sigma$.
		From \cite{xin2009improving}.
	}
\end{figure}

\begin{figure*}[h]
	\centering
	\includegraphics[width=0.7\textwidth]{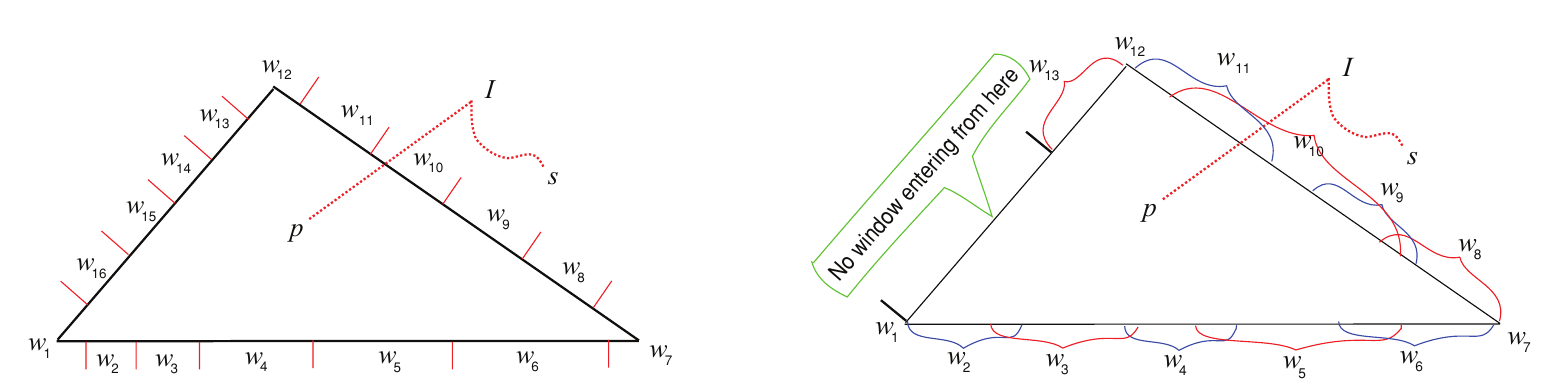}
	\caption{\label{fig:Exact_backtracing_point}
		Illustration of backtracing from a point of $\Sigma$. In this scenario, propagated windows are required to be stored on edges. However, they can be organized in different ways. Left: windows on edges are trimmed into non-overlapping ones as the MMP algorithm do \cite{Surazhsky:2005}. Right: redundant windows on edges are filtered as the ICH, VTP algorithms do \cite{xin2009improving,Qin:2016}. The remaining windows may overlap with each other.
		From \cite{xin2009improving}.
	}
\end{figure*}

\paragraph*{Shortest Path Construction.} 
All the reviewed methods employed a straightforward \textit{backtracing} strategy to construct geodesic paths, thus supporting SSSP, too. 
In general, there are two related approaches to answer two queries respectively, which can be easily adapted to all polyhedral methods:
\begin{itemize}
	\item \textbf{Shortest paths to vertices:} As Figure \ref{fig:Exact_backtracing_vertex} shows, for a vertex $v \in \Sigma$ whose shortest distance is obtained from window $w$, let $I$ be the image of $w$'s pseudosource on the plane defined by $v$ and $w$. 
	Then, the entering point $p$ on edge $e$ can be computed easily by intersecting $\overline{vI}$ and edge $e$. 
	The entire path can be constructed by backtracing in a similar way. 
	Note that the image $I$ for each vertex of $\Sigma$ can be recorded during window propagation and do not require any extra data structure, which is space-efficient.
	\item \textbf{Shortest paths to generic points:} As Figure \ref{fig:Exact_backtracing_point} shows, propagated windows must be stored on edges of $\Sigma$ as the data structure supporting geodesic path queries to points. For an interior point $p$ of a triangle, the window containing its shortest path can be computed by $w_{p} = \arg \min_{w \in W}{\| p - p' \| + D(p')}$, where $W$ is a set containing all windows on the three edges of the triangle, $p'$ is a point in $w$, $D(p')$ is the geodesic distance of $p'$. Then, the shortest path of $p$ can be computed by backtracing from $p$ to $p'$ in $w_p$ until reaching the source. Note that this approach can be space-consuming since it requires storing all propagated windows on edges.
\end{itemize}

\subsubsection{Polyhedral Methods -- Approximated}
\label{sec:approx}
Methods for exactly solving polyhedral geodesic problems can be time-consuming for large-scale applications.
Thus, some methods approximate the polyhedral distance, which can be applied in scenarios insensitive to accuracy.

To solve the SSGD problem efficiently, Surazhsky et al. \cite{Surazhsky:2005} proposed an approximated version of the MMP algorithm to reduce the time and memory costs. 
This algorithm works just as the MMP algorithm. 
The only difference is that: in the approximated version, the algorithm tries to merge a window with adjacent windows on the same edge before every propagation step (Figure \ref{fig:exact_MMP_merge_windows}).
To make sure that the merged window is valid and with bounded error, the two windows to be merged must satisfy five conditions:
\begin{figure}[t]
	\centering
	\includegraphics[width=0.5\linewidth]{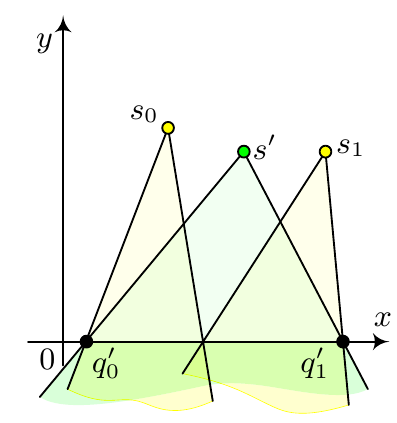}
	\caption{\label{fig:exact_MMP_merge_windows}
		Merging two windows into a new window.
		$s'$ is the pseudosource of the new window, which covers the visible regions of both merged windows.
		From \cite{Surazhsky:2005}
	}
\end{figure}
\begin{itemize}
	\item \textbf{Directionality.} The two windows must have the same propagation direction with respect to the edge.
	\item \textbf{Visibility.} To guarantee that the approximated distance field have no gaps, the new window must cover the visible regions of the merged windows.
	\item \textbf{Continuity.} The distance field along an edge must be continuous. Thus, the distances at the endpoints of the new window must be the same as the corresponding ones of the merged windows.
	\item \textbf{Accuracy.} The algorithm bounds the local error of each merging step by a user-specified threshold.
	\item \textbf{Monotonicity.} To guarantee the algorithm's termination, the new window must have larger distances than the merged windows. However, the authors did not encounter any infinite loops in practice and thus they did not perform the corresponding checks in their implementation.
\end{itemize}
In their experiments, the local error threshold is set to be 10\%.
The test results show that the merging operation is very effective and reduces the WPE (Window Per Edge) to a low number slightly larger than 1.
Thanks to it, their approximated MMP algorithm runs in $O(n \log n)$ time and outperforms the Fast Marching method \cite{Kimmel98computinggeodesic} (see Section \ref{sec:fem}) in both running time and accuracy.

\begin{figure}[t]
	\centering
	\includegraphics[width=0.6\linewidth]{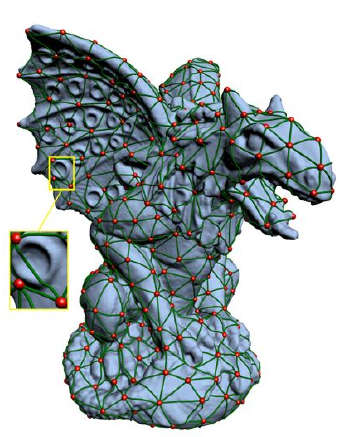}
	\caption{\label{fig:exact_Xin2012_delaunay}
		The Delaunay triangulation of a mesh according to polyhedral geodesic metric.
		From \cite{Xin:2012}
	}
\end{figure}

To solve the APGD problem efficiently, Xin et al. \cite{Xin:2012} proposed an algorithm which uses the ICH algorithm \cite{xin2009improving} as a subroutine.
The main idea is to embed pre-computed geodesic triangles in $\mathbb{R}^2$ so that the geodesic distance between two points can be approximated by the Euclidean distance.
Their algorithm contains two steps:
\begin{itemize}
	\item \textbf{Pre-processing step.} In this step, $m$ points are first sampled on $\Sigma$. Then, the ICH algorithm \cite{xin2009improving} is used to compute the geodesic distance field on $\Sigma$ with the $m$ samples as sources. Utilizing the computed distance field, Delaunay triangulation is applied on $\Sigma$ according to polyhedral geodesic metric (Figure \ref{fig:exact_Xin2012_delaunay}). Finally, the distances between each pair of the $m$ sample points, together with the distances between each mesh vertex and the vertices of the geodesic triangle containing it, are saved for later use.
	\item \textbf{Query step.} For any two points on $\Sigma$, the two geodesic triangles containing them are first found.
	Then, the two triangles together with the two points are unfolded onto $\mathbb{R}^2$ while preserving the geodesic edge lengths.
	After the unfolding, the geodesic distance between the two points is computed as the Euclidean distance between them.
\end{itemize}

From the time cost of the ICH algorithm, it can be easily concluded that the pre-processing step consumes $O(mn^2 \log n)$ time.
After it, the distance query between any two points on $\Sigma$ can be answered in $O(1)$ time.
Intuitively, there is a trade-off between accuracy and computational cost in this algorithm.
That is, the more sample points, the slower the pre-processing and the larger the space requirements, but more accurate.

\paragraph*{Shortest Path Construction.} 
The approximate MMP algorithm \cite{Surazhsky:2005} employed a similar backtracing strategy as the original MMP method with only one key difference.  Since the windows in it are merged, the positions of their pseudosources are inaccurate hence not eligible for backtracing.  Thus, the authors proposed to maintain a list of the successively merged windows and use the original pseudosources for backtracing, which yields a more accurate result.  Xin et al. \cite{Xin:2012} employed a quite different strategy to construct shortest paths: since the data structure of their method contains only geodesic distances but not windows, they are only equipped with an approximated shortest distance field.
Thus, they proposed to construct shortest paths using a gradient descent strategy similar to the ones used in \cite{Kimmel98computinggeodesic,crane2013geodesics} (see Section \ref{sec:fem}).


\subsubsection{Graph-based Methods}
\label{sec:graph}
Graph-based methods rely on the assumption that the shortest geodesic distance/path between any pair of points $p_s$ and $p_t$ can be approximated with a chain of shortest distances/paths $(p_s,v_0,\ldots,v_k,p_t)$, where $v_0,\ldots,v_k$ belong to a finite set $V_G$ of points of $S$ such that the shortest distance/path between pairs of points of $V_G$ is precomputed and encoded in the edges $E_G$ of a graph $G=(V_G,E_G)$.
Methods differ for the choice of points of $V_G$ end edges of $E_G$ and the strategy to build graph $G$. 
Once $G$ is given, the PPGP and SSGD problems are easily resolved through shortest path queries on $G$, most frequently with standard Dijkstra search \cite{Dijkstra:1959}. Several methods in this class provide data structures that can also support efficient solutions to SSSP or APGD/APSP.

\begin{figure}[tb]
	\centering
	\includegraphics[width=0.7\linewidth]{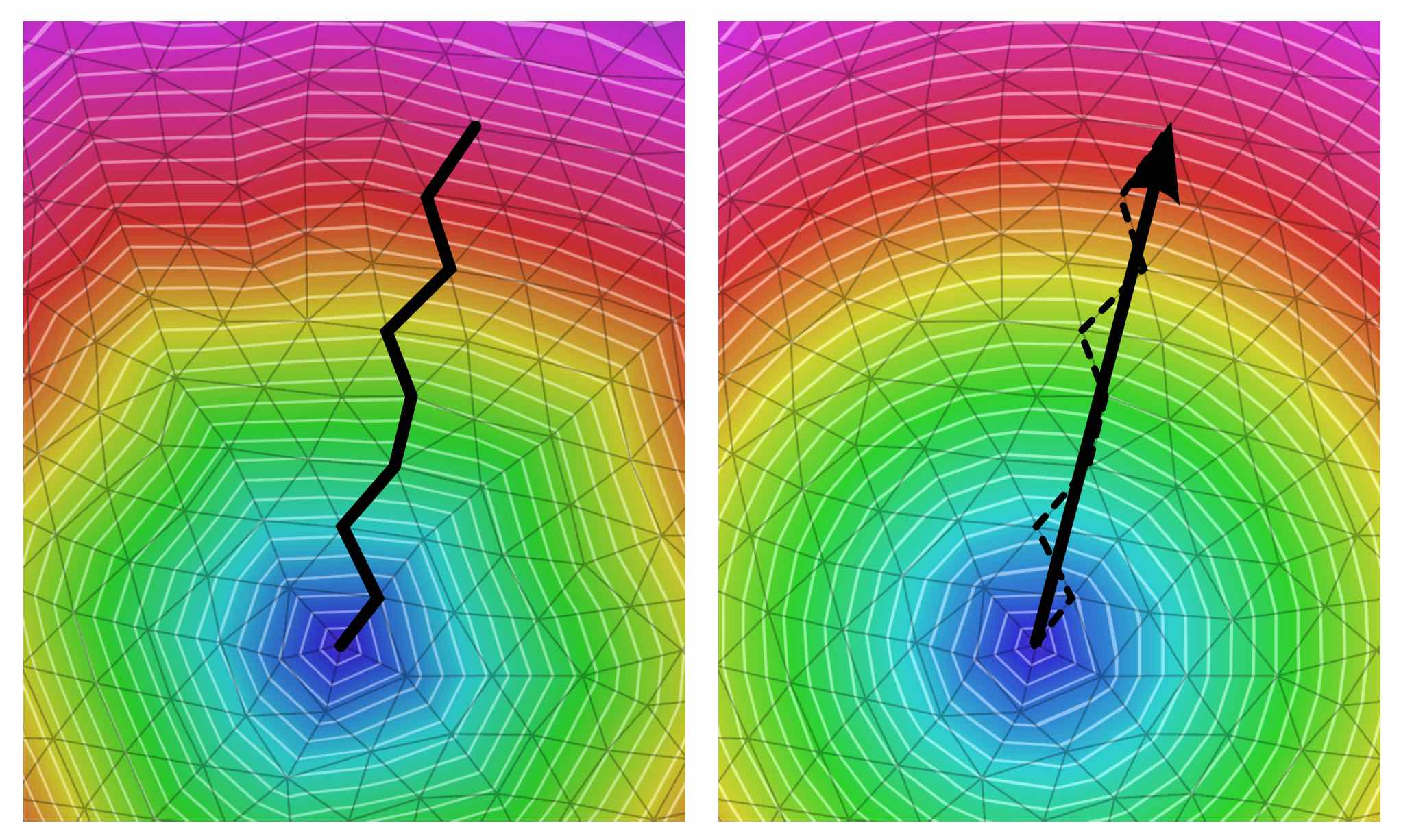}
	\caption{\label{fig:zigzag}
		A shortest path (black line) and a corresponding distance field (color map) computed on the edge graph (left); and the exact solution (right). From \protect\cite{Campen:2013}.
		}
\end{figure}

The quest for graph-based methods probably started with the PhD thesis of Lanthier \cite{Lanthier:1999}. 
The bottom line comes from the observation that the shortest geodesic path between a pair of vertices of mesh $\Sigma$ can be approximated by a shortest path on the graph of edges of $\Sigma$. 
This approximation, however, turns out to be poor. 
In fact, such a path is not allowed to cross the faces of $\Sigma$, while it is forced to meander about their boundaries, thus resulting in a zigzag walk that most of the times is far more wiggly than the exact geodesic path (see Figure \ref{fig:zigzag} for an example).
Starting at this observation, Lanthier et al.\ \cite{Lanthier:1997bg,Lanthier:2001} present three strategies to extend the edge graph to a graph $G$ that incorporates paths across the faces of $\Sigma$.
They build the set of nodes $V_G$ by distributing Steiner points along the edges of $\Sigma$, and interconnect them with arcs $E_G$ that walk either along edges, or across faces:
\begin{itemize}
\item The \emph{fixed} scheme distributes a fixed number of Steiner points uniformly along each edge and defines $V_G$ as the union of all the vertices of $\Sigma$ and all the Steiner points. Next, for each triangle $t$, it interconnects all the vertices and Steiner points on the boundary of $t$ to form a so-called \emph{face graph}, which connects each node (either a vertex or a Steiner point) with all the nodes belonging to the other two edges of $t$ and to its two neighbors along the edge it belongs to. In practice, the face graph of $t$ is a complete graph, except for omitting arcs between collinear nodes that are not immediately adjacent along edges of $t$ (see Figure \ref{fig:facegraph} left).
Graph $G$ is then obtained by collecting all the face graphs.
Now we known that an exact polyhedral geodesic path is a polyline having its nodes at either vertices or edges of $\Sigma$ \cite{Mitchell:1987}.
The segment $ab$ of one such path traversing a given triangle $t$ is approximated by an edge $cd$ in the face graph of $t$, where $c$ and $d$ are the closest nodes in the face graph of $t$ along the edges that contain $a$ and $b$, respectively (see Figure \ref{fig:facegraph} right).   
\begin{figure}[tb]
	\centering
	\includegraphics[width=0.4\linewidth]{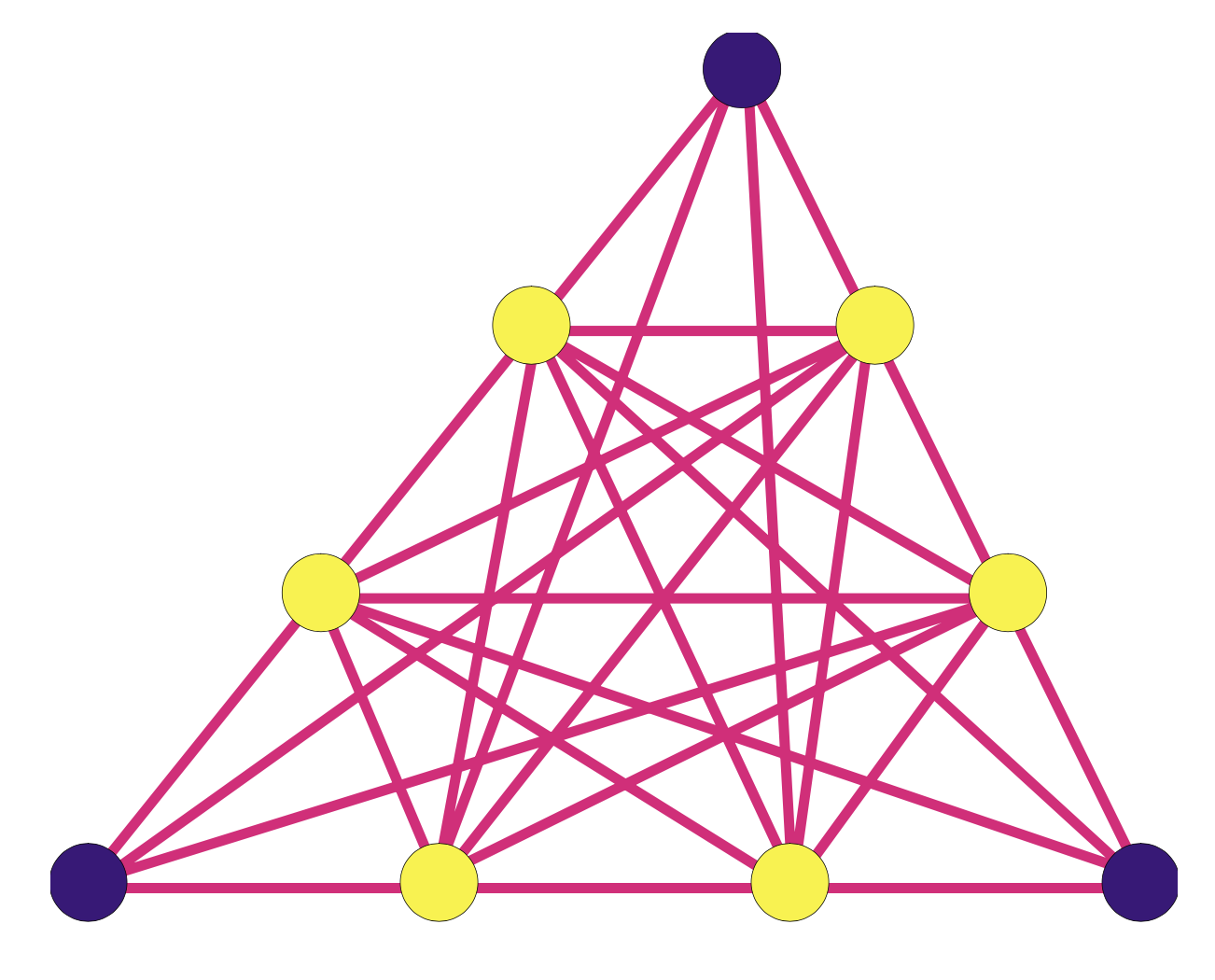} \hspace{1cm}
	\includegraphics[width=0.4\linewidth]{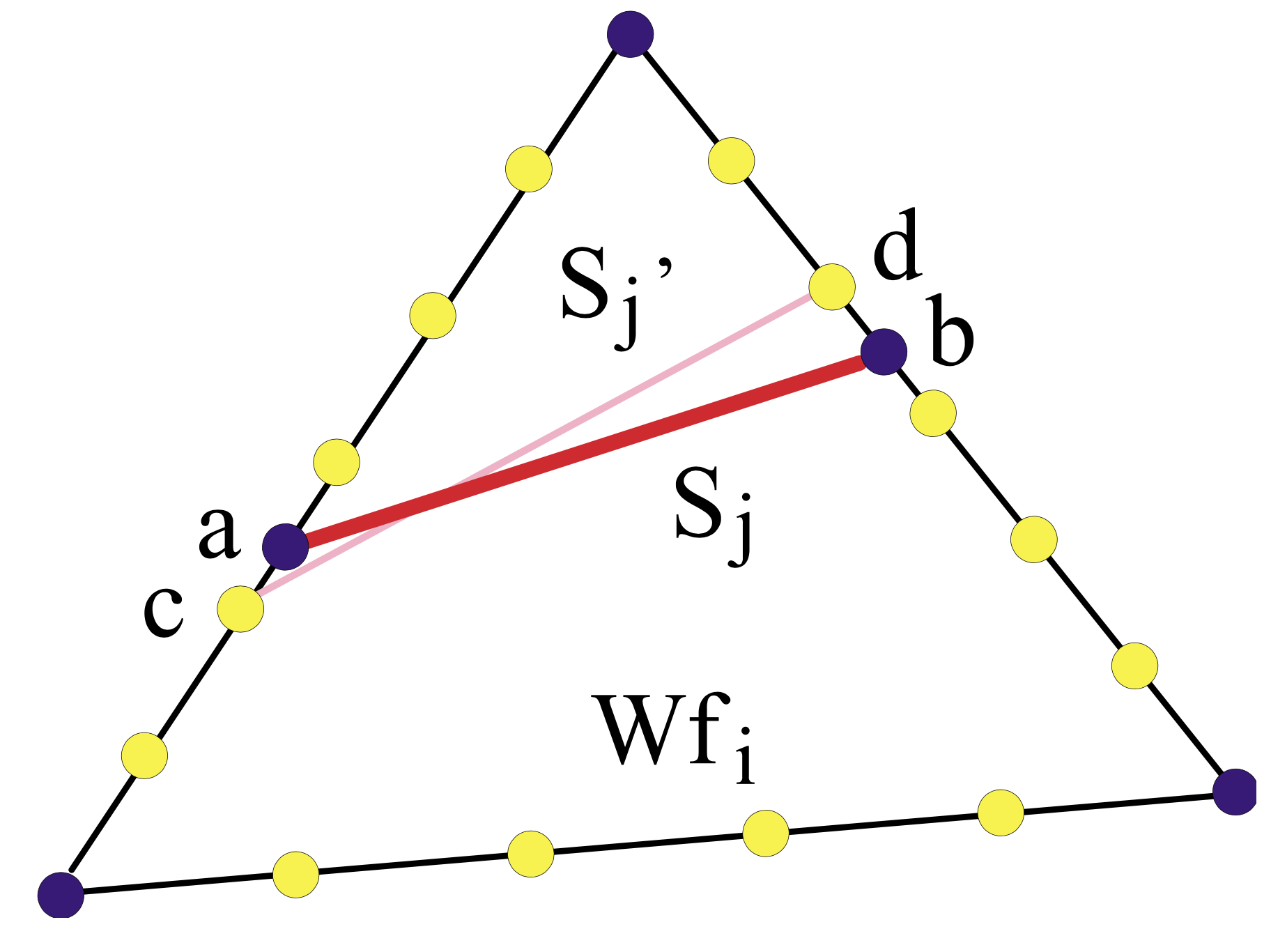}	
	\caption{\label{fig:facegraph}
		The face graph of a triangle for the fixed scheme with two Steiner points per edge (left). 
		The portion of shortest path $ab$ crossing triangle $Wf_i$ is approximated with edge $cd$ of $E_G$ (right). From \protect\cite{Lanthier:2001}.
		}
\end{figure}

\item The \emph{interval} scheme is similar to the previous one, but Steiner points are distributed at uniform distance along each edge, so that the number of Steiner points per edge depends on edge length. Lanthier at al.\ show that this scheme can achieve the same accurcay of the previous one with a smaller number of Steiner points, \ie, a more compact graph $G$ (see Figure \ref{fig:fixed-interval}).

\begin{figure}[htb]
	\centering
	\includegraphics[width=\linewidth]{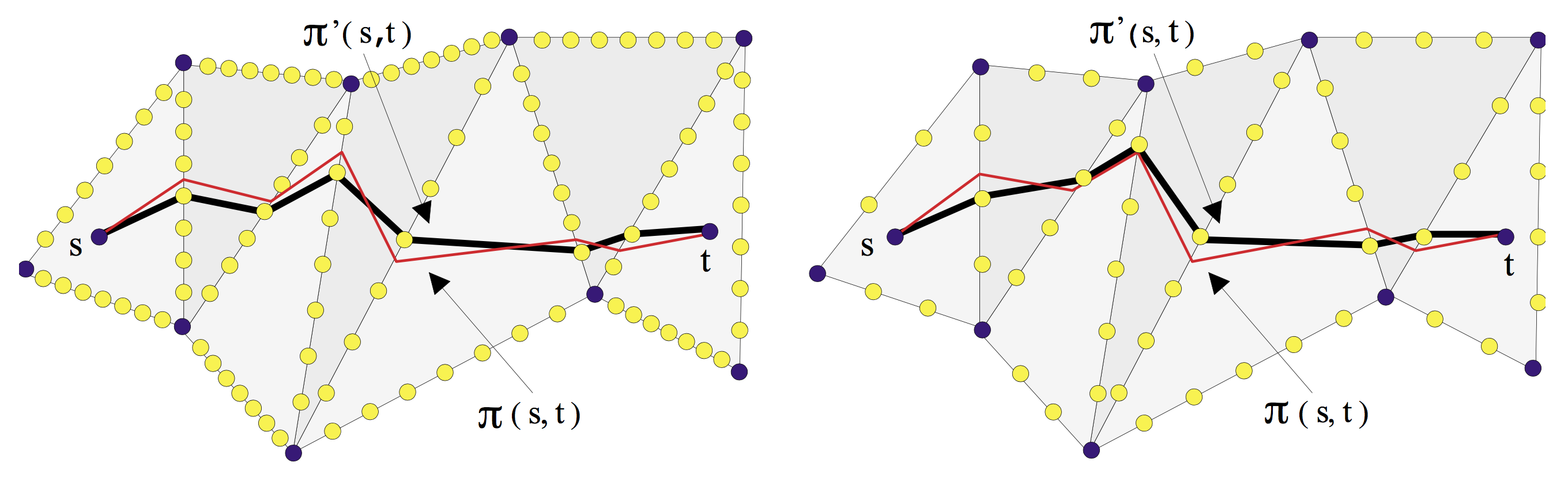}
	\caption{\label{fig:fixed-interval}
		Approximated paths computer with the fixed scheme (left) and the interval scheme (right): $\pi(s,t)$ is the shortest geodesic path, $\pi'(s,t)$ is the approximated path. From \protect\cite{Lanthier:2001}.
		}
\end{figure}

\item The \emph{spanner} scheme uses the same Steiner points as the interval scheme, but it builds a more sparse graph $G$. Instead of connecting each node $v$ in the face graph to all other visible nodes in the triangle, a predefined set of cones of given width is considered, which emanate from $v$, and $v$ is connected to at most one other node per cone (see Figure \ref{fig:spanner}). In this way, graph $G$ is guaranteed to have a number of arcs that is at most a fixed multiple of  $|V_G|$. 
\end{itemize}

\begin{figure}[htb]
	\centering
	\includegraphics[width=0.5\linewidth]{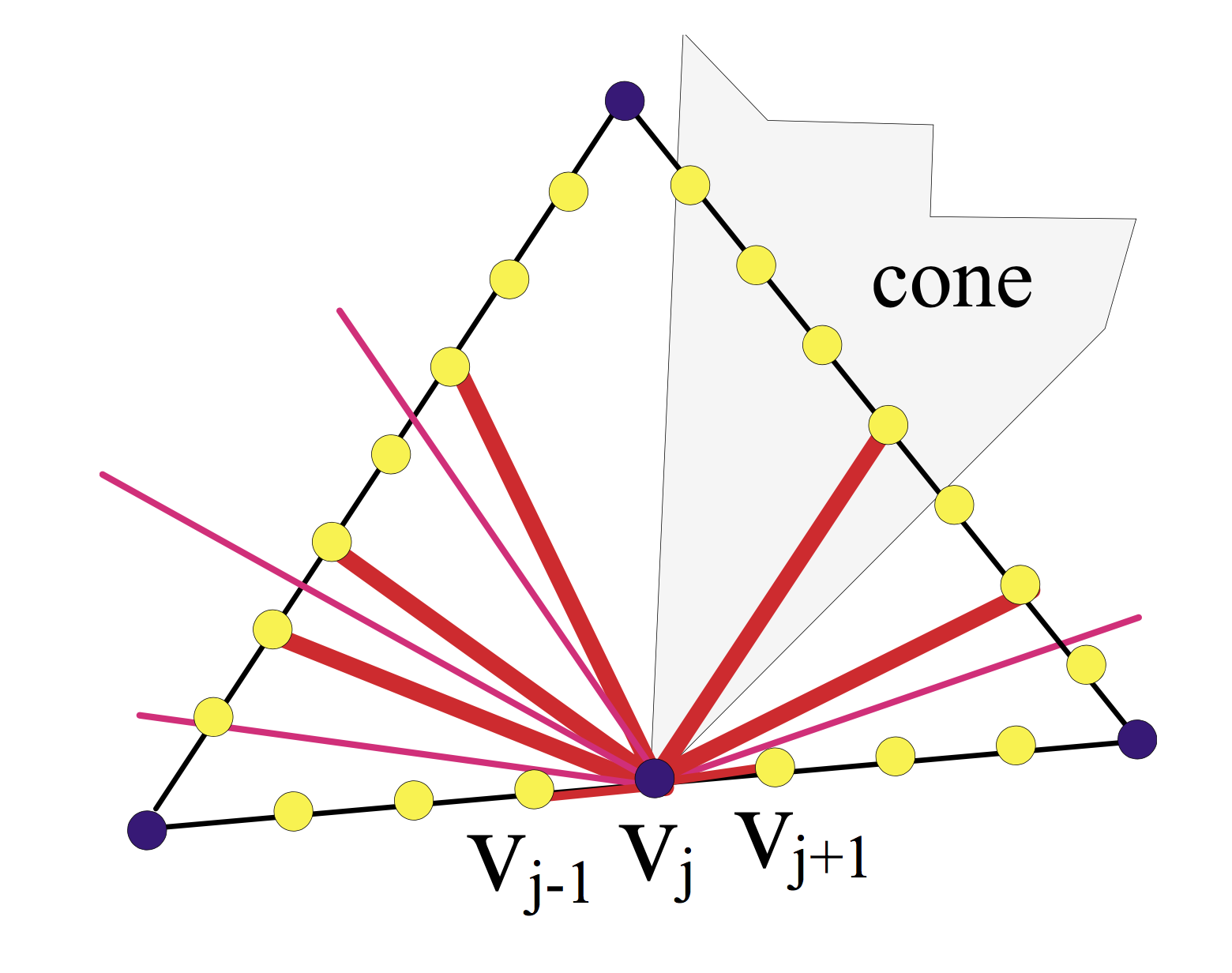}
	\caption{\label{fig:spanner}
		Spanner edges from point $v_j$ (thick red lines) for cones of width $30^{\circ}$; thin red lines bound cones. From \protect\cite{Lanthier:2001}.
		}
\end{figure}

In all schemes, shortest paths are computed by a Dijkstra search on $G$: SSGD can be resolved with a complete visit of $G$, while for PPGP search can be pruned as soon as distance at the target point becomes definite. 
SSSP is supported by recording during search the predecessor of each node in the path to the source: in this way, shortest paths can be retrieved by back-tracing from any vertex to the source. 
 
Lanthier et al. \cite{Lanthier:2001} prove that if the number of Steiner points is large enough, they can approximate the geodesic shortest path within an additive bound that is a function of the length of the longest edge in $\Sigma$. 
The interval scheme results more accurate, while the spanner scheme results more compact and faster, at the cost of some loss in accuracy. 
Theoretical bounds are conservative, though, and reasonable bounds can be guaranteed only with a very large number of Steiner points, which is impractical for the applications. 
On the other hand, they perform extensive testing of the different schemes on polyhedral terrains and provide empirical evidence that very accurate results are achieved with just an average of 5-6 Steiner points per edge. 
More recent comparative tests given in \cite{Campen:2013} suggest that the situation may be less favorable if the distance metric is anisotropic: in that case, they report the need of 10-20 Steiner points per edge on average to achieve accurate results. 
Nevertheless, Lanthier's techniques remain a reference in the context of graph-based methods, because they join practical effectiveness to ease of implementation. 

\begin{figure}[tb]
	\centering
	\includegraphics[width=0.6\linewidth]{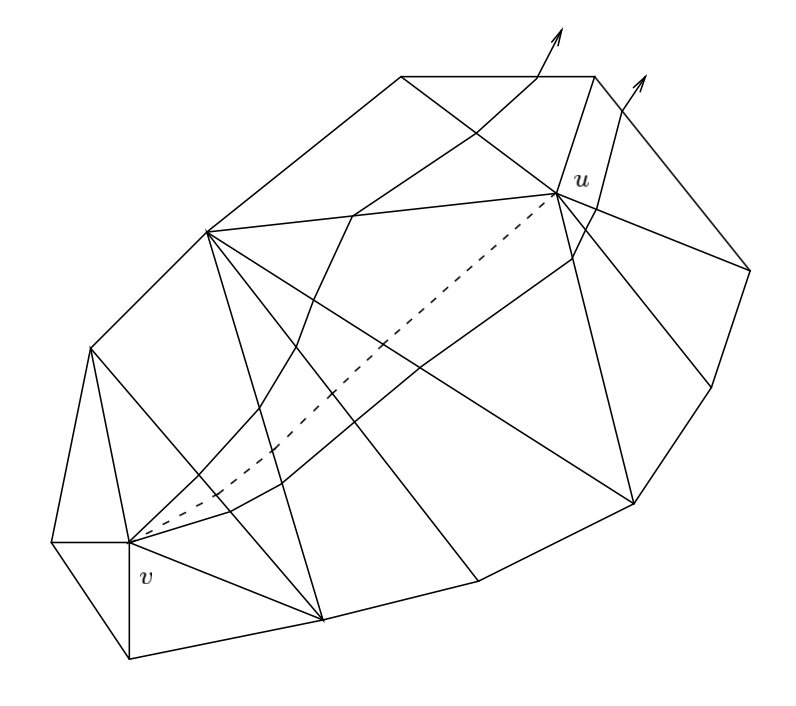}
	\caption{\label{fig:mata}
		A cone determined by two rays (continuous paths) from $v$ is split at $u$; the path from $v$ to $u$ (dashed) is added as an edge in graph $G$. From \protect\cite{Mata:1997kt}.
		}
\end{figure}

In \cite{Mata:1997kt}, Mata and Mitchell propose a scheme that is similar in spirit to the spanner scheme of Lanthier. 
Differently from Lanthier, they do not add Steiner points, but they enrich the edge graph of $\Sigma$ with new arcs connecting vertices of $V$, in such a way that for each vertex $v$ of $\Sigma$ a rather uniform set of directions radially arranged about $v$ is explored. 
They build $k$ equally spaced cones about $v$ and they propagate the boundaries of such cones according to an exact polyhedral geodesic tracing (\eg, as in MMP).
Propagation is stopped as soon as the two boundaries of a cone intersect different edges of $\Sigma$: this means that there exist a vertex $u$ within the cone, which splits it into two; then they determine the shortest path between $v$ and $u$  and add it as an edge to $E_G$. 
See Figure \ref{fig:mata}.
Note that in this case an edge of $E_G$ is not a simple line segment but rather a polyline.
Shortest paths in $G$ are computed through Dijkstra, as in Lanthier's method.

Following the same line of Lanthier's approach, Aleksandrov et al. \cite{aleksandrov1998varepsilon,Aleksandrov:2000gh} proposed different techniques to sample Steiner points on edges in geometric progression (\ie, more dense close to the vertices and coarser towards the mid edge). 
The density -- hence the number -- of sampled points depends on a user specified parameter $\epsilon$ and they guarantee that the length of the approximated paths is within a factor $(1+\epsilon)$ of the shortest geodesic path.   
The same authors, in \cite{Aleksandrov:2005dq}, propose a different strategy that samples Steiner points along the bisectors of triangles, rather than on triangle edges, with a similar geometric progression. They show that this strategy achieves the same  $(1+\epsilon)$  approximation factor with a better time complexity.

The results in   \cite{aleksandrov1998varepsilon,Aleksandrov:2000gh,Aleksandrov:2005dq,Mata:1997kt} have mostly a theoretical interest, as the number of Steiner vertices necessary to warrant a small $\epsilon$ is too large for practical purposes.  
Unfortunately, no experimental results were provided to show whether the sampling techniques proposed in such works can provide better results than the original Lanthier's  techniques in practice.  

\begin{figure}[tb]
	\centering
	\includegraphics[width=\linewidth]{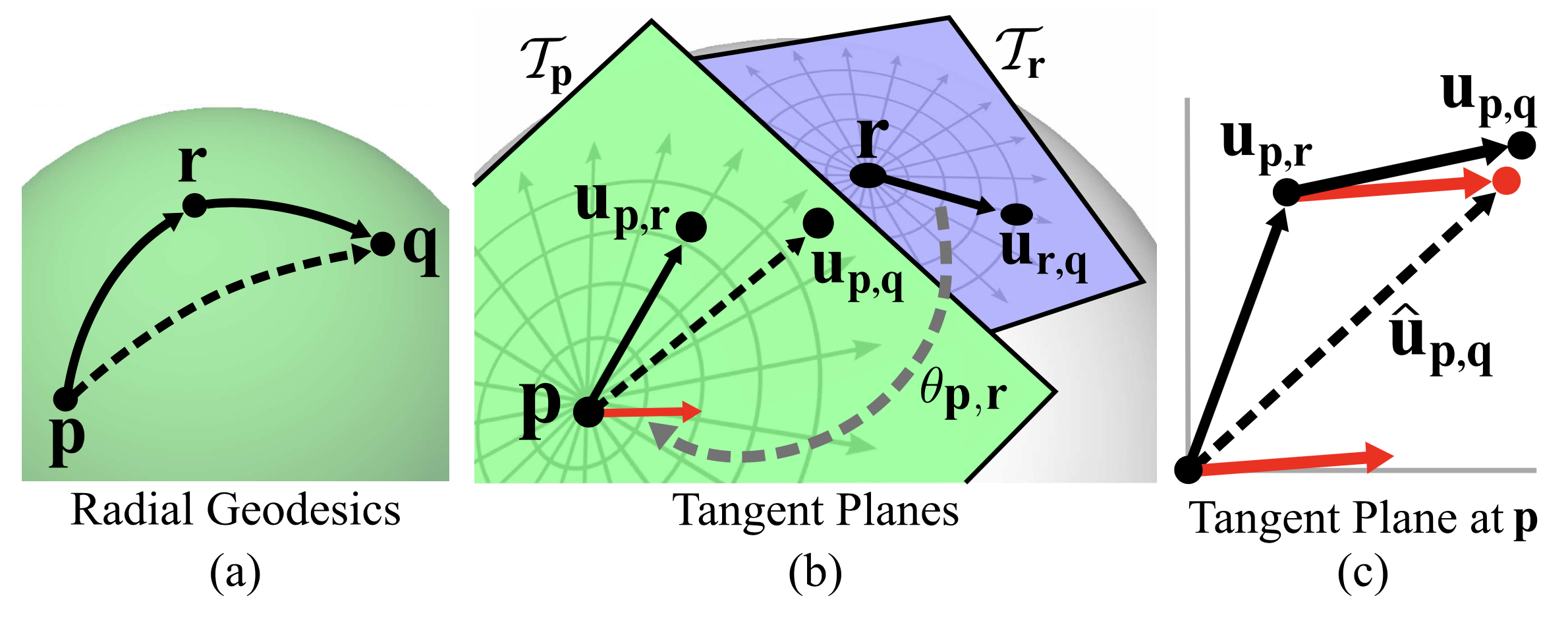}
	\caption{\label{fig:Schmidt-transport}
		The normal coordinates $\mathbf{u}_{p,q}$ of the unknown radial geodesic from $p$ to $q$ in (a) can be approximated using the known geodesics from $p$ to $r$ and $r$ to $q$. The vector to $\mathbf{u}_{r,q}$ (in normal coordinates at $r$) is transferred to the tangent plane at $p$ using a 2D rotation with angle $\theta_{p,r}$, producing the red vector in (b). This vector is an approximation to $(\mathbf{u}_{p,q} - \mathbf{u}_{p,r})$ and can be added to $\mathbf{u}_{p,r}$ (c) to get the approximate result $\hat{\mathbf{u}}_{p,q}$. From \protect\cite{Schmidt:2006tt}.
		}
\end{figure}

\begin{figure}[tb]
	\centering
	\includegraphics[width=\linewidth]{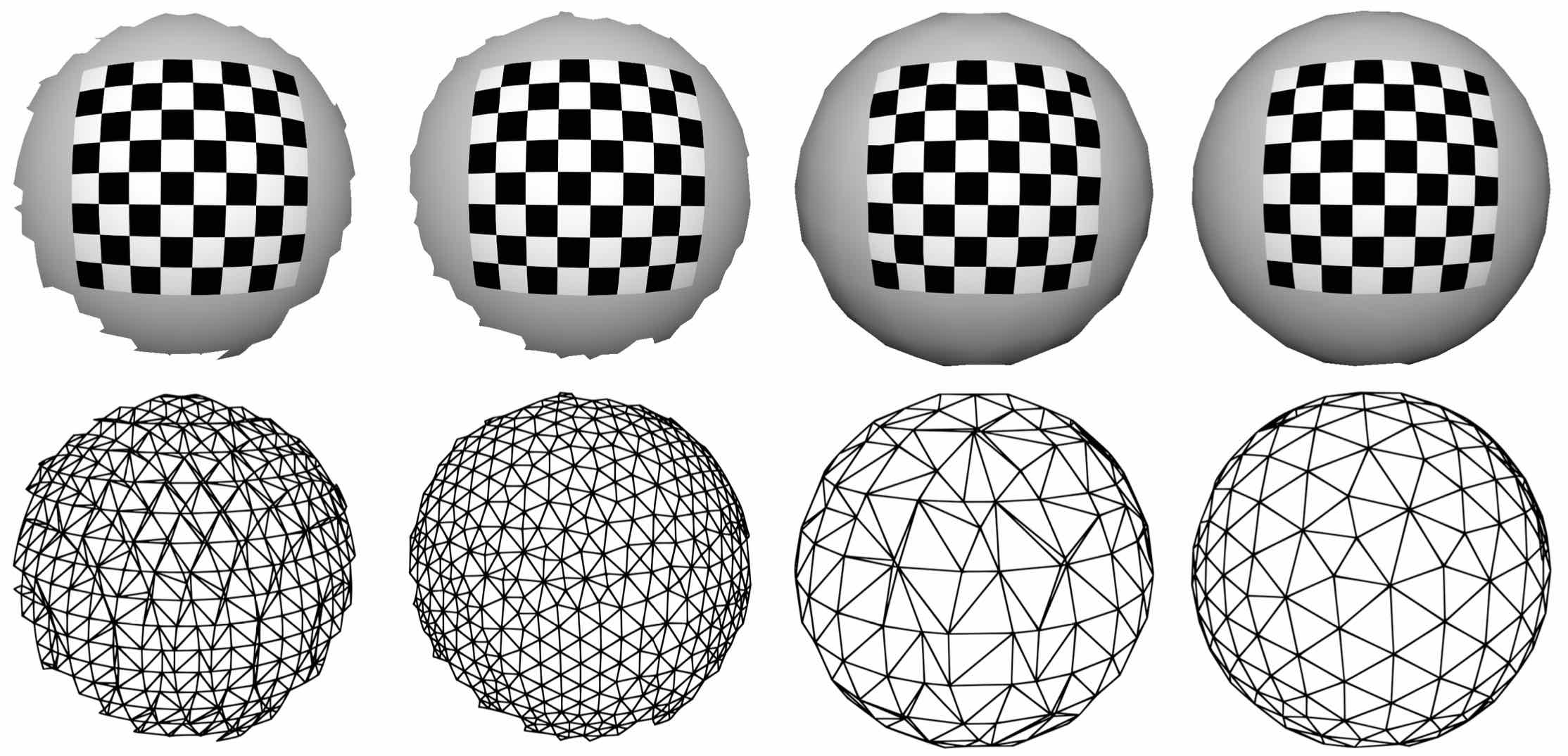}
	\caption{\label{fig:Schmidt-sphere}
		The DEM with origin at the center of the checkerboard is computed on different discretizations of the sphere (below);  sampling rate and mesh quality have small effects on this decal. From \protect\cite{Schmidt:2006tt}.
		}
\end{figure}

Schmidt et al. \cite{Schmidt:2006tt} build a \emph{discrete exponential map} (DEM) approximation in order to parametrize normal patches of surface. 
In the context of the problems we analyze here, they are interested in resolving SSGD (not SSSP) within the limited scope of a normal neighborhood of the source.  
In order to estimate geodesic distances, they rely solely on the edge graph of $\Sigma$. 
Once a shortest path from source $p$ to a generic vertex $v$ has been found in the edge graph, they consider edges in the path as displacement vectors and they use parallel transport to bring all such vectors to a common frame. 
Parallel transport is implemented with a pair of rotations that align local frames at the vertices along the path. 
See Figure \ref{fig:Schmidt-transport}.
Next they estimate the geodesic distance between $p$ and $v$ as the length of the vectorial sum of all such displacements. 
The length and orientation of the resulting vector in fact provide the polar coordinates of $v$ in the DEM centered at $p$.
In practice, this mechanism attempts to ``straighten" the wiggly path from the edge graph. 
Note that, however, they do not compute the straightened path, but only estimate its length. 
No theoretical analysis of accuracy is provided in \cite{Schmidt:2006tt}, but empirical tests show good results and resiliency to the quality of meshing on reasonably smooth shapes. See Figure \ref{fig:Schmidt-sphere}.

\begin{figure}[tb]
	\centering
	\includegraphics[width=\linewidth]{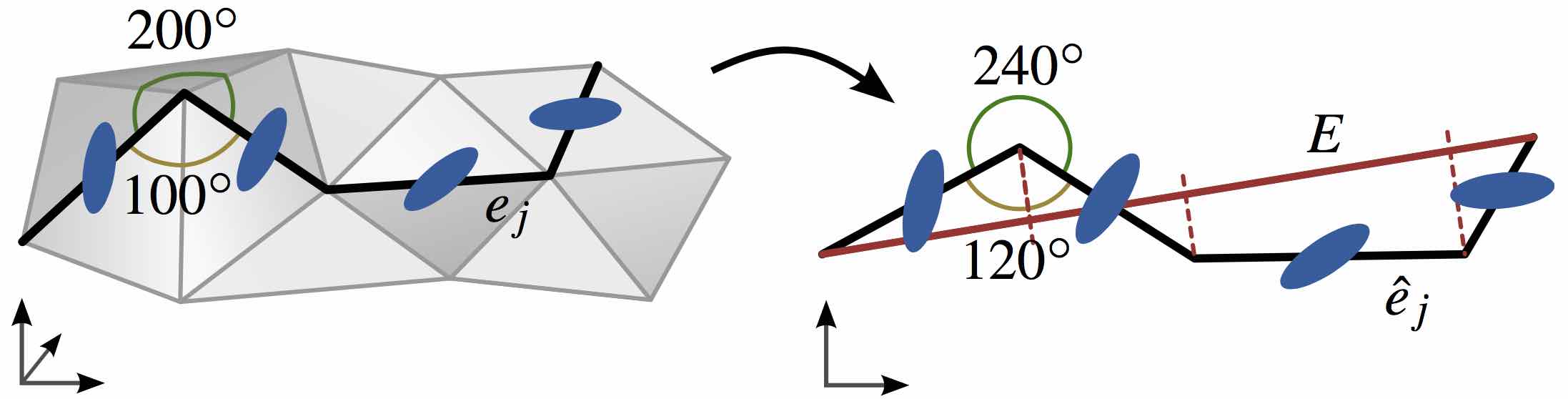}
	\caption{\label{fig:anisotropic}
		Unfolding of edge chains to the plane. Edge lengths and relative 1-ring angles are preserved. The sum vector $E$ (red) is then subdivided according to the orthoprojection of the individual edges. The resulting portions are measured by the respective norms ? here visualized as tensor ellipses (blue) ? and their lengths summed to get the weight of $E$. From \protect\cite{Campen:2013}.
		}
\end{figure}

Campen et al.\ \cite{Campen:2013} elaborate on the same approach in the presence of an anisotropic metric, aiming at a method to resolve SSGD on the whole surface. 
They start from the observation that the straightened distance from a path in the edge graph introduces a shortcut that cannot take into account either the presence of holes (boundaries) in the domain, or geometric variations of the surface (bumps). 
Hence, the straightening approach by Schmidt et al.\ cannot be accurate outside a normal neighborhood.
In order to overcome this limitation, they propose a \emph{Short-Term Vector Dijkstra} algorithm (STVD) that applies shortcuts locally in the context of a standard Dijkstra search in the edge graph. 
During the relax phase of Dijkstra algorithm, distance from source $s$ to vertex $w$ is updated by taking the minimum between the distances of each of the $k$ predecessors of $w$ along the path, summed to the shortcut, computed by edge unfolding, from $w$ to that predecessor.
In practice, the vector shortcut of Schmidt et al. is applied just locally, on a sliding window of length $k$ that moves along the path. 
Values of $k$ in the range between 5 and 10 are reported in the experiments, with higher values used for higher anisotropies of the input metric.
Shortcuts are computed with a technique different from \cite{Schmidt:2006tt}: the edge path is unfolded to the plane by preserving the length of edges, while mapping the angles between pairs of consecutive edges according to the exponential map at the common vertex, computed as in \figref{SmoothStructure}. Note that this procedure is fully intrinsic, as the rescaling of angles depends only on the total angle about each vertex in $\Sigma$.
The length of a shortcut is measured by the Euclidean distance between the source and the target in the unfolded path. 
In the presence of anisotropic distance, that shortcut may be segmented in order to apply different weights in the segments corresponding to different edges. See Figure \ref{fig:anisotropic}.
Campen et al.  \cite{Campen:2013} compare this approach on a single example with anisotropic metric, with respect to several other methods, among which FMM, HM, OUM, and Lanthier's, by taking as reference a solution computed with Lanthier's method with 200 Steiner vertices per edge (which is assumed to be nearly exact).
They report  that only Lanthier's with 10-20 Steiner points per edge beats STDV in terms of accuracy. 
They also comment that the advantage of STDV is much less evident for an isotropic metric; in particular, they report that for low anisotropies, the FMM applied to the intrinsic Delaunay triangulation of a subdivided version of the input mesh is very competitive; no experiment is shown for this case. 

Ying et al.\ \cite{Ying:2013} start from a basic fact stated in \cite{Mitchell:1987}: each exact (polyhedral) geodesic path is a polyline having its internal joints either on edges, or at saddle vertices of $\Sigma$ (a saddle vertex is a vertex having negative Gaussian curvature $K$ - see Section \ref{sec:polyhedral}). A path is said to be \emph{direct} if it does not contain any intermediate saddle vertex. 
On the basis of this observation, they build the \emph{Saddle Vertex Graph} (SVG) $G$ having the vertices of $\Sigma$ as set of nodes $V_G$,  and one arc in $E_G$ for any direct path between a pair of vertices. Note that each arc in $E_G$ is not a single line segment but a polygonal path. Such paths need not be stored as they can be back-traced by unfolding triangles starting at the destination: only a starting direction per edge needs to be stored.
If the SVG contains all direct paths, then the \emph{exact} shortest geodesic path between any pair of vertices can be found as a shortest path on the SVG.
In this respect, the SVG can be considered a data structure to support the APSP problem, as well as all the other boundary problems. 
In principle, the SVG could be a complete graph though (hence intractable in the applications, at least on large meshes).
Ying et al.\ show empirically that real world models contain a large fraction of saddle vertices (between $40\%$ and $60\%$) and most paths 
$(91\% - 99.9 \%)$ are not direct.  
Nevertheless, a full SVG may still be too large for practical purposes. 
Ying et al.\ overcome this limitation by computing a sparsified SVG as follows: for each vertex $v$ they consider only a geodesic disc containing $K$ other vertices and only direct paths to such vertices contribute to the SVG. In this way, $K$ is the maximum degree of nodes in $G$. 
The SVG is built by running an exact method (ICH/MMP) from each vertex, pruning the search as soon as the first $K$ vertices have finalized their distance. 
Variations of Dijkstra search are discussed in the paper, which resolve different boundary problems after the SVG has been built. 
They run experiments with values of $K$ between 8 and 1000 and they report the mean approximation error of shortest paths from SVG to be about $0.1\%$ for $K=50$.  
The construction of the SVG can be quite time expensive for high values of $K$, but they can achieve reasonable times by using a GPU implementation that computes independently in parallel paths starting at different sources, by sharing just the mesh in read-only mode among the different threads.
Concerning query times, Ying et al.\ compare SVG search with the heat method (with pre-conditioned matrix): they report that the SVG is faster for $K<100$, comparable for $100<K<500$ and slower, but more accurate, for larger values of $K$; they also observe that the SVG is better scalable to large meshes.  
They also compare SVG to the GTU method \cite{Xin:2012}, reporting that SVG is faster and requires less memory; no comparison about accuracy is reported.

 \begin{figure}[htb]
	\centering
	\includegraphics[width=\linewidth]{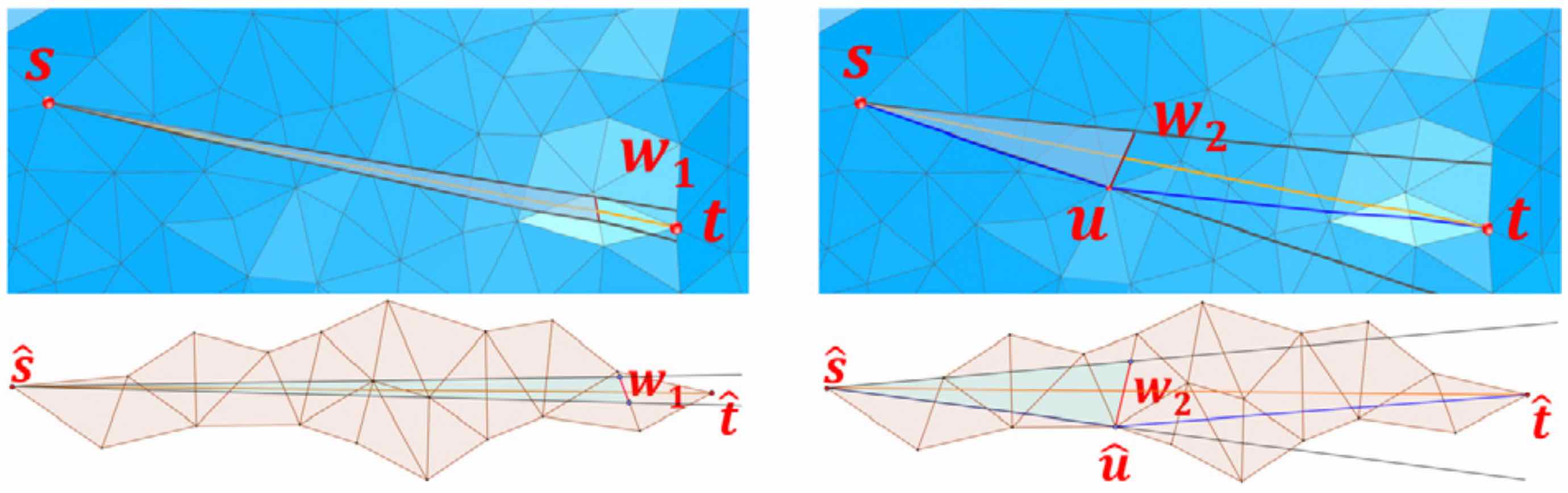}
	\caption{\label{fig:Wang-cone}
		Window propagation in the exact method (left) and in the construction of DGG (right): window $W_2$ is small enough to stop propagation and allow the path from $s$ to any vertex $t$ in the cone beyond $W_2$ to travel through vertex $u$. From \protect\cite{Wang:2017}.
		}
\end{figure}

Wang et al.\ \cite{Wang:2017} define the \emph{Discrete Geodesic Graph} DGG, which improves over the SVG by allowing approximated paths to go also through non-saddle vertices.
They start from the key observation that cones resulting from window propagation in the exact MMP/CH method become progressively narrower. 
If a cone is long enough, than any direct path from the source $s$ to a vertex $t$ in the cone can be approximated by the sum of two shortest paths $su$ and $ut$, where $u$ is a vertex preceding $t$ and lying on the boundary of the cone. See figure \ref{fig:Wang-cone}.
In particular, they show that if the geodesic distance between $s$ and $t$ is larger than a certain threshold, which depends on the parameters defining a window $W$ that bounds the cone and on a tolerance threshold $\epsilon$, then the approximation error is within a $(1+\epsilon)$ multiplicative factor. So, while generating direct paths as in the SVG, they set an early termination of window propagation when the length of the cone reaches the above threshold. Note that a direct path may be eventually approximated with a chain of paths through non-saddle vertices, therefore the approximation rate of a generic path extracted from the DGG will not be within $(1+\epsilon)$, but it remains $O(\epsilon)$ in any case. 
The DGG is built through two fundamental steps, for which detailed pseudo-code is provided in  \cite{Wang:2017}: first the standard MMP window propagation, with early termination as described above, is run from all vertices; an arc in the DGG is generated for each vertex reached by a window; next, the resulting set of candidate arcs is pruned by deleting all arcs that can be approximated with sequences of other arcs in the graph within the given tolerance; this latter step is performed by running a standard Dijkstra search from each vertex. For anisotropic meshes, in which the fan of arcs incident at some vertices might have gaps, an additional step is run to generate additional arcs (see \cite{Wang:2017} for details).
Shortest path queries on the DGG are resolved with a SLF/LLL heuristic \cite{Bertsekas:1998vt}, combined with a technique that restricts the neighbors to visit during the relax phase, based on the fact that the angle between consecutive edges in a shortest path cannot be smaller than $\pi-\arcsin \sqrt{\epsilon}$ (see \cite{Wang:2017} for details). The latter pruning technique is reminiscent of an analogous technique presented in \cite{Aleksandrov:2000gh}, based on Snell's law, for the case of weighted distances. Extensive experiments are presented in \cite{Wang:2017} on meshes up to 3M vertices and with accuracies up to $\epsilon = 0.001\%$; they make extensive comparison with the SVG: DGG beats it in terms of space occupancy, construction time, query time and accuracy, thus resulting globally better on all experiments. They also compare to other methods, including HM on an intrinsic Delaunay triangulation and FWP, which is taken as a reference:  DGG beats all such methods in terms of time performance and all approximated methods in terms of accuracy, resulting the best graph-based algorithm to date. 

Aiello et al.\ \cite{Aiello2015} adopt a hierarchical approach to support the computation of geodesic distances (not explicit paths) between any pair of vertices. 
Similarly to SVG and DGG, they aim at building a graph that contains many shortcuts between far vertices. 
They start from a hierarchical partition of the surface into patches: the uppermost level of the hierarchy is a Voronoi partition of the input shape, and each lower level is formed by Voronoi partitions of the patches in their upper level. Then a hierarchical graph is built, which contains all shortcuts among vertices on the boundary of each patch (i.e., one complete graph of boundary vertices per patch) and a complete graph per patch only at the lowest level of the hierarchy (\ie, both boundary and internal vertices of the patch are connected via shortcuts). 
The shortcuts in the graph are computed with an exact method (in the ICH/MMP family). 
The distance between a pair of vertices is found by a Dijkstra search that is pruned by visiting the graph in a hierarchical manner: search starts from the patch containing the source at the lowest level of the hierarchy, until it meet  its boundary; then it traverses either its sibling patch beyond the boundary, or shortcuts from patches in the upper levels, until a patch containing the target is found; the path is concluded by moving down the hierarchy  until the patch at the lowest level, which contains the target, is found. See \cite{Aiello2015} for details. The authors report quite slow preprocessing times for building the graph, but fast performances and empirically good approximations during queries.


\subsection{Local Methods}
\label{sec:local}

Local methods start from an initial path connecting two endpoints and aim to refine it to a geodesic path. These methods are usually  iterative, and work either updating one vertex at a time (\secref{iterative_point}) or the whole path at each iteration (\secref{iterative_path}). Point-to-point problems of this kind (PPGP) could in principle be solved as a byproduct of global methods that strive for SSGP (\secref{global}). Nevertheless, there is a number of algorithms that are specifically tailored for it. One good reason to prefer local methods to compute PPGP is that they are in general faster, easier to code, and require less memory due to the smaller domain they consider. On the negative side, they only converge to a local minimum, producing a locally optimal geodesic path at best. It follows that their ability to find the global optimum depends on the initialization, which may come from the Dijkstra's algorithm, from heuristics, or from user interaction.

For all methods, the input is a polyhedral mesh and a piece-wise linear path $P = p_0,p_1,\dots,p_n$ connecting a source $p_0$ with a destination $p_n$. The computational domain amounts to all the mesh vertices, edges and faces traversed by $P$.
The scalability of local methods does not depend on mesh size, but rather on the number of elements traversed by $P$, which in turn depends on its length and the local mesh density. A particular instance of the problem occurs when source and destination coincide (i.e., the path is a closed loop). Most of the local methods natively support this special case, but there are also methods which are specifically designed to shrink closed loops on a mesh~\cite{xin2012efficiently}. Shrinking loops is useful in a number of scenarios, for example to refine previously computed homology bases or the set of handles and loops of a mesh (see~\cite{dey2013efficient} and references therein).

\subsubsection{Local assessment of discrete geodesic paths}
\label{sec:local_criteria}
\begin{figure}[tb]
\centering
\includegraphics[width=0.9\linewidth]{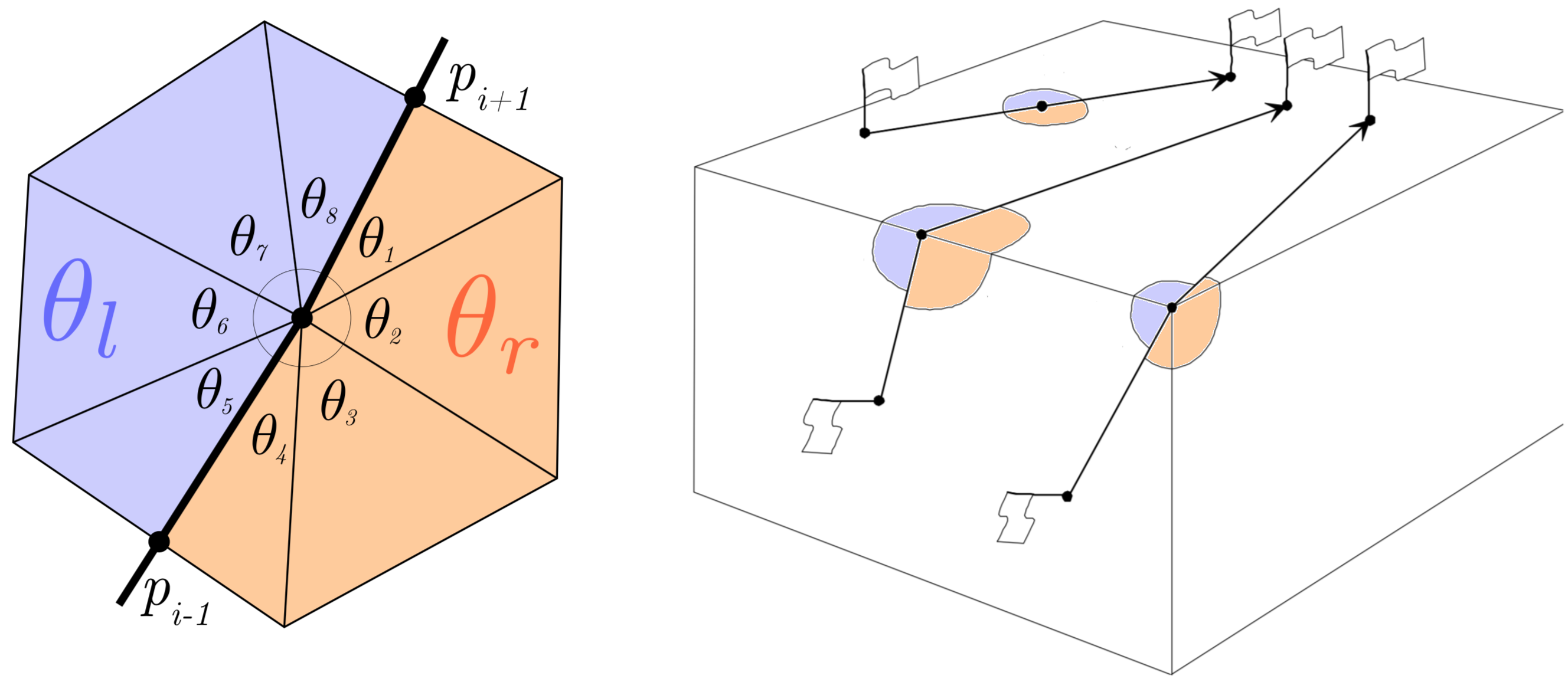}
\caption{Left: the total angle $\theta$ at point $p$ is divided into left angle $\theta_l = \sum_{i=5}^8\theta_i$ and right angle $\theta_r = \sum_{i=1}^4\theta_i$ by the path that traverses it. The path is: (i) locally shortest if $\theta_l \geq \pi$ and $\theta_r \geq \pi$; (ii) locally straightest if $\theta_l = \theta_r$. Right: three different paths connecting pairs of vertices in a cuboid. The path passing through the spherical vertex is locally straightest, but cannot be locally shortest as its total angle $\theta$ is less than $2\pi$. The path on a facet and the one crossing an edge are both locally shortest and straightest, because $\theta_l = \theta_r = \pi$. Right image from \cite{polthier1998straightest}.}
\label{fig:locally_shortest}
\end{figure}

Key to many methods is the ability to locally assess whether a path is a geodesic or not. To this end, 
the most important thing to remember is that while in the continuous the concepts of straightest and shortest paths coincide, in the discrete setting the two concepts are not always equivalent (\secref{ShortestVsStraightest}).
Polthier and Schmies~\cite{polthier1998straightest} observed that a path passing through a vertex $p$ divides its total angle $\theta$ into two components ($\theta_{l}$ and $\theta_{r}$) and defined geodesic paths in terms of these quantities (\figref{locally_shortest}). In particular, they state that a path is:
\begin{itemize}
\item[-] locally straightest, if $\theta_{l} = \theta_{r}$
\item[-] locally shortest, if $\theta_{l} \geq \pi$ and $\theta_{r} \geq \pi$
\end{itemize}
One can easily observe that if the path traverses a flat area or an edge (i.e., if $\theta$ is $2\pi$) the two definitions above are equivalent, hence the parallel with the smooth theory still holds. Considering non euclidean vertices, we obtain that: (i) a straightest path passing through a spherical vertex cannot be locally shortest (if $\theta_l + \theta_r < 2\pi$, then $\theta_l$ and $\theta_r$ cannot be both greater or equal than $\pi$); (ii) there are infinite shortest paths passing through a hyperbolic vertex (any solution of $\theta_{l} + \theta_{r} = \theta$, with $\theta_{l} \geq \pi$ and $\theta_{r} \geq \pi$ defines a shortest path).
Motivated by the problems of non existence and non uniqueness of locally shortest paths, Polthier and Schmies advocate the use of locally straightest paths as a way to trace geodesics on discrete surfaces, and propose both Euler and Runge-Kutta integration schemes based on them~\cite{polthier1998straightest,polthier1999geodesic}. The beauty of locally straightest geodesics is that they exist and are uniquely defined at any point of a polyhedral mesh. However, there is a price to pay for this. In particular, it must be observed that straightest geodesics do not converge to geodesic paths on smooth surfaces under mesh refinement~\cite{lieutier2009convergence,kumar2003geodesic}, and that locally straightest distances do not satisfy the triangular inequality, therefore the straightest geodesic distance is not a metric~\cite{LIU2017}. Last but not least, we remind the reader that these local geodesic criteria apply only to the Euclidean embedding, and do not extend to alternative metrics (e.g., weighted or anisotropic), thus limiting the applicability of the algorithms that are based on them.

\subsubsection{Iterative point update}
\label{sec:iterative_point}
Early methods to transform a general path into a geodesic path are heavily based on the local geodesic criteria introduced in \secref{local_criteria}. Given a path $P = p_0,p_1,\dots,p_n$, these methods work by iteratively flattening the mesh around each point $p_i$, and updating its position in order to locally straighten or shorten the path. 
The whole procedure is repeated until convergence, that is, until each point locally satisfies its geodesic criterion. Since at each iteration only one point is updated, these methods tend to require a high number of iterations to reach convergence. However, the computational cost of each iteration is usually low, as both the local flattening and the vertex update do not involve time consuming operations.


\begin{figure}[tb]
\centering
\includegraphics[width=\linewidth]{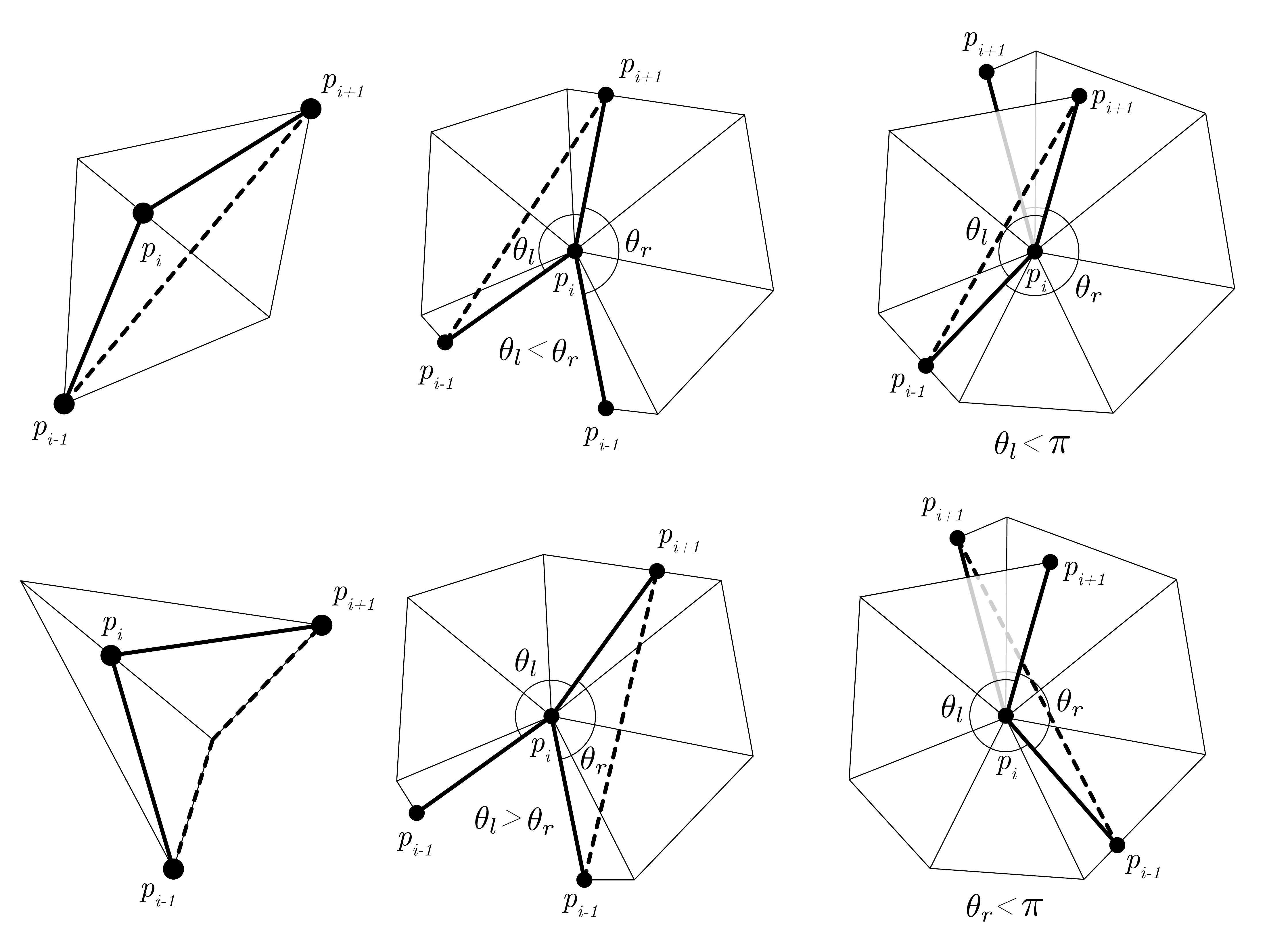}
\caption{The local operators used in \cite{wang2004cybertape} to locally shorten a path crossing: an edge (left column), a spherical vertex (middle column), and a hyperbolic vertex (right column). If straightening the line the path goes outside of the area span by the flattening, the path is forced to pass through the local border (see bottom left corner). Vertex operators are chosen by querying a look-up table indexed according to the vertex type (euclidean, spherical, hyperbolic) and the relation between $\theta_l$ and $\theta_r$.}
\label{fig:push_operators}
\end{figure}

CyberTape~\cite{wang2004cybertape} strives for locally shortest geodesics, flattening the mesh around each point $p_i$ and tracing a straight line going from $p_{i-1}$ to $p_{i+1}$. If the line crosses new edges or vertices, intersection points are added to the path.  Depending on where you are, there are multiple ways to locally shorten a path. \figref{push_operators} shows all the update  operators 
used by the algorithm to handle points at edges, spherical vertices, and hyperbolic vertices. Corner cases are also handled: if the local update would move the path outside the area spanned by the flattening, it is automatically constrained to adhere to the border of the local domain. This may happen any time the border of the flattening is non convex (see an example at the top right corner of \figref{push_operators}). 
Vertex operators are chosen by querying a look-up table indexed according to the type (euclidean, spherical, hyperbolic) and the relation between $\theta_l$ and $\theta_r$. The algorithm converges when all the points satisfy the geodesic criterion, which means that the path is locally shortest everywhere. It is interesting to note that, however, the final path 
may unnecessarily deviate from 
the input path $P$. This is because local operators move the path away from hyperbolic points even if locally shortest paths may traverse them, thus introducing an unnecessary deviation from $P$.
One year later, Mart\'\i nez et al. \cite{martinez2005computing} published a similar method, which substitutes the lookup table with a more faithful implementation of the ideas of \cite{polthier1998straightest}. Specifically, the path is not locally updated at hyperbolic vertices if both $\theta_l$ and $\theta_r$ are greater than $\pi$, as the path cannot be locally shortened (in~\cite{wang2004cybertape} the path was pushed away from the vertex anyways). This allows~\cite{martinez2005computing} not only to converge to a path that is locally shortest everywhere, but also to find the one that deviates from the input path $P$ only if strictly necessary.
Finally, Xin et al.~\cite{Xin:2007} strive for locally shortest geodesics, but rather than using the local angle criteria expressed in~\cite{polthier1998straightest}, they adopt an equivalent concept based on the Fermat principle, which states that light always follows the shortest optical path. As in~\cite{wang2004cybertape} and \cite{martinez2005computing}, the visibility-based method is guaranteed to converge to a path that is locally shortest everywhere in the sense of \cite{polthier1998straightest}.

\noindent\textbf{Remark: } A subtle problem affecting all the methods presented in this section is that they do not guarantee convergence to the closest local minimum. In other words, among all the locally shortest paths connecting source with destination, the output may not be the path which is closest to the input. Let us consider a path that traverses a spherical vertex in a way that perfectly halves its total angle (i.e., $\theta_l = \theta_r$). A locally shortest paths never traverses spherical vertices, therefore it should move sideways. Depending on which side it moves, the algorithm will converge to a different local minimum. There is no way to locally assess which side is best: algorithms use either consistent (e.g. always left) or randomized choices. In both cases, this may not lead to the closest solution. Since the initial path may traverse many spherical vertices, it is difficult to provide bounds on how much the geodesic path will deviate from it. 


\subsubsection{Iterative path update}
\label{sec:iterative_path}
\begin{figure}[htb]
\centering
\includegraphics[width=\linewidth]{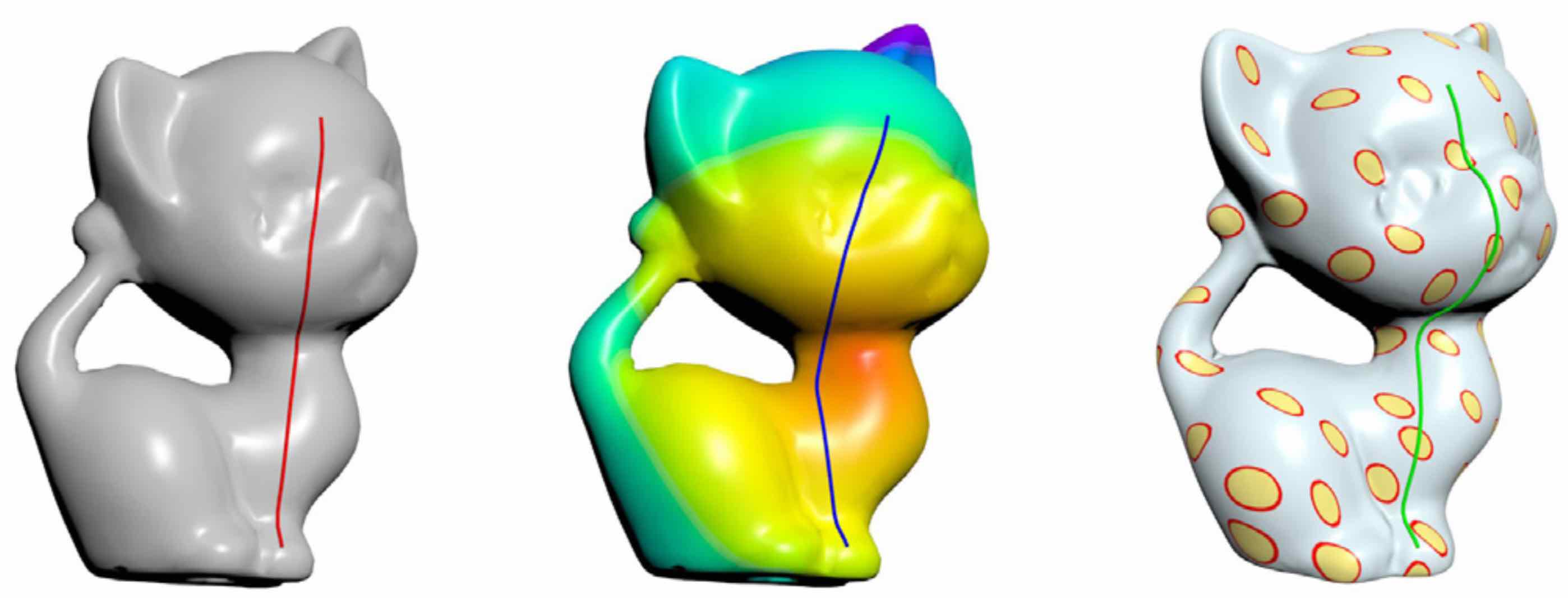}
\caption{The optimization based formulation used in \cite{LIU2017} allows to generate shortest paths with respect to the Euclidean (left), weighted (middle) and anisotropic (right) metrics. This is the only local method able to support such a variety of different metrics. From \cite{LIU2017}.}
\label{fig:local_liu}
\end{figure}

Methods that iteratively move one vertex at a time require many iterations to converge. A way to speed up convergence consists in solving a more complex problem at each iteration, thus requiring fewer iterations. 
Liu et al.~\cite{LIU2017} address the problem by using constrained optimization over the whole path. 
In their algorithm, each point $p_i$ is assigned to the mesh edge $e_i$ containing it. 
For points at vertices, only one of the incident edges is considered. 
The position of $p_i$ is expressed as a linear combination of the endpoints $v_0,v_1$ of $e_i$
$$
p_i = \lambda_i \cdot v_0 + (1-\lambda_i)\cdot v_1
$$
where $\lambda_i$ controls the linear interpolation between the two endpoints. The length of the whole path $P$
$$
\vert P \vert = \Vert p_0 - p_1 \Vert + \dots + \Vert p_{n-1} - p_n \Vert
$$
can now be expressed as a function of $\lambda_1,\dots,\lambda_n$, and minimized by forcing the vector of its partial derivatives to zero. This would in general lead to a linear system, but in order to keep each point on its edge and avoid extrapolation the problem is constrained, making sure that $\lambda_i \in [0,1]$, for $i=0,\dots,n$. The solution of this optimization problem gives the shortest path from $p_0$ to $p_n$, constrained on the edges traversed by the original path. Note that unknowns $\lambda_0, \lambda_n$ allow the solver to deviate from source and destination. One could in principle hard constrain them, but the authors opted for highly weighted soft constraints followed by snapping, as this ensures better numerical performances from the solver. In order to obtain the optimal path, the problem is solved multiple times, moving each $p_i$ from its current edge to some adjacent edge whenever the corresponding $\lambda_i$ goes to $0$ or $1$. The algorithm converges when none of the points $p_i$ wants to move to another edge. Since at each iteration the whole optimal path is re-computed from scratch, this algorithm requires far less iterations if compared to the methods discussed in \secref{iterative_point}.
In the paper, the authors show that this algorithm exhibits super-linear convergence, outperforming~\cite{martinez2005computing}.
An additional outcome of this formulation is that the Euclidean distance can be easily substituted with any other metric, such as weighted,
or anisotropic (\figref{local_liu}).
The same does not hold for local methods based on the criteria of Polthier and Schmies~\cite{polthier1998straightest}, which do not extend to non Euclidean embeddings (\secref{local_criteria}). 

\subsubsection{Graph-based methods}
\label{sec:graph_based_local}
Graph-based methods can also be designed to work in a local fashion. 
Similarly to the methods described in \secref{graph}, local graph-based methods find the shortest path connecting two points running Dijkstra's algorithm on a graph, which is initialized with the vertices and edges of the input mesh, and progressively refined adding Steiner points at mesh edges. 
Differently from global methods, Kanai and Suzuki \cite{kanai2000approximate} refine the graph only around the initial edge path connecting source to destination, thus producing a smaller graph and leading to a faster and better scalable implementation. 
Being based on Dijkstra's algorithm, weighted metrics can be easily incorporated by associating different weights to the edges in the graph. 
Lanthier et al.~\cite{Lanthier:2001} present a local heuristic method to refine a given path to an approximated locally shortest path. 
Their method is a local variant of the global one presented at the beginning of \secref{graph}, and can be used with both the fixed and the interval scheme. 
As for the other local methods, the path is restricted to a sub-domain composed of all the mesh faces traversed by the initial path. 
If the initial path passes through a mesh vertex, the authors propose a heuristic to decide whether the new path should pass it to the left or to the right. 
As already observed at the end of \secref{iterative_point}, these heuristic choices impact the output result, possibly resulting in a geodesic line that converged to a bad local minimum.
Also exact algorithms can be adapted to the PPGP problem. In the second part of their article, Surazhsky et al. \cite{Surazhsky:2005} proposed a local variant of their MMP implementation optimized to compute exact point to point shortest paths connecting a source $s$ and a target $t$. This variant is based upon an aggressive pruning strategy, which avoids propagating windows that have at least one point $p$ not satisfying the inequality
$$
L_s(p) + L_t(p) \leq U_{st} \:,
$$
where $L_s(p), L_t(p)$ are the lower bounds on the distances $d(s,p)$ and $d(s,t)$ measured on the mesh, respectively, while  $U_{st}$ is the upper bound on the path length, obtained with Dijkstra. The algorithm works in two steps: at the first step the pruning strategy is used to obtain a minimum amount on windows; in the second step the exact algorithm uses the windows to compute an exact shortest path between $s$ and $t$ (details in Section~\ref{sec:exact}) . As observed by the authors, alternative pruning schemes could potentially be used, possibly providing more performing implementations of point to point MMP.

\section{Methods for Geodesics Tracing}
\label{sec:tracing}
In GT, the problem of computing a geodesic curve on a domain is formulated as an \emph{initial value problem}. As discussed in \secref{smooth_theory}, a curve $\gamma$ on a surface patch $S$ is a geodesic if its geodesic curvature $\kappa_g$ is zero everywhere. Geodesic curvature vanishes when its projection on the binormal vector is zero, which in turn means that the curvature vector $\gamma\:''$ is parallel with the surface normal of $S$ at any point of $\gamma$. Given this premise, tracing the geodesic curve that starts from point $p \in S$ and proceeds as straight as possible in direction $v$ amounts to solving a second order ordinary differential equation, subject to the following boundary conditions~\cite{polthier1999geodesic}
\begin{align}
\gamma\:(0)   &= p\\
\gamma\:'(0)&= v \:.
\end{align}

There is a variety of methods that aim to solve this problem, which mostly differentiate to each other for the type of domain they admit as input. Early works in the field were designed to compute geodesic paths on parametric surfaces, such as NURBS and B\'ezier patches~\cite{patrikalakis1989offsets,maekawa1996computation}. This is not surprising, as this was the dominant representation for curved surfaces in the design and manufacturing industry. In this survey we do not provide details on these methods, partly because the article is focused on polyhedral meshes, but also because the reference literature dates back to '80s and '90s, and has already been covered in previous surveys and books. We point the reader to the book of Patrikalakis and Maekawa~\cite{patrikalakis2009shape} for further details on the topic.

\subsection{Tracing on polyhedral meshes}
Restricting to polyhedral surfaces, the topic was pionereed by Polthier and Schmies~\cite{polthier1998straightest,polthier1999geodesic}, who laid the theoretical bases for tracing geodesics on polyhedral surfaces (\secref{local_criteria}). 
In 
\cite{polthier1998straightest} they propose two alternative methods to integrate a field on a discrete mesh, one based on Euler integration, and the other based on the fourth order Runge-Kutta method.
It should be noted that tracing a path by numerical integration on a surface immersed in ambient space $\mathbb{R}^3$ requires an extra effort to keep the path on the surface. Some sort of projection method must be devised in order to guarantee it. The concept of straightest geodesics implicitly solves this problem, as the path can be integrated rotating around vertices by some angular distance, thus never leaving the polyhedral surface. This avoids the tedious and error prone implementation of geometric projection (e.g.  shooting rays or computing intersections with the surface), resulting in a more robuste software.

Kumar and colleagues~\cite{kumar2003geodesic} observed that the angle criterion proposed by Polthier and Schmies does not take into account surface normals, and suffers when there are abrupt orientation changes between adjacent facets. They proposed an alternative technique to trace a locally straightest geodesic on polyhedral surfaces. Given a point $p$ on a surface $S$, and a tracing direction $v$, they define the locally straightest path as the intersection between $S$ and the plane passing through $p$ and containing both $v$ and the surface normal at $p$. The procedure is repeated iteratively, until the path hits the boundary of the mesh, or a certain length is reached. 
The article also offers a very informative comparison between this strategy, the method of Polthier and Schmies~\cite{polthier1998straightest}, and a state of the art method for geodesic tracing on NURBS. The outcome of their experiments is that: (i)  discrete methods suffer from the bias introduced by the tessellation and are consistently less accurate than geodesic tracing on NURBS; (ii) considering the surface normal orientation increases the accuracy of discrete methods. In particular, the authors observed that while for simple developable surfaces all three methods perform equally well and converge on the smooth geodesic curve, for doubly curved non developable surfaces the method proposed in~\cite{kumar2003geodesic} is comparatively closer to NURBS-based methods, whereas the one of Polthier and Schmies~\cite{polthier1998straightest} converges to a different curve.

\begin{figure*}[h]
\centering
\includegraphics[width=\linewidth]{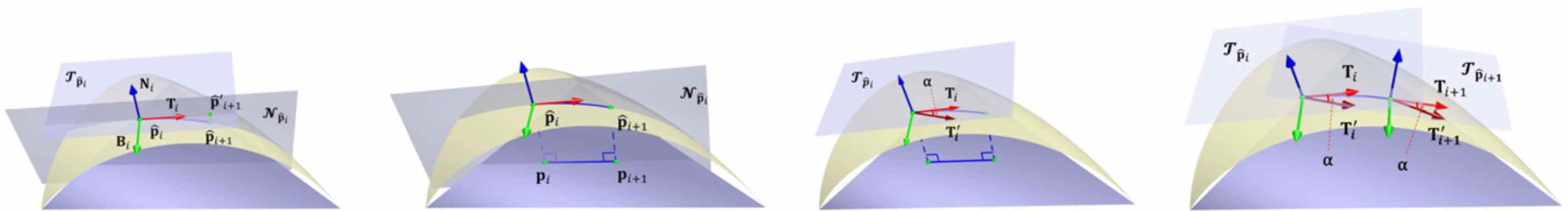}
\caption{Illustration of an iteration of the discrete tracing method described in~\cite{Cheng:2016}. Left: starting from point $p_i$ and tangent vector $\textbf{T}_i$, the subsequent point $p_{i+1}$ is computed on a B\'ezier patch (yellow) that interpolates a triangle facet (violet). Point $p_{i+1}$ is then projected back on the polyhedral surface (middle-left) and the tangent vector updated (middle-right). Tangent vector $\textbf{T}_{i+1}$ is parallel transported from $p_i$ to $p_{i+1}$. The hat is used to distinguish between entities living in the polyhedral mesh (no hat) from entities living in the B\'ezier patch (angular hat). Image taken from~\cite{Cheng:2016}.}
\label{fig:cheng_et_al}
\end{figure*}

\subsection{Hybrid approaches}
In order to fill the gap between continuous and discrete approaches, recent literature proposes hybrid solutions, where the discrete geodesic path $\gamma$ is computed on a piece-wise smooth meta mesh, and then projected onto the underlying polyhedral surface. In~\cite{Cheng:2016} the authors fit a three sided B\'ezier patch that interpolates both vertex positions and normals of each facet of a  triangular mesh, and use it as a domain to solve the geodesic tracing problem. 
Patches are glued side by side, guaranteeing $G^1$ continuity across common edges. This ensures that the path can seamlessly travel across mesh edges and vertices. Differently from~\cite{polthier1998straightest}, where path integration is intrinsically linked to the polyhedral surface, the communication between the mesh and the atlas of B\'ezier patches is provided by shooting rays along surface normals. While this is in general not very robust, the authors handle bad cases by shrinking the integration step size any time the new point does not project onto the same facet as the previous one.
The method works by iteratively repeating the following four steps (\figref{cheng_et_al}):
\begin{itemize}
\item Starting from a point $p_i$ and a tangent direction, the next point $p_{i+1}$ is computed in the B\'ezier patch by moving along the tangent direction with a user prescribed step size;
\item The new point $p_{i+1}$ is projected on the polyhedral surface;
\item A new tangent vector is numerically computed by solving a first order ODE with the Runge-Kutta method;
\item The new tangent vector is parallel transported from $p_i$ to $p_{i+1}$;
\end{itemize}
The algorithm stops when the path hits the boundary of the domain, or when some prescribed path length is reached. 
Compared to classical tracing methods on parametric surfaces, this method has the advantage to solve a first order ODE on the tangent vector, thus requiring only $G^1$ continuity. Classical methods solve a second order ODE on point positions, and therefore require a $G^2$ continuous domain~\cite{patrikalakis2009shape}. Furthermore, when compared to purely discrete methods such as \cite{polthier1998straightest,polthier1999geodesic}, this hybrid approach avoids the problems about existence and uniqueness of the solution around spherical and hyperbolic vertices (\secref{local_criteria}), and about convergence to the smooth geodesic curve under mesh refinement. 

%

\subsection{Tracing streamlines of a vector field}
\begin{figure}[h]
\centering
\includegraphics[width=\linewidth]{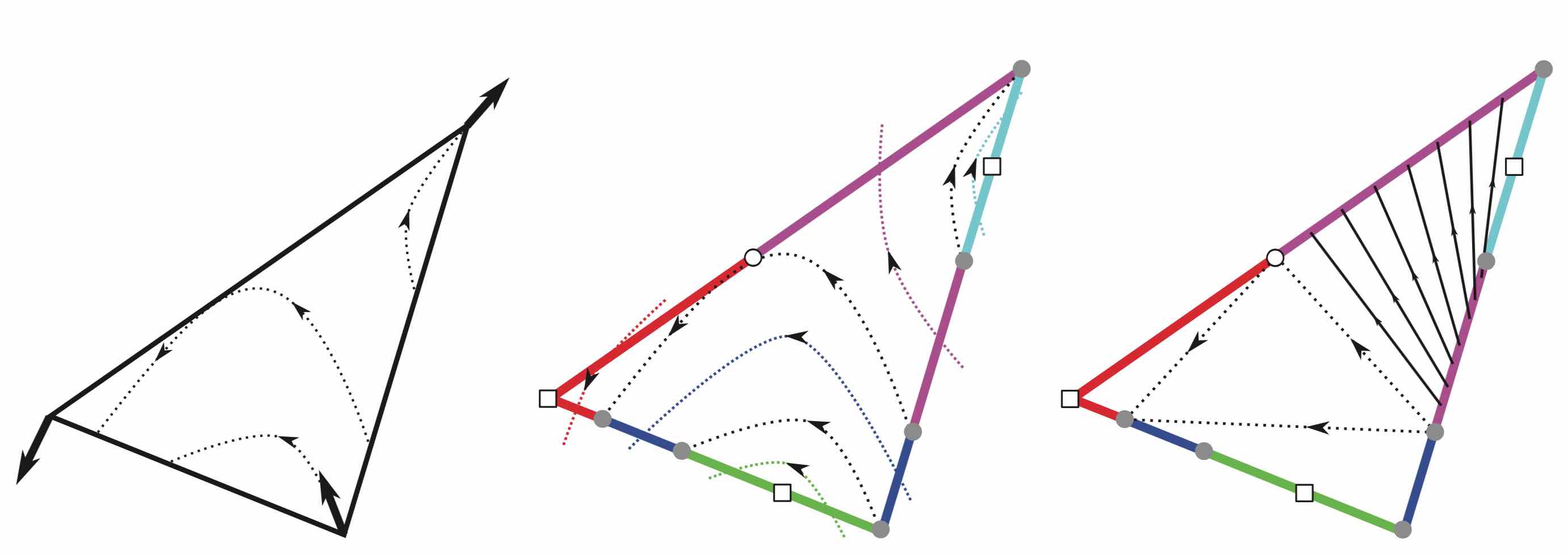}
\caption{EdgeMaps start from a vector field defined on vertices and linearly interpolated inside each triangle (left), and split triangle edges into slabs where the flow is either incoming or outcoming (middle). The information is eventually transformed into a linear map, which allows to easily pair entry and exit points (right). In~\cite{jadhav2012consistent} it was shown that for triangles there exist only 23 possible flow  configurations, therefore all possible maps can be efficiently encoded into a lookup table.}
\label{fig:edgemaps}
\end{figure}

Methods to trace streamlines of a general vector field can also be used for GT. Typically, these methods start from a field defined at the vertices of a discrete mesh, and extend it inside each facet by linear interpolation. 
For geodesic tracing, the starting point is a distance field, then transformed in a piece-wise linear vector field by computing its gradient (see~\cite{MLP18} for an overview of different techniques for gradient field computation).
Typical tracing techniques are based on numerical integration, performed with Euler or the fourth order Runge-Kutta method~\cite{polthier1998straightest}. These numerical approaches however are problematic, because the approximation error accumulates along the path, possibly producing drifting streamlines that fail to encode the correct global structure of the field. Recent research has shown thath cumulative error propagation can be avoided by integrating the path one facet at a time, and imposing proper constraints along edges. Bhatia and colleagues~\cite{bhatia2011edge} introduced the concept of EdgeMaps, replacing classical integration with linear maps defined at the edges of each facet. Edges are segmented into portions where the flow can be either inward or outward. Given an entry point to a facet, the corresponding exit point can be computed by approximating the flow linearly, which means that the curved paths of the vector field are replaced with straight lines that robustly indicate where the flow leaves the facet (\figref{edgemaps}). In~\cite{jadhav2012consistent} it was observed that there exist only 23 possible configurations for a triangle, therefore tracing the flow within a facet can be conveniently reduced to a map lookup. EdgeMaps do not completely substitute numerical integration, which is still used at initialization time to find the split points for edges. Furthermore, edge intervals guarantee no overlap of streamlines within the triangle only if their map is monotonic, which is true only up to numerical precision. Ray and Sokolov~\cite{ray2014robust} introduced the concept of stream meshes, which extends the idea of EdgeMaps. They completely avoid numerical integration, and rather rely on the field direction only along edges, providing an interval pairing system which makes heavy use of arbitrary precision floating point numbers to resolve precision issues. The result is a more robust tracing method, which is guaranteed to avoid crossing and collapse between close-by curves. It should be noted that, by avoiding integration within triangles, the lines may deviate from the guiding field, while still preserving the correct global behavior.

\subsection{Point to Point tracing}
\begin{figure}[h]
\centering
\includegraphics[width=\linewidth]{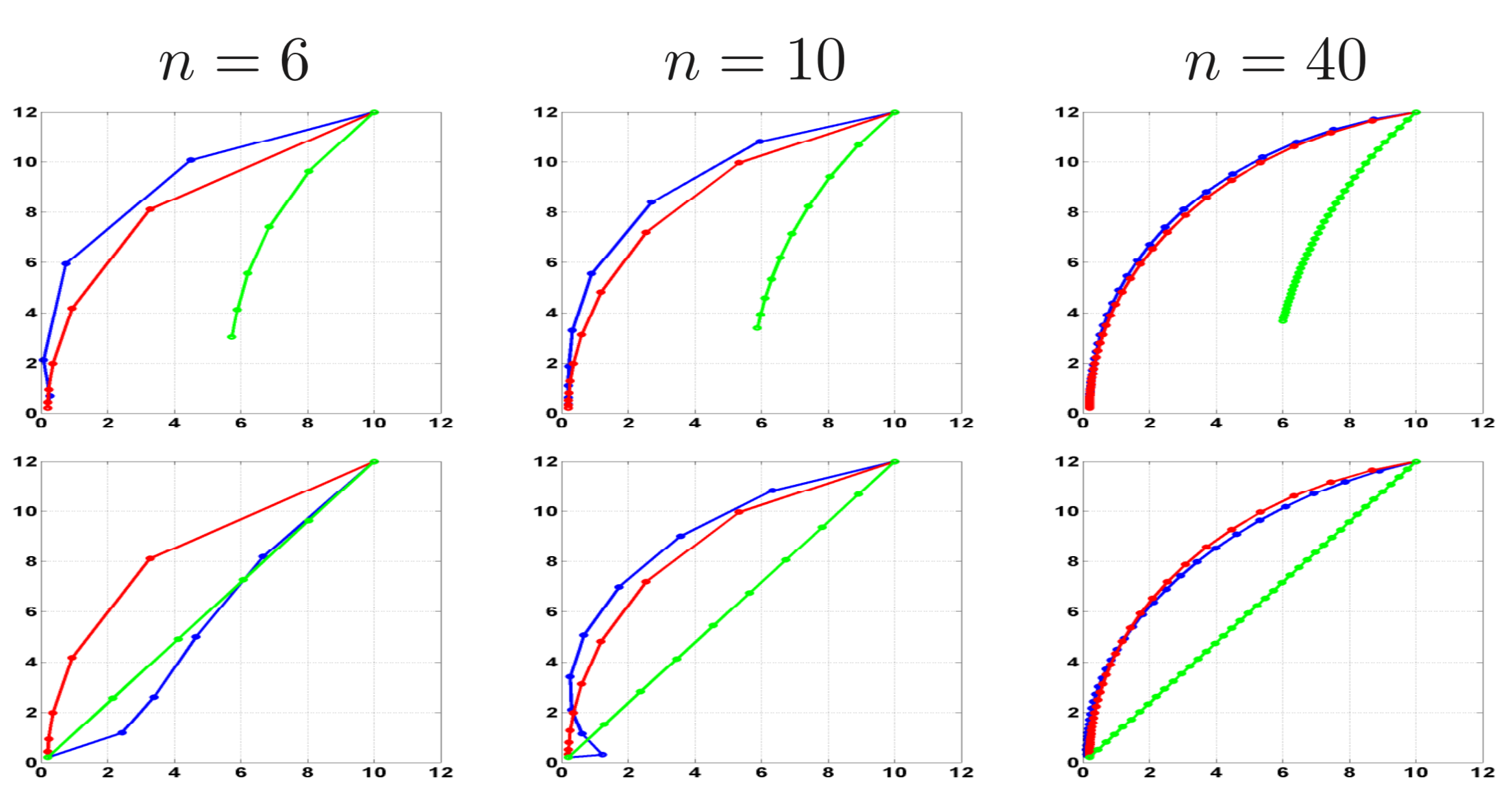}
\caption{Xie et al.~\cite{Xie:2013ju} propose two iterative methods that use geodesic tracing to solve a PPGP problem. Top line shows the shooting method. Bottom line shows the straightening method. The user chooses the number of sample points the path should be composed of ($n= 6, 10, 40$), and the algorithms iteratively update an initial path (green) to make it converge (blue) to the exact geodesic obtained from analytical expressions (red). All geodesics are traced in the 2D hyperbolic space. Image taken from~~\cite{Xie:2013ju}.}
\label{fig:gt_as_ppgp}
\end{figure}

Xie et al.~\cite{Xie:2013ju} offer a different view on geodesic tracing, which is employed to parallel transport deformations across surfaces, essentially solving a PPGP problem where the starting point $p_0$ represents the input shape, the final point $p_1$ the target shape, and the geodesic path $\gamma$ connecting them realizes a morphing between the two. The authors propose two alternative methods to compute $\gamma\:$: the \emph{shooting} method and the  \emph{straightening} method. Both methods are iterative, and generate a sequence of paths $\gamma\:^0, \dots, \gamma\:^n$ that progressively approximate the exact geodesic path connecting $p_0$ and $p_1$. 
The shooting method works by shooting a discrete geodesic path $\gamma\:^0$ with finite number of steps starting from $p_0$ along a random direction $v^0$, and to iteratively update the shooting direction $v$ in order to converge to the target path. Assuming all $\gamma\:^i$ are parameterized by arc length, the update amounts to computing the vector $w = p_1 - \gamma\:^i(1)$, parallel transport it from $\gamma\:^i(1)$ to $\gamma\:^i(0)$, and set $v^{i+1}= v^i + \delta w$. The process is repeated until convergence (i.e. until $\Vert w \Vert$ goes to zero). A visual example of the shooting procedure is given in \figref{gt_as_ppgp} (top line).
The straightening method works by initializing $\gamma\:^0$ as an arbitrary path connecting $p_0$ and $p_1$, and iteratively straightening it using a gradient descent approach, minimizing the energy function
$$
E(\gamma) = \int_0^1 \left\langle \frac{d}{dt} \gamma, \frac{d}{dt} \gamma \right\rangle dt \:.
$$
The inner product inside the integral is defined using the Riemannian metric of the shape space. Given the gradient $\nabla E$, the path is updated as $\gamma\:^{i+1} = \gamma\:^i - \delta \nabla E$, where $\delta > 0$ is a step size. A visual example of the straightening procedure is given in \figref{gt_as_ppgp} (bottom line).


\section{Meshing}
\label{sec:meshing}

A key issue that arises in all methods for computing geodesic distance (including those for computing the exact polyhedral distance) is that the surface must be meshed appropriately in order to be able to carry out the necessary computations.  The meshing of the surface will have an impact on things like numerical accuracy, guarantees about properties of the solution, as well as the time and memory needed to execute the algorithm.  In the context of PDE-based methods (\secref{fem}), the mesh is explicitly given to the algorithm and should ideally satisfy familiar properties demanded by finite element methods for hyperbolic and elliptic problems.  In the context of computational geometry methods (\secref{cg}), the mesh is implicitly constructed during the course of the algorithm -- viewed through this lens, many of the strategies discussed in \secref{cg} can be viewed as different strategies for keeping the size and quality of this mesh under control.  This viewpoint also helps to understand the trade offs between different classes of methods: PDE-based methods are often more efficient because they can generate a single, high-quality mesh ahead of time and re-use this mesh for many different distance queries; however, this mesh cannot be adapted to a particular distance query \apriori{}.  In contrast, methods from computational geometry can provide exact solutions to the polyhedral geodesic distance problem because the mesh implicitly constructed by the algorithm can be explicitly tailored to a particular query point, \ie, the surface is meshed exactly along characteristics of the solution; however, this same mesh cannot easily be re-used for subsequent queries, leading to a large number of caching and approximation schemes (as discussed in \secref{cg}).  In this section we consider the relationship between meshing and accurate geodesic distance computation.

\subsection{PDE-Based Methods}
\label{sec:MeshingPDEBasedMethods}

The two basic classes of methods we have studied (hyperbolic and elliptic) naturally lead to two different basic criteria on meshes needed to generate accurate results: hyperbolic methods generally perform better on \emph{acute} triangulations; elliptic methods perform better on \emph{Delaunay} triangulations.  On the whole, these criteria should be viewed as ``rules of thumb'' -- for instance, there are plenty of non-Delaunay meshes on which elliptic methods still perform well.  However, in specific cases using a mesh with the desired characteristics will provide absolute guarantees about properties of the solution, \eg, absence of spurious local minima.  Interestingly, acute meshes are a strict subset of Delaunay meshes; hence, meshes that work well for hyperbolic methods will also work quite well for elliptic methods.

\subsubsection{Hyperbolic Methods}
\label{sec:MeshingHyperbolicMethods}

As noted in \secref{FastMarching}, methods based on wavefront propagation must update distance values at vertices in an order that respects causality, \ie, the active node of smallest value must always be the one closest to the source.  On a triangulated surface, this property can be violated in the presence of obtuse angles -- consider, for instance, executing the fast marching method on the triangulation pictured in \figref{FMM_mesh}, \figloc{left}.  If we place a source at vertex 1, then the distance at vertices 2 and 3 will be computed (and finalized) prior to vertex 4, even though 4 is closer to 1; as a result, we will get an overestimate of the distance at 4.  This phenomenon cannot occur unless the mesh has obtuse angles; a solution, therefore, is to mesh the domain such that it has only acute (or at least nonobtuse) angles, \ie, interior angles no greater than \(\pi/2\).

Generating nonobtuse triangulations has traditionally been challenging, though recent years have seen significant developments.  Zamfirescu provides a nice survey of results in the planar setting~\cite{Zamfirescu:2013:STA}; Erten \& \"{U}ng\"{o}r developed some of the first practical algorithms for acute triangulations of planar regions, for which software is readily available~\cite{Erten:2009:QTL,Erten:2009:CTS}.  Obtaining acute triangulations of curved surfaces is more challenging: Burago \& Zallgaller~\cite{Burago:1960:PEN} show that an acute \emph{geodesic} triangulation can be obtained for any polyhedron (\ie, a triangulation where edges are required only to be geodesic arcs along the polyhedron), though this triangulation may not include the edges of the polyhedron itself.  Saraf~\cite{Saraf:2009:ANT} later proves the existence of acute triangulations which contain the edges by constructing compatible acute triangulations on each face; in either case, no upper bound is given on the number of triangles.  More recently, Maehara~\cite{Maehara:2011:OPA} give an upper bound on the size of the triangulation in terms of the local geometry of the given polyhedron, though no practical algorithm is provided.  In higher dimensions (\eg, for computing geodesics on triangulated regions of \(\mathbb{R}^3\)) even less is known, and in general acute triangulations may not even exist; Brandts~\etal~\cite{Brandts:2009:NSP} provide a survey.  The upshot of this discussion is that (at the time of writing of this article) there is no way to generate meshes that are guaranteed to satisfy the requirements of wavefront-based methods on polyhedral models, \ie, no absolute guarantee that the upwind ordering will be monotonic with respect to distance.  From here, there are essentially three options: (i) apply an \emph{unfolding} procedure, though as noted in \secref{FastMarching} this procedure may not always terminate; (ii) iteratively update the solution via multiple sweeps (\ala{Bornemann:2004:FED}), though at this point one may simply consider other optimization strategies such as ADMM~\cite{belyaev2015variational}, or (iii) simply ignore the fact that the solution may exhibit spurious local minima.  In practice, wavefront-based methods perform quite well even on meshes with a some mildly obtuse triangles, and for many applications (\eg, visualization) small violations of monotonicity do not present major problems.

\subsubsection{Elliptic Methods}
\label{MeshingEllipticMethods}

For geodesic distance methods based on solving elliptic questions, the key issue is no longer causality but rather satisfaction of a \emph{smooth maximum principle}.  In particular, let \(L \in \mathbb{R}^{|V| \times |V|}\) be a matrix representing any discrete Laplace operator, \ie, a weighted graph Laplacian with edge weights \(w_{ij} \in \mathbb{R}\).  We say that a function \(\phi \in \mathbb{R}^{|V|}\) is \emph{discrete harmonic} (with respect to \(L\)) if it is in the kernel of \(L\) (\ie, if \(L\phi = 0\)), in analogy with smooth harmonic functions, which sit in the kernel of the Laplace-Beltrami operator (\(\Delta\phi = 0\)).  In the smooth setting, the maximum principle says that a harmonic function has no local extrema.  In the discrete setting, the (local) maximum principle likewise asks that the value of a discrete harmonic function at any interior vertex be a convex combination of the neighboring values.  An elementary calculation shows that this property will be satisfied if and only if the weights \(w_{ij}\) associated with interior edges are positive; in particular, when \(L\) is the cotangent Laplacian positivity of edge weights is equivalent to the \emph{intrinsic Delaunay condition}
\[
   \alpha_{ij} + \beta_{ij} > \pi,
\]
where \(\alpha_{ij},\beta_{ij}\) are the interior angles opposite an interior edge \(ij\).  This condition is necessary (but not sufficient) to ensure a \emph{global} maximum principle---see Wardetzky~\etal{}~\cite{Wardetzky:2007:DLO} for further discussion.  Violation of the maximum principle can (as with causality) yield spurious local minima in the distance function computed by elliptic methods.

\begin{figure}
   \label{fig:IntrinsicDelaunay}
   \includegraphics[width=\columnwidth]{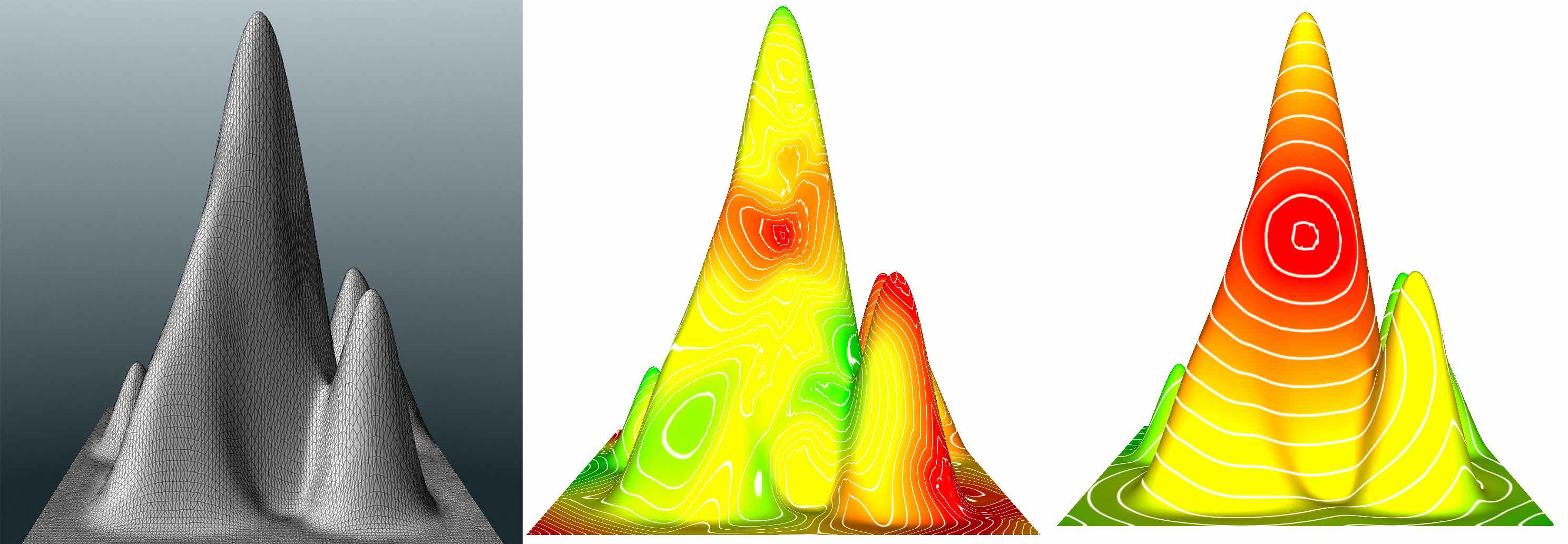}
   \caption{A mesh with many non-Delaunay triangles \figloc{(left)} may yield inaccurate solutions that depend on the cotan-Laplace operator \figloc{(right)}, but is easily remedied by using a Delaunay triangulation \figloc{(right)}.  Here we show the heat method as implemented in CGAL~\cite{Vaz:2018:HMI}, which uses an intrinsic Delaunay triangulation to avoid increasing the mesh size.
   }
\end{figure}

Fortunately, Delaunay triangulations are far easier to obtain than acute triangulations, and there is a large body of knowledge about both theoretical guarantees and practical algorithms -- see for instance the book by Cheng, Dey, and Shewchuk~\cite{Cheng:2012:DMG}.  Less is known about Delaunay triangulation of polyhedral surfaces, though even here there are a variety of practical algorithms, including Delaunay refinement~\cite{Cheng:2005:WDR,Dey:2010:PSR,Dey:2010:LDR} and edge splits~\cite{Dyer:2007:DMC,Liu:2015:ECS}.  Such methods can either exactly triangulate the input geometry, or reduce the size of the triangulation by allowing small modifications to the geometry.  Alternatively, one can construct an \emph{intrinsic Delaunay triangulation}~\cite{Bobenko:2005:DLB,Fisher:2006:ACI} where the new triangulation is represented only by the mesh connectivity and a collection of edge lengths; this data is sufficient to construct the Laplace operator \(L\), and allows greater flexibility in the definition of the triangulation while still preserving the input geometry exactly.  \figref{IntrinsicDelaunay} shows one example of how Delaunay triangulations can significantly improve the accuracy of elliptic methods -- in this case, without even increasing the size of the mesh.

\subsection{Computational Geometry Methods}
\label{sec:ComputationalGeometryMethods}

One appeal of the methods considered in \secref{cg} is that they can work directly on a given input without remeshing, though in reality these methods still effectively re-mesh the surface via the construction of windows (or other auxiliary data).  In other words, they incrementally modify a topological mesh data structure (often, just a standard half edge mesh~\cite{Surazhsky:2005}) that describes a new tessellation of the original mesh.  From this point of view, the basic approach of computational geometry methods is not entirely different from PDE-based methods: in either case, one must construct a mesh that supports accurate computation of a solution.  However there are some notable differences, namely (i) window-based methods remesh the domain \emph{during} computation of geodesic distance, rather than doing it ahead of time and (ii) the tessellation implicitly generated by this process depends on a particular choice of query point (or points).  In particular, the surface is effectively ``meshed along geodesics,'' \ie{}, it explicitly encodes geodesics from pseudosources to window boundaries (see for instance \figref{exact_window_structure}).  This strategy presents some clear trade offs.  On the one hand, it provides a great deal of accuracy since geodesics are directly encoded in the mesh -- from the PDE point of view, this is akin to meshing the domain along characteristics of the eikonal equation.  It is also therefore much easier to recover geodesic paths.  For this same reason, however, the mesh implicitly generated by a window-based scheme is difficult to re-use for computing distance to a different query point, which will induce an entirely different collection of windows.  This situation explains the large number of different variants on the basic MMP and CH algorithms that are needed to cache or otherwise accelerate new queries, as discussed in \secref{cg}.

One topic that has not received much attention is the impact of the \emph{input} triangulation on the performance of windowing algorithms.  One rather surprising observation is that the MMP algorithm sometimes runs much \emph{faster} on noisy inputs -- see in particular \cite[Figure 7]{Surazhsky:2005}.  Why this should be true, and more broadly, what kind of input triangulations lead to the best performance is a very interesting question, possibly leading to a hybrid approach where the mesh is first preprocessed (\eg, to be Delaunay or acute) prior to initiating the window insertion process.


\section{Evaluation}
\label{sec:eval}

Implementations of several methods reviewed in the previous sections have been released to the public domain, either by their authors, or by others.  Table~\ref{tab:implementations} provides a summary of the software we could find. 
Some such implementations come in the form of API's, some consist of source code for stand alone applications, and some others are just executables. 
Running such software in a consistent framework that allows us to compare their performances is not always possible. 
Several methods involve a setup phase (preprocessing) which is run just once per dataset, and is separated from the query phase; the costs of such two phases must be evaluated separately and this is not always possible in the implementations we collected. 
Moreover, as we already discussed previously, methods working in the smooth and in the discrete setting should not be compared directly, as they address different problems. 

\begin{table*}
\centering
\scriptsize
\begin{tabular}{|l|l|l|l|}
\hline
\textbf{Method} & \textbf{Implem.} & \textbf{URL} & \textbf{Type}\\
\hline
MMP & \cite{Surazhsky:2005} & \scriptsize{\url{http://hhoppe.com/proj/geodesics/}} & C++ API \\
\hline
Chen-Han & \cite{Chen:1990,xin2009improving} & \scriptsize{\url{https://doc.cgal.org/latest/Surface_mesh_shortest_path/index.html}} & C++ API\\
& & \scriptsize{\url{https://sites.google.com/site/xinshiqing/knowledge-share}} & C++ API\\
\hline
Heat & \cite{crane2013geodesics} & \scriptsize{\url{https://www.cs.cmu.edu/~kmcrane/Projects/HeatMethod/code.zip}} & C API$^{(*)}$\\
& & \scriptsize{\url{https://github.com/dgpdec}} & C++ API$^{(*)}$\\
& & \scriptsize{\url{https://github.com/mlivesu/cinolib}} & C++ API$^{(*)}$ \\
& & \scriptsize{\url{https://github.com/OpenGP/starlab}} & C++ App\\
& & \scriptsize{\url{http://www.numerical-tours.com/matlab/meshproc_7_geodesic_poisson/}} & Matlab \\
& & \scriptsize{\url{https://mathematica.stackexchange.com/questions/129207/how-to-estimate-geodesics-on-discrete-surfaces/}} & Mathematica\\
& & \scriptsize{\url{https://geometrycollective.github.io/geometry-processing-js/projects/geodesic-distance/index.html}} & JavaScript \\
\hline
Biharm. &\cite{Lipman:2010:BD} & \scriptsize{\url{https://pixl.cs.princeton.edu/pubs/Lipman_2010_BD/index.php}} & Matlab \\
\hline
VTP & \cite{Qin:2016} & \scriptsize{\url{https://github.com/YipengQin/VTP_source_code}} & C++ API \\
\hline
FWP & \cite{XuWLL015} & \scriptsize{\url{http://geometry.cs.ucl.ac.uk/projects/2015/fast_wavefront/paper_docs/FWPcode.zip}} & Executable\\
\hline
\end{tabular}
\small{(*) supports solving by back-substitution on a pre-factored matrix}
\caption{An overview of available implementations. For each method we report: method's name, reference to the original papers, link to the code, and its format.}
\label{tab:implementations}
\end{table*}

In the following two subsections, we provide just some comparisons between methods at the state-of-the-art, focusing on the SSGD problem and showing separate results for the PDE-based methods (smooth setting) and the CG methods (discrete setting).

%

\subsection{Computational geometry methods}
\label{sub:evalcg}
As we reviewed in \secref{cg}, most global methods for resolving the SSGD problem belong to two distinct classes: the exact methods that propagate windows (\secref{exact}), and the graph-based methods (\secref{graph}). 
In Table \ref{tab:VTPvsGraph} we compare the performances of the author's implementation of the VTP method \cite{Qin:2016}, which represents the state-of-the-art in the first class, and our basic implementation of Lanthier's method \cite{Lanthier:1997bg,Lanthier:2001} with the \emph{interval} scheme and SLF/LLL heuristic \cite{Bertsekas:1998vt} to compute shortest paths.\footnote{The state-of-the-art for graph-based methods is probably the DGG method \cite{Wang:2017}. Unfortunately, we only have an executable available for this method, which prevents us from profiling performances to compare with the other methods. The authors said that an improved version of DGG is under minor revision for publication in a journal, and they will release the source code as soon as it is accepted. We plan to include further experiments with their code in the final version.}

We have run experiments on meshes of increasing size, from a few thousands to about 14 million triangles; all meshes are manifold and watertight. 
Trials were run on a laptop equipped with 2.9 Ghz Intel Core i7 CPU and 16 GB RAM by using a single core.   
For each dataset and method, we did the setup once, and then ran queries on 100 random seeds.
In Table \ref{tab:VTPvsGraph}, for each method we report the time to setup and the average time for a single query.  
The VTP method, which is exact, provides the ground truth for the discrete problem. 
We compare the accuracy of the approximated graph-based method by measuring the average RMS error, evaluated as
\[ \mbox{Error} = \sqrt{\sum_{v\in V\setminus\{s\}}\left(\frac{\mbox{dist}_{VTP}(v)-\mbox{dist}_G(v)}{\mbox{dist}_{VTP}(v)}\right)^2}\]
where $V$ is the set of vertices, $s$ is the source, and $\mbox{dist}_{VTP}(v)$ and $\mbox{dist}_{G}(v)$ denote the distances of a vertex $v$ from $s$ computed with the two methods, respectively.
For Lanthier's method, there is an obvious trade-off between the time performance and the accuracy, which depends on the average number of Steiner points per edge.
Thus, to make an exhaustive comparison, we ran the experiments by varying such number between 1 and 10. 
Higher values were not considered due to their inefficiency: although the error becomes tiny, the time performance becomes not competitive with respect to the exact methods.

\begin{table*}
\centering
\scriptsize
\begin{tabular}{| l | r | r | r | r | r | r | r | r | r | r | r | r | r | r | r |}
\hline
\multicolumn{2}{|c|}{ } & \multicolumn{2}{c|}{VTP} & \multicolumn{3}{c|}{Lanthier's $S=1$} & \multicolumn{3}{c|}{Lanthier's $S=3$} & \multicolumn{3}{c|}{Lanthier's $S=5$} & \multicolumn{3}{c|}{Lanthier's $S=10$}\\
Dataset &  \multicolumn{1}{l|}{Size} & Setup & SSGD & Setup & SSGD & Error & Setup & SSGD & Error & Setup & SSGD & Error & Setup & SSGD & Error\\
\hline
Hand	&	8k	&	0.002	&	0.027	&	0.031	&	0.002	&	0.013	&	0.082	&	0.008	&	0.005	&	0.108	&	0.013	&	0.003	&	0.245	&	0.028	&	0.001	\\
Joker	&	27k	&	0.016	&	0.109	&	0.130	&	0.015	&	0.015	&	0.263	&	0.033	&	0.005	&	0.418	&	0.053	&	0.003	&	0.877	&	0.116	&	0.001	\\
Bunny	&	70k	&	0.042	&	0.359	&	0.250	&	0.035	&	0.022	&	0.626	&	0.086	&	0.007	&	0.857	&	0.133	&	0.004	&	1.984	&	0.294	&	0.001	\\
David	&	99k	&	0.055	&	0.490	&	0.394	&	0.050	&	0.014	&	0.839	&	0.116	&	0.005	&	1.312	&	0.185	&	0.003	&	2.842	&	0.432	&	0.001	\\ 
Armadillo	&	346k	&	0.392	&	2.942	&	1.876	&	0.215	&	0.015	&	3.569	&	0.500	&	0.006	&	5.441	&	0.754	&	0.003	&	11.945	&	1.667	&	0.001	\\ 
Gargoyle	&	700k	&	0.402	&	5.560	&	2.849	&	0.372	&	0.016	&	6.399	&	0.902	&	0.006	&	8.096	&	1.415	&	0.003	&	20.956	&	3.068	&	0.001	\\ 
Blade	&	2M	&	2.098	&	41.600	&	9.268	&	1.298	&	0.012	&	18.759	&	2.992	&	0.005	&	26.163	&	4.538	&	0.003	& \multicolumn{1}{c|}{(*)} & \multicolumn{1}{c|}{(*)} & \multicolumn{1}{c|}{(*)} \\ 
Happy	&	2.6M	&	2.968	&	37.449	&	12.777	&	2.071	&	0.015	&	23.994	&	4.903	&	0.006	&	45.299	&	8.021	&	0.004	& \multicolumn{1}{c|}{(*)} & \multicolumn{1}{c|}{(*)} &\multicolumn{1}{c|}{(*)} \\ 
Neptune	&	2M	&	3.600	&	66.840	&	17.564	&	2.660	&	0.015	&	35.956	&	6.649	&	0.006	& \multicolumn{1}{c|}{(*)} & \multicolumn{1}{c|}{(*)} &\multicolumn{1}{c|}{(*)} & \multicolumn{1}{c|}{(*)} & \multicolumn{1}{c|}{(*)} &\multicolumn{1}{c|}{(*)} \\
Lucy7M	&	14.5M	&	18.831	&	567.655	&	149.795	&	14.206	&	0.014 	&	\multicolumn{1}{c|}{(*)} & \multicolumn{1}{c|}{(*)} & \multicolumn{1}{c|}{(*)} & \multicolumn{1}{c|}{(*)} & \multicolumn{1}{c|}{(*)} & \multicolumn{1}{c|}{(*)} & \multicolumn{1}{c|}{(*)} & \multicolumn{1}{c|}{(*)} & \multicolumn{1}{c|}{(*)} \\
\hline
\end{tabular}
\caption{Comparison between VTP \protect\cite{Qin:2016} and Lanthier's \protect\cite{Lanthier:1997bg,Lanthier:2001} on various datasets. From the left: dataset name and size (number of triangles in thousands and millions); VTP times for setup and SSGD query; Lanthier's times for setup and SSGD query, and RMS error wrt ground truth. Experiments with Lanthier's are repeated with 1, 3, 5, and 10 Steiner points per edge (average). Times are in seconds; (*) means that the data structure exceeds physical RAM: the program runs, but with very slow times due to paging.}
\label{tab:VTPvsGraph}
\end{table*}

Overall, given a\com{ loaded} dataset, we assume that the setup is done once and the SSGD queries are performed many times, which is the common practice in various applications.
Therefore, as long as the setup remains within reasonable time bounds, query times are most relevant.

For the VTP method, the query times dominate (Table~\ref{tab:VTPvsGraph}). 
The setup phase of VTP takes negligible time since it only involves building the basic data structures to support the window propagation.
In spite of the theoretically superlinear complexity, the practical query times of VTP appear to increase linearly with the size of the dataset.
The only exception is the 
larger time consumed on the Blade dataset, which contains large flat regions.
Such performance degeneration may reveal the inherent challenge of the window-based methods: on a flat region, almost all the propagated windows are valid and there is little space for acceleration through redundancy removal.
In this case, all the window-based methods degenerate and behave similarly.

On the contrary, the setup phase dominates in Lanthier's method, which is devoted to building the graph containing both vertices and Steiner points. 
Note that with 3 or more Steiner points per edge, the cost of this phase is comparable or even more expensive than the whole cost of running VTP. 
The times for the setup phase increase linearly with the size of the mesh and quadratically with the number $S$ of Steiner points per edge (indeed, the size of the graph is quadratic in $S$, as shown in \figref{facegraph}).
Once the graph has been built, query times for meshes of moderate size are between one and almost two orders of magnitude faster than VTP, up to the case of 5 Steiner points per edge, while they become comparable or slower than VTP by using 10 Steiner points per edge. 
Speed is paid at the cost of some error, which is about $1.5\%$ with 1 Steiner point and about $0.1\%$ with 10 Steiner points; these figures are consistent in all experiments. 
Times for the query phase increase super-linearly, though: with large meshes, the performance is not competitive with respect to VTP for $S=5$ already.
With several Steiner points on large meshes 
the memory foortprint for the graph becomes too large, the computer starts paging and the performances collapse; on the largest mesh with over 14M triangles, collapse occurs with $S=3$ already. 
This limit could be probably improved with a better implementation, but the increase in the number nodes of the graph, which is quadratic in $S$, is inherent in the method.




\section{Concluding Remarks}
\label{sec:conc}

Algorithms for solving various versions of the geodesic problem are quite mature and ready to be used in applications.
We have reviewed methods at the state of the art, which are divided into two broad classes: PDE-based methods that resolve the problem in the smooth setting, and computational geometry methods that resolve the problem in the discrete setting. The latter class is subdivided further into methods that propagate windows, all stemming from the seminal MMP method, and graph methods that introduce one further level of discretization. We have also reviewed local methods that address single point-to-point queries and methods for geodesic tracing, as well as
discussed practical aspects such as the criteria that meshes should satisfy in order to squeeze the maximal power from each of the surveyed methods. 

Several methods, in both classes, support the separation of the setup phase from the query phase, where the former is performed once after loading the data, and the latter may be executed an arbitrary number of times on the same setup, to amortize the computational cost.
Depending on methods, this approach permits to deplete much of the computational burden during pre-processing, thus achieving high speed-ups at query time.
Overall, there is no best method for all purposes, as each class of methods has its own characteristics.

PDE-based methods are based on a global approach to the problem, and are largely aimed at solving the smooth geodesic problem, \ie, finding the best approximation of the true geodesic distance on the sampled surface.  These methods involve the resolution of large sparse linear systems and may benefit from modern numerical solvers, which in turn allows them to easily exploit features like parallelism.  Matrix pre-factorization can be done during the setup phase, thus reducing queries to fixed-order back-substitution, which is extremely fast. As with any finite element method, the approximation quality of PDE-based methods will be influenced by the quality of the input mesh, though as discussed in \secref{meshing} quality can often be improved via straightforward meshing strategies (such as use of an intrinsic Delaunay mesh).  Use of direct solvers can fail for very large meshes (due to limits on memory); in this case a practical solution is to switch to a large-scale iterative solver (such as multi-grid or preconditioned conjugate gradient).  Looking forward, recent work on computing \emph{localized} solutions to large linear system opens the door to using PDE-based methods for local geodesic queries (\eg, just at a single point) while still leveraging the benefits of direct solvers~\cite{herholz2017localized,Herholz:2018:cup}.  Pushing these methods toward higher accuracy via use of higher-order elements is also an interesting area for future work~\cite{DeGoes:2016im,Nguyen:2016:CFE}.

Computational geometry-based methods are based on local window propagation, and are most appropriate for solving the polyhedral geodesic problem.  Although the seminal algorithms in this category had poor performance due to a large number of windows, careful management of window construction in recent work has led to algorithms with a small memory footprint, thus supporting efficient geodesic queries on large datasets. On the other hand, such methods cannot easily separate the setup and query phases, largely due to the phenomenon discussed in \secref{meshing}: splitting edges into windows effectively constructs a different mesh of the surface for each geodesic distance query.  Thus, the amortized performance for repeated queries is not immediately as good as, say, PDE-based methods, leading to a focus in recent work on caching or otherwise pre-computing information useful for multiple queries~\cite{Ying:2013}.  A key question in future work is how to balance performance and accuracy with the typically high degree of implementation complexity associated with polyhedral algorithms.

Graph-based methods provide an approximated solution to the discrete problem and permit to trade-off accuracy for speed. They excel in numerical stability and they are very efficient in separating setup from query phase, thus achieving a relevant speed-up with respect to window propagation methods and times comparable with the fastest PDE-based methods. On the other hand, the memory footprint to store graphs may become relevant, and even prohibitive if high accuracy is required on large meshes, thus making such methods not competitive with the exact ones.



\begin{table*}
\footnotesize
\centering
\begin{tabular}{|l|c|c|c|c|c|c|}
\hline
\textbf{Source} & \textbf{Name} & \textbf{Domain} & \textbf{Class} & \textbf{Accuracy} & \textbf{Metrics} & \textbf{Problem}\\
\hline
\cite{Kimmel98computinggeodesic}	&	FM	&	S	&	D (wavefront)	&	A		&	E	&	SSGD\\
\cite{Belkin:2003:LED}			&		&	S	&	D (diffusion)	&	A		&	E	&	SSGD\\
\cite{Coifman:2006:DM}	&	Diffusion		&	S	&	D (diffusion)	&	A		&	E	&	SSGD\\	
\cite{Gurumoorthy:2009:SEF}		&		&	S	&	D (wavefront)	&	A		&	E	&	SSGD\\
\cite{Lipman:2010:BD}	&	Biharmonic	&	S	&	D (diffusion)	&	A		&	E	&	SSGD\\
\cite{Luo:2013ff}		&	Fast sweep	&	S	&	D (wavefront)	&	A		&	E	&	SSGD\\
\cite{crane2013geodesics,Crane:2017:HMG}&Heat&	S	&	D (diffusion)	&	A		&	E	&	SSGD\\
\cite{Sinha:2016:GWC}			&		&	S	&	D (wavefront)	&	A		&	E	&	SSGD\\
\cite{Mitchell:1987,Surazhsky:2005,Liu:2007,Liu:2013}   & MMP & P &  W	(priority queue) 	& 	E 		&	E 	& 	SSGD\\
\cite{Chen:1990,Kaneva:2000}     	& 	CH 	& 	P 	& 	W (FIFO queue)	& 	E		& 	E 	& 	SSGD\\
\cite{Surazhsky:2005} 			& Approx.MMP & P 	& 	W (priority queue) & E/A (local bounds) & E 	& 	SSGD\\
\cite{balasubramanian2009exact} 	& 		& 	P 	& 	W (FIFO queue) & 	E 		& 	E 	& 	APSP\\
\cite{xin2009improving}  			& 	ICH 	& 	P 	& 	W (priority queue)	& 	E 		& 	E 	& 	SSGD\\
\cite{Xin:2012} 					&	GTU& 	P 	& 	W (priority queue) & A (implicit bounds) & E 	& 	APSP\\
\cite{Ying:2014} 				& 	PCH & 	P 	& 	W (parallel) & 	E 		& 	E 	& 	SSGD\\
\cite{XuWLL015} 				& 	FWP	& 	P 	& 	W (bucket queue)	& 	E 		& 	E 	& 	SSGD\\
\cite{Qin:2016} 					& 	VTP & 	P 	& 	W (priority queue)			& 	E 		& 	E 	& 	SSGD\\
\cite{Lanthier:1997bg,Lanthier:2001}	& Lanthier's & 	P 	&G (Steiner) 		& A (additive bounds)& E, W	& SSSP\\
\cite{Mata:1997kt}				&		&	P	&G (vertex-to-vertex arcs) & A ($(1+\epsilon)$ bound) & E, W	& SSSP\\
\cite{aleksandrov1998varepsilon,Aleksandrov:2000gh,Aleksandrov:2005dq}	& Lanthier's var. & P & G (Steiner) & A ($(1+\epsilon)$ bound) & E, W	& SSSP\\
\cite{Schmidt:2006tt}				& 		& 	P 	&G  (Exp.map, local)  & 	A   		& 	E 	& 	SSGD\\
\cite{Campen:2013}				& STDV 	&	P 	& G (edge graph) 	& 	A   		& 	A 	& 	SSGD\\
\cite{Ying:2013}					& SVG 	& 	P 	& G (vertex-to-vertex arcs) & E/A  	& 	E, W & 	APSP\\
\cite{Wang:2017}				& DGG 	& 	P 	& G (vertex-to-vertex arcs) & A ($O(\epsilon)$ bound) & E, W & APSP\\
\cite{Lanthier:1997bg,Lanthier:2001} & 		& 	P 	& L (Steiner) 				& 	A 		& E, W 	& 	PPGP\\
\cite{kanai2000approximate}		& 		& 	P 	& L (Steiner) 		& 	A 		& 	E,W	& 	PPGP\\
\cite{wang2004cybertape} 		& CyberTape & P 	& L (locally shortest) & 	E (local) 	& 	E 	&	PPGP\\
\cite{martinez2005computing} 		& 		& 	P 	& L  (locally shortest)				& 	E (local) 	&	E 	& 	PPGP\\
\cite{Xin:2007}					& 		& 	P 	& L (Fermat's visibility) 	& 	E (local) 	& 	E 	& 	PPGP\\
\cite{LIU2017}					& 		& 	P 	& L (numerical opt.) 				& 	A 		& E, W, A 	& 	PPGP\\
\cite{polthier1998straightest,polthier1999geodesic} & & P & T (locally straightest)  				& 	A		& E 		& 	GT\\
\cite{kumar2003geodesic} 		& 		& 	P 	& T (locally straightest)				& 	A 		& E 		& 	GT\\
\cite{bhatia2011edge} 			& EdgeMaps & P 	& T (general fields) 				& 	A 		& E, W, A 	&	GT \\
\cite{ray2014robust} 				&		& 	P 	& T (general fields)				& 	A		& E, W, A 	& 	GT\\
\cite{Xie:2013ju} 				& 		& 	S 	& T (point to point)				& 	A		&  	E 	& 	GT/PPGP\\
\cite{Cheng:2016} 				& 		& 	S 	& T (B\'ezier patches)				& 	A		& E 		&	GT\\
\hline
\end{tabular}
\label{tab:summary}
\caption{Summary of the methods considered in this survey. For each method we report: the type of domain (S for sampling of a continuous surface or P for polyhedral mesh); its class (D for PDE-based methods, W for global methods based on window propagation, G for graph-based methods, L for local methods, T for tracing methods); its accuracy (E for exact, A for approximated); the distance metric is supports (E for Euclidean, W for weighted, A for anisotropic); The class of problems it addresses (PPGP, SSGD, APGD,GT); and its asymptotic complexity (if given by the authors).}
\end{table*}

%
\bibliographystyle{ACM-Reference-Format}
\bibliography{geodesics}

\end{document}